\documentclass[12pt,a4paper]{article}
\usepackage{jheppub}  
\usepackage{amssymb} 
\usepackage{amsmath}
\usepackage{mathtools}
\usepackage{amsfonts}    
\usepackage{dsfont}
\usepackage{pdfpages}
\usepackage{verbatim}
\usepackage{tensor}
\usepackage{mathrsfs}
\usepackage[mathscr]{euscript}
\usepackage{tikz}

\newcommand{\ba}{\begin{align}}

\newcommand{\be}{\begin{equation}}
\newcommand{\ee}{\end{equation}}
\def\bd{\begin{tikzpicture}}
\def\ed{\end{tikzpicture}}
\newcommand{\ket}[1]{| #1\rangle}

\allowdisplaybreaks[1]

\title{The Worldsheet Dual of the Symmetric Product CFT}

\author{Lorenz Eberhardt$^a$, Matthias R.\ Gaberdiel$^a$, and Rajesh Gopakumar$^b$} 
\affiliation{$^a$ Institut f\"ur Theoretische Physik, ETH Zurich, \\
\hspace*{0.3cm}CH-8093 Z\"urich, Switzerland}
\affiliation{$^b$ International Centre for Theoretical Sciences-TIFR, \\
\hspace*{0.3cm}Shivakote, Hesaraghatta Hobli, \\
\hspace*{0.3cm}Bengaluru North, India 560 089}
\emailAdd{eberhardtl@itp.phys.ethz.ch, \\ gaberdiel@itp.phys.ethz.ch, rajesh.gopakumar@icts.res.in}

\abstract{Superstring theory on ${\rm AdS}_3\times {\rm S}^3\times \mathbb{T}^4$ with the smallest amount of NS-NS flux (``$k=1$'') is shown to be dual to the spacetime CFT given by the large $N$ limit of the free symmetric product orbifold $\mathrm{Sym}^N(\mathbb{T}^4)$. 
To define the worldsheet theory at $k=1$, we employ the hybrid formalism in which the ${\rm AdS}_3\times {\rm S}^3$ part is described by the $\mathfrak{psu}(1,1|2)_1$ WZW model (which is well defined). 
Unlike the case for $k\geq2$, it turns out that the string spectrum at $k=1$ does {\it not} exhibit the long string continuum, and perfectly matches with the large $N$ limit of the symmetric product. 
We also demonstrate that the fusion rules of the symmetric orbifold are reproduced from the worldsheet perspective. 
Our proposal therefore affords a tractable worldsheet description of a tensionless limit in string theory, for which the dual CFT is also explicitly known.}

\begin{document}

\maketitle

%make math in all titles bold
\makeatletter
\g@addto@macro\bfseries{\boldmath}
\makeatother
%end code

%%%%%%%%%%%%%%%%%%%%%%%%%%%%%%%%%%%%%%%%%%%%%%%%%%%%%%%%%%%%%%%
\section{Introduction} \label{sec:intro}

Our understanding of the inner workings of the AdS/CFT correspondence 
has been hampered by the lack of adequate control over the string theoretic description of AdS spacetimes. This has effectively meant that much of our investigations and comparisons between the boundary CFTs  and the bulk (that go beyond supersymmetrically protected quantities) has been confined to the strong coupling regime, i.e.\ to Einstein gravity.

${\rm AdS}_3$ spacetimes have long held out hope in this regard since it is possible to consider these backgrounds without R-R flux turned on, and thus evade the complications necessitated by a worldsheet description that incorporates such fluxes. Indeed, building on a number of earlier works, 
Maldacena and Ooguri made a detailed study of bosonic strings in an ${\rm AdS}_3$ background with pure NS-NS three form flux. By looking in turn at the spectrum \cite{Maldacena:2000hw}, the one loop worldsheet partition function \cite{Maldacena:2000kv}, and at tree level three and four point functions \cite{Maldacena:2001km}, they established the consistency of the worldsheet CFT description despite it having some peculiar  features. 

These unfamiliar features had to do with the so-called long string phenomenon whereby one can have light long string excitations in ${\rm AdS}_3$ stabilised by the three form flux. This leads to there being a continuum in the perturbative spectrum (usually above a gap). The resulting singularities in the partition function and the correlators therefore have to be a feature of the dual 2d spacetime CFT as well. 

Our best guess for the 2d CFT dual to the $\mathcal{N}=(4,4)$ supersymmetric background  ${\rm AdS}_3\times {\rm S}^3\times {\cal M}_4$ is that it lies on the moduli space of the symmetric product orbifold CFT $\mathrm{Sym}^N({\cal M}_4)$ where ${\cal M}_4$ is either ${\rm K3}$ or $\mathbb{T}^4$ \cite{Maldacena:1997re}, see e.g.\ \cite{David:2002wn} for a review. (For the other maximally supersymmetric background with ${\cal M}_4={\rm S}^3\times {\rm S}^1$ there is good evidence that the dual is also a symmetric product CFT \cite{Eberhardt:2017pty}.) This is essentially a free CFT but it possesses a complicated 20 parameter space of moduli (in the case of ${\cal M}_4=\mathbb{T}^4$) where it is generically interacting. It is therefore not easy to identify the region in this moduli space which is dual to strings with pure NS-NS flux, with all its singular behaviour. Indeed, it has never even been clear which ${\rm AdS}_3$ background is described by the free symmetric product orbifold CFT itself \cite{Seiberg:1999xz}. Thus we are in the piquant position where we are fortunate to have an explicit worldsheet description of an AdS spacetime, but are unable to pinpoint the dual CFT description and make any meaningful comparisons (beyond BPS protected quantities). Conversely, we have a free CFT but do not know which string background it corresponds to. 
\smallskip

In this paper we propose a way out of this impasse. We provide strong evidence, from the worldsheet description of the superstring theory on ${\rm AdS}_3\times {\rm S}^3\times \mathbb{T}^4$ with the smallest amount of NS-NS flux (``$k=1$''), that it is dual to the spacetime CFT given by the large $N$ limit of the free symmetric product orbifold $\mathrm{Sym}^N(\mathbb{T}^4)$. This might seem like a surprising statement given the facts, recapped above, on the singularities of the NS-NS flux theory. It turns out that the $k=1$ theory is special in that it does not have the long string singularity and thus qualitatively differs from the backgrounds with more than one unit of flux. 
%Physically this is perhaps reasonable since the $k=1$ theory corresponds to a single NS5 brane which does not have a near horizon tubelike throat regime unlike the case for $k \ge 2$. 
In fact, our proposal rests on defining the worldsheet theory at $k=1$ since, as we will see, it is not obtained by a naive extrapolation of the $k\geq2$ theories. 

In previous works, we had already identified the novel 
nature of the small radius limit on ${\rm AdS}_3$ in the presence of NS-NS flux \cite{Gaberdiel:2017oqg, Ferreira:2017pgt}. Furthermore, in \cite{Gaberdiel:2018rqv} we had examined the spectrum  in a specific extrapolation, to $k=1$, of the generic spectrum of ${\rm AdS}_3\times {\rm S}^3\times \mathbb{T}^4$. This had enabled us to identify the spectrum of $\mathrm{Sym}^N(\mathbb{T}^4)$ as sitting at the bottom of the continuum that exists for $k \geq 2$. In this paper, we show, using the hybrid formulation of \cite{Berkovits:1999im}, that the perturbative string spectrum at $k=1$ actually {\it does not have any of the continuum states}, except the ones at the bottom,\footnote{It also does not have any discrete representations.} and therefore precisely agrees with that of the free symmetric product orbifold (in the large $N$ limit). Moreover, we also argue that the fusion rules of the worldsheet CFT are nontrivially consistent with those of the spacetime orbifold CFT. 

Our proposal thus answers the question of which ${\rm AdS}_3$ background is dual to the free symmetric product orbifold CFT. Namely, it is the near horizon geometry of $N$ fundamental strings and a single NS5 brane.\footnote{In principle, the background could also have a finite amount of D1 and D5 brane charge, which is not visible from our tree level analysis. We thank Ashoke Sen for raising this possibility.} Furthermore we have to take the large $N$ limit so that the string coupling constant 
\be
g_s^2 \sim \frac{Q_5 \cdot {\rm vol}(\mathbb{T}^4)}{Q_1} =  \frac{{\rm vol}(\mathbb{T}^4)}{N}
\ee
is small. This answer also fits in with the picture in higher dimensional ${\rm AdS}$ spacetimes that the free CFT dual corresponds to a small radius or tensionless limit of the string background. Here $k=1$ corresponds to the smallest radius theory that can be sensibly defined --- with a single unit of flux --- and should be viewed as the tensionless limit. However, unlike the higher dimensional cases, we now have a concrete worldsheet description of this tensionless limit! 

A possible relation of the WZW model (describing the NS-NS theory) to a symmetric orbifold CFT at $k=1$ was already mentioned in \cite{Seiberg:1999xz}, and more recently an alternative proposal to \cite{Gaberdiel:2018rqv} was made in \cite{Giribet:2018ada}. These proposals differ from our analysis in that the proposed dual CFT has a continuous spectrum in both cases, coming from noncompact directions in the symmetric product. We should also mention \cite{Argurio:2000tb} where a symmetric orbifold dual of a NS-NS background was proposed; however, their analysis was not specific to $k=1$.

As mentioned earlier, the novel worldsheet CFT that is proposed here for $k=1$ arises from the hybrid formulation of Berkovits, Vafa and Witten \cite{Berkovits:1999im}. 
The ${\rm AdS}_3\times {\rm S}^3$ \linebreak factor is described by a 
$\mathrm{PSU}(1,1|2)$ supergroup WZW model while the  $\mathbb{T}^4$ is topologically twisted.  
This description, in addition to having manifest spacetime SUSY,  overcomes a key limitation of the RNS formulation, which is not {\it a priori} well defined for $k <2$. The supergroup sigma model is, on the other hand, well-defined at $k=1$ but has special features to it. This includes the absence of the long string continuum which is essentially because a shortening condition (null vector) at $k=1$ removes the continuous representations of $\mathfrak{sl}(2,\mathds{R})_k$ (and their spectrally flowed versions), % which occur for $k >1$, 
except for the bottom of the continuum. We describe in detail the representations of the $\mathfrak{psu}(1,1|2)_1$ algebra that give rise to the string spectrum as well as their fusion rules. In doing so, we rely heavily on a free field construction for the  $\mathfrak{psu}(1,1|2)_1$ algebra. 

As we will see, our worldsheet description of this tensionless limit exhibits features that 
have often been ascribed to such limits. In particular, we will see signatures that the theory is a topological string. This is reflected in the fact that only short representations of the worldsheet 
CFT contribute to the string spectrum, and that all the oscillator degrees of freedom coming from ${\rm AdS}_3 \times {\rm S}^3$ are removed by the ghost contributions. We also see this in the worldsheet partition function which can formally be viewed as a set of delta function contributions from worldsheets which admit holomorphic maps. We plan to return to this aspect in future work.    
 \medskip
 
The plan of the paper is as follows:  In the next section we review some of the relevant aspects of both the RNS and the hybrid formulation of the worldsheet theory for ${\rm AdS}_3\times {\rm S}^3\times \mathbb{T}^4$ with NS-NS flux. In Section~\ref{sec:representations psu112} we describe the representations of the global superalgebra $\mathfrak{psu}(1,1|2)$ paying special attention to the short representations which will play a leading role. Section~\ref{sec:psu112 wzw model} describes the $\mathfrak{psu}(1,1|2)_1$ WZW theory and the complete set of affine representations (together with their spectrally flowed images) which close under fusion. We describe the free field theory realisation at level one, which makes the analysis tractable. A technical complication arises since this is a logarithmic CFT and therefore some of the representations are indecomposable. Taking these complications into account we obtain the (formally) modular invariant partition function. Section~\ref{sec:string theory partition function} involves taking this sigma model together with that on $\mathbb{T}^4$ and applying the physical state conditions to obtain the perturbative string spectrum. We also comment on how the tower of ``massless" higher spin fields, which generate the enhanced chiral algebra of the Higher Spin Square (HSS) \cite{Gaberdiel:2015mra}, arise. In Section~\ref{sec:fusion rules dual CFT} we explain how the fusion rules of the symmetric product CFT can be extracted from the worldsheet theory using the so-called $x$-basis. 
We find nontrivial agreement, thus providing further support for our identification of the spacetime CFT. There are a number of appendices which contain various technical details; in particular, we give a description of the free field realisation of $\mathfrak{psu}(1,1|2)_1$ in Appendix~\ref{app:free field representation}, and discuss the indecomposable nature of the representations in some detail, see Appendix~\ref{app:T details}. 

%%%%%%con%%%%%%%%%%%%%%%%%%%%%%%%%%%%%%%%%%%%%%%%%%%%%%%%%%%%%%%%%

\section{Strings on $\boldsymbol{\mathrm{AdS}_3 \times \mathrm{S}^3 \times \mathbb{T}^4}$} \label{sec:strings on AdS3S3T4}

\subsection{The RNS formalism} \label{subsec:RNS}
String theory on $\mathrm{AdS}_3 \times \mathrm{S}^3 \times \mathbb{T}^4$ with pure NS-NS flux can be described in the RNS formalism by the supersymmetric WZW model \cite{Giveon:1998ns, Maldacena:2000hw, Maldacena:2000kv, Maldacena:2001km}, see also \cite{Israel:2003ry,Raju:2007uj,Ferreira:2017pgt}
\be 
\mathfrak{sl}(2,\mathds{R})^{(1)}_k \oplus \mathfrak{su}(2)_{k}^{(1)}\oplus\mathbb{T}^4 \ . 
\ee
Here, the last factor is to be understood as the supersymmetric sigma model on $\mathbb{T}^4$, and the superscript $(1)$ denotes the $\mathcal{N}=1$ superconformal affine algebra. In this description, $k \in \mathds{Z}_{\ge 0}$ is interpreted as the value of the NS-NS flux in the background.
Recall that one can decouple the free fermions from these current algebras, and that 
\begin{align} 
\mathfrak{sl}(2,\mathds{R})^{(1)}_k &\cong \mathfrak{sl}(2,\mathds{R})_{k+2} \oplus \text{3 free fermions}\ , \\
\mathfrak{su}(2)^{(1)}_k &\cong \mathfrak{su}(2)_{k-2} \oplus \text{3 free fermions}\ .
\end{align}
Unitarity requires $k \ge 2$ and so the description is only valid in this regime. To treat the $k=1$ theory, we shall use the hybrid formalism which remains well-defined even for $k=1$.

\subsection{The hybrid formalism} \label{subsec:hybrid formalism}
There exists an alternative formalism for strings on $\mathrm{AdS}_3 \times \mathrm{S}^3 \times \mathbb{T}^4$, usually called the hybrid formalism \cite{Berkovits:1999im}, in which spacetime supersymmetry is manifest. In this approach one replaces the factor $\mathrm{AdS}_3 \times \mathrm{S}^3$ with its superspace analogue, the supergroup $\mathrm{PSU}(1,1|2)$. This supergroup contains, as its maximal bosonic subgroup, the group $\mathrm{SU}(1,1) \times \mathrm{SU}(2) \cong \mathrm{AdS}_3 \times \mathrm{S}^3$. In addition, there are eight fermionic coordinates. $\mathrm{PSU}(1,1|2)$ has the local superisometry group $\mathrm{PSU}(1,1|2) \times \mathrm{PSU}(1,1|2)$ given by left and right multiplication, which corresponds in the dual CFT to the global part of the (small) 
$\mathcal{N}=(4,4)$ superconformal symmetry.

In the hybrid formalism the sigma-model on $\mathbb{T}^4$ is taken to be topologically twisted and hence contributes $c=0$ to the worldsheet theory. Furthermore, the ghosts of the RNS formalism get transmuted into the so-called $\rho\sigma$-ghost system. To impose the physical state conditions one views the worldsheet theory as an $\mathcal{N}=4$ topological string, and identifies the physical states with the cohomology of the corresponding twisted $\mathcal{N}=4$ algebra \cite{Berkovits:1999im}.\footnote{This is nothing particular to $\mathrm{AdS}_3$ backgrounds; one can also view the flat space superstring as an $\mathcal{N}=4$ topological string. The $\mathcal{N}=4$ cohomology (rather than only the $\mathcal{N}=2$ cohomology) restricts then to the small Hilbert space of the theory, in which the fields have a definite picture number.}

We should mention that originally one of the main motivations for the development of the hybrid formalism was 
the fact that it is conceptually straightforward to add R-R flux to the background \cite{Berkovits:1999im}. (See also \cite{Cho:2018nfn} for a recent attempt to add R-R flux using the RNS formalism.) For generic flux, the worldsheet theory is the sigma-model on $\mathrm{PSU}(1,1|2)$, for which the kinetic term and the WZ-term can have different coefficients. In fact, this sigma-model remains conformal since the dual Coxeter number  of $\mathfrak{psu}(1,1|2)$ vanishes \cite{Berkovits:1999im, Bershadsky:1999hk}. While the resulting string theory is quite complicated --- in particular, once R-R flux is switched on, the ghost fields couple non-trivially to the sigma-model fields and are no longer free --- a (non-chiral) current symmetry survives \cite{ Ashok:2009xx, Benichou:2010rk} even away from the pure NS-NS background, and this allows one to get the resulting theory at least under some control \cite{Eberhardt:2018exh,Eberhardt:2018vho}.
%Interestingly, the modulus on the worldsheet which turns on R-R flux is no longer present for $k=1$ (since it involves the primary field transforming in the adjoint representation of $\mathfrak{psu}(1,1|2)_1$, which in particular would contain the adjoint representation of $\mathfrak{su}(2)_1$). Thus, it is not clear whether R-R flux can be added to the theory as was possible for $k \ge 2$.

Here we will only consider the case of pure NS-NS flux, where the sigma-model on $\mathrm{PSU}(1,1|2)$ becomes a WZW model at level $k$, where $k$ is the same as the level of the RNS formalism, and thus corresponds to the amount of NS-NS flux in the background. The equivalence of this description to the RNS formalism was demonstrated (for low-lying states) in \cite{Troost:2011fd,Gaberdiel:2011vf, Gerigk:2012cq}, see also \cite{Gotz:2006qp}. For us the main reason for employing the hybrid formalism (relative to the RNS formalism) is that it remains well-defined for $k=1$. Indeed, as we shall explain in detail below, the WZW model of the supergroup 
$\mathrm{PSU}(1,1|2)$ continues to make sense at $k=1$ (although it behaves somewhat differently to the case with $k>1$ since at $k=1$ there is a null-vector at $h=2$ that fixes the value of the Casimir, see Section~\ref{sec:4.2} below). The main result of our paper is that the physical spectrum it gives rise to agrees precisely with that of the symmetric product orbifold theory of ${\mathbb T}^4$. Furthermore, the fusion rules of the worldsheet theory match rather non-trivially  those of the symmetric product theory.

\section{Representations of $\boldsymbol{\mathfrak{psu}(1,1|2)}$} \label{sec:representations psu112}

For the following, it will be important to study various aspects of the representation theory of $\mathfrak{psu}(1,1|2)_k$. We will denote the generators of the bosonic affine subalgebras $\mathfrak{sl}(2,\mathds{R})_k$ and $\mathfrak{su}(2)_k$ by $J^a_m$ and $K^a_m$, respectively, while the fermionic generators are labelled by $S^{\alpha\beta\gamma}_m$.  We shall work with the conventions that the commutators and anti-commutators are given by 
\begin{subequations}
\begin{align} 
[J^3_m,J^3_n]&=-\tfrac{1}{2}km\delta_{m+n,0}\ ,  \label{eq:psu112 commutation relations a}\\
[J^3_m,J^\pm_n]&=\pm J^\pm_{m+n}\ , \label{eq:psu112 commutation relations b}\\
[J^+_m,J^-_n]&=km\delta_{m+n,0}-2J^3_{m+n}\ , \label{eq:psu112 commutation relations c}\\
[K^3_m,K^3_n]&=\tfrac{1}{2}km\delta_{m+n,0}\ , \label{eq:psu112 commutation relations d}\\
[K^3_m,K^\pm_n]&=\pm K^\pm_{m+n}\ , \label{eq:psu112 commutation relations e}\\
[K^+_m,K^-_n]&=km\delta_{m+n,0}+2K^3_{m+n}\ , \label{eq:psu112 commutation relations f}\\
[J^a_m,S^{\alpha\beta\gamma}_n]&=\tfrac{1}{2}c_a\tensor{(\sigma^a)}{^\alpha_\mu} S^{\mu\beta\gamma}_{m+n}\ , \label{eq:psu112 commutation relations g}\\
[K^a_m,S^{\alpha\beta\gamma}_n]&=\tfrac{1}{2}\tensor{(\sigma^a)}{^\beta_\nu} S^{\alpha\nu\gamma}_{m+n}\, , \label{eq:psu112 commutation relations h}\\
 \{S^{\alpha\beta\gamma}_m,S^{\mu\nu\rho}_n\}&=km \epsilon^{\alpha\mu}\epsilon^{\beta\nu}\epsilon^{\gamma\rho}\delta_{m+n,0}-\epsilon^{\beta\nu}\epsilon^{\gamma\rho} c_a\tensor{\sigma}{_a^{\alpha\mu}} J^a_{m+n}+\epsilon^{\alpha\mu}\epsilon^{\gamma\rho} \tensor{\sigma}{_a^{\beta\nu}} K^a_{m+n}\ . \label{eq:psu112 commutation relations i}
\end{align}
\end{subequations}
Here, $\alpha,\beta,\dots$ are spinor indices and take values in $\{+,-\}$. The third spinor index of the supercharges encodes the transformation properties under the outer automorphism $\mathfrak{su}(2)$ of $\mathfrak{psu}(1,1|2)$. Furthermore, $a$ is an $\mathfrak{su}(2)$ adjoint index and takes values in $\{+,-,3\}$. It is raised and lowered by the standard $\mathfrak{su}(2)$-invariant form
\be 
\eta_{+-}=\eta_{-+}=\tfrac{1}{2}\ , \qquad \eta_{33}=1\ .
\ee 
The constant $c_a$ equals $-1$ for $a=-$, and $+1$ otherwise. Finally, the $\sigma$-matrices are explicitly given by 
\begin{align}
\tensor{(\sigma^-)}{^+_-}&=2\ , & \tensor{(\sigma^3)}{^-_-}&=-1\ , & \tensor{(\sigma^3)}{^+_+}&=1\ , & \tensor{(\sigma^+)}{^-_+}&=2\ , \\
\tensor{(\sigma_-)}{^{--}}&=1\ , & \tensor{(\sigma_3)}{^{-+}}&=1\ , & \tensor{(\sigma_3)}{^{+-}}&=1\ , & \tensor{(\sigma_+)}{^{++}}&=-1\ ,
\end{align}
while all the other components vanish. 

\subsection{Representations of $\mathfrak{sl}(2,\mathds{R})$} \label{subsec:representations sl2R}

To analyse the $\mathfrak{psu}(1,1|2)_1$ WZW model it will be necessary to develop first the representation theory of the global $\mathfrak{psu}(1,1|2)$-algebra, i.e.\ of the algebra of zero-modes of eq.~\eqref{eq:psu112 commutation relations a} -- \eqref{eq:psu112 commutation relations i}. %; this is the subject of this section. 
%
% of  \eqref{eq:psu112 commutation relations a} -- \eqref{eq:psu112 commutation relations i}. 
%The commutation relations of the global $\mathfrak{psu}(1,1|2)$-algebra can be read off from the zero-modes of \eqref{eq:psu112 commutation relations a} -- \eqref{eq:psu112 commutation relations i}.
%
The bosonic subalgebra of $\mathfrak{psu}(1,1|2)$ is $\mathfrak{sl}(2,\mathds{R}) \oplus \mathfrak{su}(2)$. While the representations of $\mathfrak{su}(2)$ that appear are finite-dimensional, representations of $\mathfrak{sl}(2,\mathds{R})$ are not, and we begin by reviewing them.
There are two kinds of representations of $\mathfrak{sl}(2,\mathds{R})$ that will be relevant for us \cite{Maldacena:2000hw}.
\begin{enumerate}
\item Discrete representations. These are representations of $\mathfrak{sl}(2,\mathds{R})$ that possess a lowest (highest) weight state. The representation is characterised by the $\mathfrak{sl}(2,\mathds{R})$-spin $j$ of the lowest (highest) weight state.
\item Continuous representations. The continuous representations of $\mathfrak{sl}(2,\mathds{R})$  do not contain  a highest or a lowest weight state. These representations are characterised by their Casimir $\mathcal{C}$, as well as the fractional part of the $J^3_0$-eigenvalues which we label by $\lambda \in \mathds{R}/\mathds{Z}$.
\end{enumerate}
More specifically, the continuous representations of $\mathfrak{sl}(2,\mathds{R})$ are defined via
\begin{subequations}\label{mbasis}
\begin{align}
J^+_0 \ket{m}&=\ket{m+1}\ , \label{eq:sl2R continuous representation a}\\
J^3_0 \ket{m}&=m \ket{m}\ , \label{eq:sl2R continuous representation b}\\
J^-_0 \ket{m}&=\big(m(m-1)+\mathcal{C}\big) \ket{m-1}\ . \label{eq:sl2R continuous representation c}
\end{align}
\end{subequations}
Here, $\mathcal{C}$ is the quadratic Casimir of the $\mathfrak{sl}(2,\mathds{R})$ representation, which in these conventions takes the form
\be 
\mathcal{C}=-J_0^3J_0^3+\frac{1}{2}\big(J_0^+J_0^-+J_0^-J_0^+\big)\ ,
\ee
while $m$ takes the values $m \in \mathds{Z}+\lambda$. Provided that the Casimir satisfies $\mathcal{C}\geq \lambda(1-\lambda)$ where we take $\lambda \in [0,1]$, these representations are unitary, see \cite{Sugiura,Kitaev:2017hnr} for useful reviews. (For the case of the discrete representations, the relevant condition is $j\geq 0$.)

The discrete representations can be found as subrepresentations of the continuous representations. It is convenient to parametrise the Casimir of the continuous representations by $j$ as well, i.e.\ to write 
\be 
\mathcal{C}=-j(j-1)=\frac{1}{4}-\left(j-\frac{1}{2}\right)^2\ , \label{eq:sl2 Casimir}
\ee
where $j \in \mathds{R} \cup \big(\tfrac{1}{2}+i \mathds{R}\big)$. In this notation, $j$ and $1-j$ parametrise the same continuous representation. For real $j$, we then see by virtue of the relation \eqref{eq:sl2R continuous representation c} that $J^-_0 \ket{j}=0$. Thus, the states $\ket{m}$ for which $m-j \in \mathds{Z}_{\ge 0}$ form a subrepresentation, which is isomorphic to a lowest weight discrete representation. 

We will denote the continuous representation by $\mathscr{C}^j_\lambda$ and the discrete representation by $\mathscr{D}^j_+$. There exist also highest-weight (rather than lowest-weight) discrete representations, which we will denote by $\mathscr{D}^j_-$; they are characterised by
\be 
J^+_0 \ket{j}=0\ , \qquad J^3_0 \ket{j}= - j \ket{j}\ .
\ee

In addition to these infinite-dimensional representations, there are also the usual finite-dimensional representations of $\mathfrak{sl}(2,\mathds{R})$, although they will not be part of the worldsheet spectrum. We will simply denote the $m$-dimensional representation by $\mathbf{m}$. Except for the trivial representation, these are non-unitary. Below, we will need the Clebsch-Gordan coefficients of the tensor product of $\mathscr{C}^j_\lambda$ and $\mathscr{D}^j_\pm$ with $\mathbf{2}$. An explicit calculation shows that 
\be 
\mathscr{C}^j_\lambda \otimes \mathbf{2} \cong \mathscr{C}^{j+\frac{1}{2}}_{\lambda+\frac{1}{2}} \oplus \mathscr{C}^{j-\frac{1}{2}}_{\lambda+\frac{1}{2}}\ , \qquad
\mathscr{D}^j_\pm \otimes \mathbf{2} \cong \mathscr{D}^{j+\frac{1}{2}}_\pm \oplus \mathscr{D}^{j-\frac{1}{2}}_\pm\ . \label{eq:sl2 Clebsch Gordan}
\ee
\smallskip

The representations of $\mathfrak{sl}(2,\mathds{R})$ that appear in the worldsheet spectrum of Maldacena \& Ooguri \cite{Maldacena:2000hw} lead to the standard unitary representations of the WZW model group ${\rm SL}(2,\mathds{R})$, and the associated Lie algebra $\mathfrak{sl}(2,\mathds{R})$ is identified with that of the spacetime M\"obius group. However, this does {\it not} imply that also the representations with respect to the 
{\it spacetime} M\"obius group are the same standard representations. In fact, from the WZW perspective, the worldsheet generator $J^3_0$ is identified with the compact Cartan generator $J^3_0 = L^{\rm WZW}_1 + L^{\rm WZW}_{-1}$ of the WZW group ${\rm SL}(2,\mathds{R})$, see the discussion around eq.~(8) of \cite{Maldacena:2000hw}. 
%This generator acts diagonally on the standard representations of $\mathfrak{sl}(2,\mathds{R})$.
On the other hand, with respect to the spacetime M\"obius group, the correct identification is $J^3_0 = L^{\rm M}_0$, which corresponds to a non-compact generator of the spacetime M\"obius group. 
%, they differ from those that are usually considered in the mathematics literature, see e.g.\ \cite{Sugiura}, in an important but subtle way. From that perspective, the generator that acts diagonally is identified with $J^3_0 = L_0$, the non-compact Cartan generator of the spacetime M\"obius group ${\rm SL}(2,\mathds{R})$, while for the usual unitary representations of ${\rm SL}(2,\mathds{R})$ it is the compact Cartan generator $\Lambda_0 = L_1 + L_{-1}$ that acts diagonally.\footnote{Apart from this fact, the structure of the Lie algebra representation is however the same, i.e., this difference is not visible from the viewpoint of the Lie algebra alone.}
These two generators are {\it not} conjugate to one another in ${\rm SL}(2,\mathds{R})$ --- they are only conjugate to one another in ${\rm SL}(2,\mathds{C})$ --- and as a consequence they lead to different representations of the group ${\rm SL}(2,\mathds{R})$. In fact, as we shall explain in Appendix~\ref{app:groupreps}, the individual worldsheet representations do not form representations of the spacetime M\"obius group ${\rm SL}(2,\mathds{R})$; instead one needs to combine all representations for a given value of the Casimir, thus leading to the so-called $x$-basis of \cite{Maldacena:2000hw}. This will play a crucial role later in Section~\ref{sec:fusion rules dual CFT}.

\subsection{Long representations of $\mathfrak{psu}(1,1|2)$} \label{subsec:long representations psu112}

Next we describe the representations of $\mathfrak{psu}(1,1|2)$. We first consider the 
long (typical) representations, which come in the form of continuous and discrete representations for the $\mathfrak{sl}(2,\mathds{R})$ subalgebra. Let us concentrate on the continuous case, since the discrete representations arise as subrepresentations.

The eight supercharges of $\mathfrak{psu}(1,1|2)$, i.e.\ the generators $S^{\alpha\beta\gamma}_0$, 
generate a Clifford module. We can find a highest-weight state of the supercharges which is annihilated by half of them. Let us assume that the highest weight state transforms in the representation $(\mathscr{C}^j_\lambda,\mathbf{n})$ with respect to the bosonic subalgebra $\mathfrak{sl}(2,\mathds{R}) \oplus \mathfrak{su}(2)$, where  ${\bf n}$ here refers to the dimensionality of the $\mathfrak{su}(2)$-representation. The supercharges transform in the bispinor representation $2\cdot (\mathbf{2},\mathbf{2})$ of the bosonic subalgebra. Thus we conclude that a typical multiplet takes the form:
\begingroup
\setlength{\tabcolsep}{-.7em}
\be 
\begin{tabular}{ccccccccc}
& & & & $(\mathscr{C}^{j}_\lambda,\mathbf{n})$ & & & & \\
& $(\mathscr{C}^{j+\frac{1}{2}}_{\lambda+\frac{1}{2}},\mathbf{n+1})$ & & $(\mathscr{C}^{j+\frac{1}{2}}_{\lambda+\frac{1}{2}},\mathbf{n-1})$ & & $(\mathscr{C}^{j-\frac{1}{2}}_{\lambda+\frac{1}{2}},\mathbf{n+1})$ & & $(\mathscr{C}^{j-\frac{1}{2}}_{\lambda+\frac{1}{2}},\mathbf{n-1})$ & \\
$(\mathscr{C}^{j+1}_{\lambda},\mathbf{n})$ & & $(\mathscr{C}^{j}_{\lambda},\mathbf{n+2})$& & $2 \cdot (\mathscr{C}^{j}_{\lambda},\mathbf{n})$& & $(\mathscr{C}^{j}_{\lambda},\mathbf{n-2})$ & & $(\mathscr{C}^{j-1}_{\lambda},\mathbf{n})$ \\
& $(\mathscr{C}^{j+\frac{1}{2}}_{\lambda+\frac{1}{2}},\mathbf{n+1})$ & & $(\mathscr{C}^{j+\frac{1}{2}}_{\lambda+\frac{1}{2}},\mathbf{n-1})$ & & $(\mathscr{C}^{j-\frac{1}{2}}_{\lambda+\frac{1}{2}},\mathbf{n+1})$ & & $(\mathscr{C}^{j-\frac{1}{2}}_{\lambda+\frac{1}{2}},\mathbf{n-1})$ & \\
& & & & $(\mathscr{C}^{j}_\lambda,\mathbf{n})$ & & & & \label{eq:long representation}
\end{tabular} 
\ee
Here, the top state is the highest weight state of the Clifford module, and the action of the supercharges moves between the different bosonic representations.

For the important cases of $\mathbf{n}=\mathbf{1}$ and $\mathbf{n}=\mathbf{2}$ some shortenings occur. For $\mathbf{n}=\mathbf{2}$, the representation involving $\mathbf{n-2}$ is absent, i.e.\ 
\setlength{\tabcolsep}{-.5em}
\be 
\begin{tabular}{ccccccccc}
& & & & $(\mathscr{C}^{j}_\lambda,\mathbf{2})$ & & & & \\
& $(\mathscr{C}^{j+\frac{1}{2}}_{\lambda+\frac{1}{2}},\mathbf{3})$ & & $(\mathscr{C}^{j+\frac{1}{2}}_{\lambda+\frac{1}{2}},\mathbf{1})$ & & $(\mathscr{C}^{j-\frac{1}{2}}_{\lambda+\frac{1}{2}},\mathbf{3})$ & & $(\mathscr{C}^{j-\frac{1}{2}}_{\lambda+\frac{1}{2}},\mathbf{1})$ & \\
$(\mathscr{C}^{j+1}_{\lambda},\mathbf{2})$ & & $(\mathscr{C}^{j}_{\lambda},\mathbf{4})$& & $2 \cdot (\mathscr{C}^{j}_{\lambda},\mathbf{2})$& & \hspace*{1.2cm}& & $(\mathscr{C}^{j-1}_{\lambda},\mathbf{2})$ \\
& $(\mathscr{C}^{j+\frac{1}{2}}_{\lambda+\frac{1}{2}},\mathbf{3})$ & & $(\mathscr{C}^{j+\frac{1}{2}}_{\lambda+\frac{1}{2}},\mathbf{1})$ & & $(\mathscr{C}^{j-\frac{1}{2}}_{\lambda+\frac{1}{2}},\mathbf{3})$ & & $(\mathscr{C}^{j-\frac{1}{2}}_{\lambda+\frac{1}{2}},\mathbf{1})$ & \\
& & & & $(\mathscr{C}^{j}_\lambda,\mathbf{2})$ & & & & \label{eq:long representation n2}
\end{tabular} 
\ee
while for $\mathbf{n}=\mathbf{1}$ even more representations are missing,
\setlength{\tabcolsep}{-.2em}
\be 
\begin{tabular}{ccccccc}
& & & $(\mathscr{C}^{j}_\lambda,\mathbf{1})$ & & \\
 & & $(\mathscr{C}^{j+\frac{1}{2}}_{\lambda+\frac{1}{2}},\mathbf{2})$ & & $(\mathscr{C}^{j-\frac{1}{2}}_{\lambda+\frac{1}{2}},\mathbf{2})$ & & \\
$(\mathscr{C}^{j+1}_{\lambda},\mathbf{1})$ & \hspace*{1cm} & $(\mathscr{C}^{j}_{\lambda},\mathbf{3})$& & $(\mathscr{C}^{j}_{\lambda},\mathbf{1})$&\hspace*{1cm} & $(\mathscr{C}^{j-1}_{\lambda},\mathbf{1})$ \\
 & & $(\mathscr{C}^{j+\frac{1}{2}}_{\lambda+\frac{1}{2}},\mathbf{2})$ & & $(\mathscr{C}^{j-\frac{1}{2}}_{\lambda+\frac{1}{2}},\mathbf{2})$ & & \\
 & & & $(\mathscr{C}^{j}_\lambda,\mathbf{1})$ & & \label{eq:long representation n1}
\end{tabular} 
\ee
All of these representations exist also in the discrete version; they can be obtained by replacing the continuous by the corresponding discrete representations, 
\be 
\mathscr{C}^j_\lambda \longrightarrow \mathscr{D}^j_\pm\ .
\ee
\endgroup

\subsection{Short representations of $\mathfrak{psu}(1,1|2)$} \label{subsec:short representations psu112}

Below we will be interested in the affine algebra of $\mathfrak{psu}(1,1|2)_k$ at level $k=1$. Then the 
$\mathfrak{su}(2)_k$ factor also has level $k=1$, and as a consequence, the affine highest weight states are only allowed to transform in the ${\bf n}={\bf 1}$ and ${\bf n}={\bf 2}$ representations of $\mathfrak{su}(2)$.\footnote{In this section we are discussing the representations of the finite-dimensional Lie superalgebra $\mathfrak{psu}(1,1|2)$. The affine highest weight states of the corresponding affine algebra will therefore transform in representations of this algebra. We will also see the shortening of the  $\mathfrak{psu}(1,1|2)$ representations at $k=1$, from the affine viewpoint, in the next section.} Thus it is clear that all of the long representations we have presented above are not allowed at $k=1$. Let us therefore look systematically for short multiplets. Specifically, we will consider shortening conditions for the multiplets \eqref{eq:long representation n2} and \eqref{eq:long representation n1}.

Starting with \eqref{eq:long representation n2}, we require that the two representations with a $\mathbf{3}$ in the second line are null. This will remove also all other representations that appear further below in the multiplet. Thus, the multiplet would reduce to
\be 
\begin{tabular}{ccc}
& $(\mathscr{C}^{j}_\lambda,\mathbf{2})$ & \\
$(\mathscr{C}^{j+\frac{1}{2}}_{\lambda+\frac{1}{2}},\mathbf{1})$ & & $(\mathscr{C}^{j-\frac{1}{2}}_{\lambda+\frac{1}{2}},\mathbf{1})$
\end{tabular} \label{eq:psu112 short multiplet}
\ee
Actually, we will see below that this requires $j=\tfrac{1}{2}$.
Similarly, for the multiplet \eqref{eq:long representation n1}, the only way to eliminate the representation involving the $\mathbf{3}$ is to require one of the representations in the second line to be null. This gives then the following two possibilities:
\be 
\begin{tabular}{ccc}
$(\mathscr{C}^{j}_\lambda,\mathbf{1})$ & & \\
& $(\mathscr{C}^{j-\frac{1}{2}}_{\lambda+\frac{1}{2}},\mathbf{2})$ & \\
& & $(\mathscr{C}^{j-1}_\lambda,\mathbf{1})$
\end{tabular}
\qquad\text{or}\qquad
\begin{tabular}{ccc}
& & $(\mathscr{C}^{j}_\lambda,\mathbf{1})$ \\
& $(\mathscr{C}^{j+\frac{1}{2}}_{\lambda+\frac{1}{2}},\mathbf{2})$ & \\
$(\mathscr{C}^{j+1}_\lambda,\mathbf{1})$ & &
\end{tabular}
\ee
However, after redefining $j \to j \pm \tfrac{1}{2}$ (and rearranging the picture),\footnote{`Rearranging' means here that we change which state we regard as the highest weight state of the Clifford module.} these representations become equivalent to \eqref{eq:psu112 short multiplet}. Thus, there is at most one such multiplet, and we shall describe it using the conventions of \eqref{eq:psu112 short multiplet}. The above discussion works similarly for the discrete case where we replace $\mathscr{C}^j_\lambda$ by $\mathscr{D}^j_\pm$.

Next, we want to analyse the conditions under which this shortening can happen. For the discrete case
$\mathscr{D}^j_+$, these multiplets are well-known in the context of superconformal field theories with small ${\cal N}=4$ superconformal symmetry, since $\mathfrak{psu}(1,1|2)$ is the global subalgebra of this superconformal algebra.  In particular, this algebra has the well-known BPS bound $h=j \ge \ell$, where $\ell$ is the $\mathfrak{su}(2)$-spin. In this context, the $\mathfrak{sl}(2,\mathds{R})$-spin is identified with the conformal weight. Thus, for a BPS-representation, we need $j=\ell=\tfrac{1}{2}$ in the discrete case. (Formally, $j=\ell=0$ is also possible, but this just corresponds to the lower representations in  
\eqref{eq:psu112 short multiplet} with $j=\ell=\tfrac{1}{2}$.)

Since there is no $\mathfrak{sl}(2,\mathds{R})$ highest weight state for the continuous representation, the analysis for the continuous case is a bit more involved. First, we note that the $\mathfrak{psu}(1,1|2)$-Casimir decomposes into its bosonic and its fermionic components as
\begin{align}
\mathcal{C}^{\mathfrak{psu}(1,1|2)}&=\mathcal{C}_\text{bos}^{\mathfrak{psu}(1,1|2)}+\mathcal{C}_\text{ferm}^{\mathfrak{psu}(1,1|2)}\ , \\
\mathcal{C}_\text{bos}^{\mathfrak{psu}(1,1|2)}&=\mathcal{C}^{\mathfrak{sl}(2,\mathds{R})}+\mathcal{C}^{\mathfrak{su}(2)}\ , \\
\mathcal{C}_\text{ferm}^{\mathfrak{psu}(1,1|2)}&=-\frac{1}{2} \epsilon_{\alpha\mu} \epsilon_{\beta\nu}\epsilon_{\gamma \rho} S_0^{\alpha\beta\gamma} S_0^{\mu\nu\rho}\ .
\end{align}
The fermionic component of the Casimir commutes by construction with the bosonic subalgebra. It is not difficult to compute its value on the different constituents of the short representation \eqref{eq:psu112 short multiplet}. For instance, on the representation $(\mathscr{C}^j_\lambda,\mathbf{2})$, we determine its value on the highest weight state of the $\mathfrak{su}(2)$-algebra, which we denote by $\ket{m,\uparrow}$. (Here $m$ labels the state in the $\mathfrak{sl}(2,\mathds{R})$ representation $\mathscr{C}^j_\lambda$, see eq.~(\ref{mbasis}).) We have
\begin{align}
\mathcal{C}_\text{ferm}^{\mathfrak{psu}(1,1|2)}\ket{m,\uparrow}&=-\frac{1}{2} \epsilon_{\alpha\mu} \epsilon_{\beta\nu}\epsilon_{\gamma \rho} S^{\alpha\beta\gamma}_0 S^{\mu\nu\rho}_0\ket{m,\uparrow} \\
&=-\frac{1}{2} \epsilon_{\alpha\mu} \epsilon_{\gamma \rho} \{S^{\alpha+\gamma}_0, S^{\mu-\rho}_0\} \ket{m,\uparrow} \\
&=-\frac{1}{2} \epsilon_{\alpha\mu} \epsilon_{\gamma \rho}\big(-\epsilon^{\gamma\rho} c_a\tensor{\sigma}{_a^{\alpha\mu}} \, J^a_0+\epsilon^{\alpha\mu}\epsilon^{\gamma\rho} \tensor{\sigma}{_a^{+-}} \, K^a_0\big)\ket{m,\uparrow} \\
&=-2 K^3_0 \ket{m,\uparrow} =- \ket{m,\uparrow}\ .
\end{align}
Thus, 
\be 
\mathcal{C}_\text{ferm}^{\mathfrak{psu}(1,1|2)}\big(\mathscr{C}^j_\lambda,\mathbf{2} \big)=-1\ , \qquad \mathcal{C}_\text{ferm}^{\mathfrak{psu}(1,1|2)}\big(\mathscr{C}^{j\pm \frac{1}{2}}_{\lambda+\frac{1}{2}},\mathbf{1} \big)=0\ ,
\ee
where the second equality follows by a similar computation. Since the complete $\mathfrak{psu}(1,1|2)$-Casimir must be equal on all the representations appearing in (\ref{eq:psu112 short multiplet}), we conclude that the $\mathfrak{sl}(2,\mathds{R})$-Casimir must satisfy
\be
\mathcal{C}^{\mathfrak{sl}(2,\mathds{R})} \bigl( \mathscr{C}^{j\pm \frac{1}{2}}_{\lambda+\frac{1}{2}}\bigr)  = \mathcal{C}^{\mathfrak{sl}(2,\mathds{R})} \bigl(\mathscr{C}^{j}_{\lambda}\bigr) - \frac{1}{4} \ . 
\ee
(Here we have used that the Casimir of $\mathfrak{su}(2)$ equals $\mathcal{C}^{\mathfrak{su}(2)} = 0$ on ${\bf n}=1$ and  $\mathcal{C}^{\mathfrak{su}(2)} = \frac{3}{4}$ on ${\bf n}=2$.) Together with \eqref{eq:sl2 Casimir}, this then implies that $j=\tfrac{1}{2}$, see the comment after eq.~(\ref{eq:psu112 short multiplet}) above. (Incidentally, this is also the same condition as for the discrete case.) Note that, as a consequence, 
\be 
\mathcal{C}^{\mathfrak{psu}(1,1|2)}=0 \label{eq:psu112 Casimir vanishing}
\ee
on these representations. 

This is the only condition  for the shortening to occur. The details of this short representation are spelled out in Appendix~\ref{app:short representation}. As we also explain there, the case $\lambda=\frac{1}{2}$ is special since then the $\mathfrak{sl}(2,\mathds{R})$ representation $\mathscr{C}^{j}_\lambda$ (with $j=\frac{1}{2}$) becomes indecomposable.

%%%%%%%%%%%%%%%%%%%%%%%%%%%%%%%%%%%%%%%%%%%%%%%%%%%%%%
\section{The $\boldsymbol{\mathfrak{psu}(1,1|2)_1}$ WZW model} \label{sec:psu112 wzw model}
This section is devoted to a detailed study of the $\mathfrak{psu}(1,1|2)_1$ WZW model. Our main aim is to show how to define a consistent CFT for this chiral algebra. We will discuss, in particular, the fusion rules and modular invariance. Subtleties appear due to the fact that this CFT is logarithmic.

\subsection{$\mathfrak{psu}(1,1|2)_k$ for $k \ge 2$} \label{sec:psu112k for kge2}
Let us first review the WZW model based on $\mathfrak{psu}(1,1|2)_k$ for $k \ge 2$. The super Wakimoto representation states the equivalence \cite{Bars:1990hx, Berkovits:1999im, Gotz:2006qp}
\begin{multline}
\mathfrak{psu}(1,1|2)_k \cong \mathfrak{sl}(2,\mathds{R})_{k+2} \oplus \mathfrak{su}(2)_{k-2} \\ \oplus \text{8 topologically twisted fermions in the $2 \cdot (\mathbf{2},\mathbf{2})$}\ . \label{eq:psu112 Wakimoto}
\end{multline}
This looks then similar to what one would obtain from the RNS  formulation upon rewriting it in GS-like language, i.e.\ applying the abstruse identity. %, see also \cite{}. 
Here, the 8 fermions transform in the $2 \cdot (\mathbf{2},\mathbf{2})$ with respect to the bosonic zero-mode algebra $\mathfrak{sl}(2,\mathds{R}) \oplus \mathfrak{su}(2) \subset \mathfrak{psu}(1,1|2)$. 
%Hence, representations can be build starting from $\mathfrak{sl}(2,\mathds{R})_{k+2} \oplus \mathfrak{su}(2)_{k-2}$-representations. 
The fermions will lead to an $2^{\frac{8}{2}}=16$-dimensional Clifford module, and there cannot be any shortenings in the Clifford module since the fermions are free. Thus only long representations of $\mathfrak{psu}(1,1|2)$ appear in the theory. Furthermore, for the continuous representations, every $\mathfrak{sl}(2,\mathds{R})$-Casimir $\mathcal{C} \ge \tfrac{1}{4}$ (corresponding to $j=\frac{1}{2} + is$ with $s$ real) is allowed in the spectrum; the corresponding states describe the continuum of long strings in the spectrum of string theory on $\mathrm{AdS}_3 \times \mathrm{S}^3 \times \mathbb{T}^4$ \cite{Maldacena:2000hw}.

\subsection{Representations of $\mathfrak{psu}(1,1|2)_1$}\label{sec:4.2}
Let us now consider the case of $k=1$. Then the equivalence \eqref{eq:psu112 Wakimoto} does not hold any longer since the affine $\mathfrak{su}(2)_{k-2}$ algebra has negative level $-1$, leading to an additional non-unitary factor. This is also reflected by the fact that the long representations lead to ${\bf n}={\bf 3}$ representations for $\mathfrak{su}(2)$, that are not allowed for $\mathfrak{su}(2)_1 \subset \mathfrak{psu}(1,1|2)_1$. (In particular, the ${\bf n}={\bf 3}$ representation is non-unitary at $\mathfrak{su}(2)_{1}$.) 

As we have explained before, there is a natural way around this problem: at $k=1$ we need to consider short representations of $\mathfrak{psu}(1,1|2)$ that do not involve the ${\bf n}={\bf 3}$ representation of $\mathfrak{su}(2)$. Such short representations exist, and they take the form of (\ref{eq:psu112 short multiplet}) with $j=\frac{1}{2}$ 
%This is in conflict with the representation theory of $\mathfrak{su}(2)_1$. The corresponding vertex operator algebra does \textit{not} possess an adjoint representation. 
%This is a sign that the considered representations \eqref{eq:long representation} -- \eqref{eq:long representation n1} do not exist in the $k=1$ case and we have to use different representations. 
\be \label{shortrep}
\begin{tabular}{ccc}
& $(\mathscr{C}^{\frac{1}{2}}_\lambda,\mathbf{2})$ & \\
$(\mathscr{C}^{1}_{\lambda+\frac{1}{2}},\mathbf{1})$ & & $(\mathscr{C}^{0}_{\lambda+\frac{1}{2}},\mathbf{1})$
\end{tabular} 
\ee
Since the shortening condition fixes $j$ to $j=\frac{1}{2}$, in particular also the Casimir is fixed. Thus the continuum of states (corresponding to arbitrary values of the Casimir for the continuous representations) is not allowed any longer, but only the bottom component (corresponding to $j=\frac{1}{2} + is$ with $s=0$) survives.  These $\mathfrak{psu}(1,1|2)$-representations can then be extended to consistent affine representations.

This is the main mechanism for how the problem with the RNS formalism at $k=1$ is circumvented in the hybrid description. In the RNS formalism only long representations of $\mathfrak{psu}(1,1|2)$ appear, since the fermions are free and transform in the adjoint representation of $\mathfrak{su}(2)$. This is reflected in the equivalence \eqref{eq:psu112 Wakimoto}. On the other hand, in the hybrid formalism it is possible to consider instead the short representations of $\mathfrak{psu}(1,1|2)$.\footnote{In \cite{Gaberdiel:2018rqv}, a proposal was made for how to make sense of the theory at $k=1$ in the RNS formalism. It was proposed there that the $\mathfrak{su}(2)_{-1}$-factor can be represented by symplectic bosons, which effectively cancel half of the fermions. This prescription yields essentially the same spectrum as the introduction of the short representations in the hybrid description.} 
The introduction of short representations has another drastic consequence: it makes the string spectrum significantly smaller than in the generic case. In particular, the final spectrum will seem to have effectively only four bosonic and fermionic oscillators on the worldsheet, instead of the usual eight oscillators.
\smallskip

We should mention that the structure of the representations can also be deduced from the null-vector of $\mathfrak{psu}(1,1|2)_1$. The generating null-vector may be taken to be the vector $K^+_{-1} K^+_{-1} |0\rangle = 0$, which sits in the same $\mathfrak{psu}(1,1|2)$ multiplet as the null-vector 
\be\label{null2}
\Bigl( L^{\mathfrak{psu}(1,1|2)}_{-2} - L^{\mathfrak{sl}(2,\mathds{R})}_{-2} - L^{\mathfrak{su}(2)}_{-2} \Bigr) |0\rangle = 0 \ . 
\ee
This equation just means that we have a conformal embedding 
\be
\mathfrak{sl}(2,\mathds{R})_1 \oplus \mathfrak{su}(2)_1 \subset \mathfrak{psu}(1,1|2)_1 \ , 
\ee
see also \cite{Bowcock:1999uy} for a related discussion. Evaluated on affine highest weight states, we therefore conclude that 
\be
\mathcal{C}^{\mathfrak{psu}(1,1|2)} = - \mathcal{C}^{\mathfrak{sl}(2,\mathds{R})} + \frac{1}{3} \, \mathcal{C}^{\mathfrak{su}(2)} \ , 
\ee
which together with the condition that the only possible $\mathfrak{su}(2)$ representations are $\mathbf{n}=\mathbf{1}$ and $\mathbf{n}=\mathbf{2}$, fixes the allowed representations. On the other hand, for $k\geq 2$, the null-vector, i.e.\ the analogue of (\ref{null2}),  appears at higher mode number (conformal dimension), and hence we do not get a constraint on the quadratic Casimir.
\smallskip

We shall denote the affine representations that are generated from the affine highest weights in \eqref{shortrep} by $\mathscr{F}_\lambda$. These representations will be the main focus of study. For $\lambda=\frac{1}{2}$, the affine representation is not irreducible (see below), and we also need the affine representations associated to \eqref{shortrep} where $\mathscr{C}^{j}_\lambda$ has been replaced by  $\mathscr{D}^j_\pm$ with $j=\frac{1}{2}$ (see also Appendix~\ref{app:short representation}); the corresponding affine representations will be denoted by $\mathscr{G}_\pm$. Finally, we also need the affine representation based on the trivial representation (i.e.~the vacuum representation), which is also consistent; it will be denoted by $\mathscr{L}$.

%\medskip
%
%In summary we conclude that only short representations of $\mathfrak{psu}(1,1|2)$ can be part of the string theory spectrum at $k=1$. Furthermore, these short representations have a unique Casimir and hence a unique conformal weight, namely zero by \eqref{eq:psu112 Casimir vanishing}. We stress that this representation content is very different from the $k \ge 2$ representation content. Not only are the relevant representations all short, also the `continuum' of states is missing. Usually, continuous representations have an arbitrary Casimir, whose momentum describes long strings. For $k=1$, we find that only the bottom of the continuum is allowed to survive.

\subsection{Spectral flow} \label{subsec:spectral flow}
$\mathfrak{psu}(1,1|2)_k$ possesses a spectral flow automorphism $\sigma$. On the bosonic subalgebra $\mathfrak{sl}(2,\mathds{R})_k \oplus \mathfrak{su}(2)_k$, it acts by a simultaneous spectral flow on both components. Explicitly, we have
\begin{subequations}
\begin{align}
\sigma^w (J^3_m)&=J^3_m+\tfrac{kw}{2}\delta_{m,0}\ , \label{eq:spectral flow a}\\
\sigma^w(J^\pm_m)&=J^\pm_{m\mp w}\ , \label{eq:spectral flow b}\\
\sigma^w(K^3_m)&=K^3_m+\tfrac{k w}{2}\delta_{m,0}\ , \label{eq:spectral flow c}\\
\sigma^w(K^\pm_m)&=K^\pm_{m \pm w}\ , \label{eq:spectral flow d}\\
\sigma^w(S^{\alpha\beta\gamma}_m)&=S^{\alpha\beta\gamma}_{m+\frac{1}{2}w(\beta-\alpha)}\ .\label{eq:spectral flow e}
\end{align}
\end{subequations}
In addition, the energy-momentum tensor transforms as 
\be 
\sigma^w(L_m)= L_m+w(K_m^3-J_m^3)\ . \label{eq:spectral flow f} 
\ee
Notice that the simultaneous spectral flow in $\mathfrak{sl}(2,\mathds{R})_k \oplus \mathfrak{su}(2)_k$ keeps the supercharges integer moded. As we shall see, it will be necessary to include also spectrally flowed representations into the theory so that the fusion rules close, see also \cite{Maldacena:2000hw}.\footnote{The necessity to add spectrally flowed representations of $\mathfrak{sl}(2,\mathds{R})$ was first noticed, using arguments based on modular invariance, in  \cite{Henningson:1991jc}.}  Thus, we are considering the set of representations (for $k=1$)
\be 
\sigma^w(\mathscr{F}_\lambda)\ , \qquad w \in \mathds{Z}\ .
\ee
As regards the discrete representations and the vacuum representation, we have in fact the identity
\be 
\sigma(\mathscr{L}) \cong \mathscr{G}_+\ , \qquad \sigma^{-1}(\mathscr{L}) \cong \mathscr{G}_-\ . \label{eq:spectral flow identification}
\ee
Thus, it suffices to consider the spectrally flowed versions of the vacuum,
\be 
\sigma^w(\mathscr{L})\ , \qquad w \in \mathds{Z}\ .
\ee
There is one final complication: since the CFT is actually logarithmic,\footnote{In fact, the same phenomenon also appears for $\mathfrak{psu}(1,1|2)_k$ %and even $\mathfrak{sl}(2,\mathds{R})_{k+2}$ 
with $k>1$, see \cite{Troost:2011fd,Gaberdiel:2011vf}.}
%This set of representations will turn out to still not be enough in order for the fusion rules to close. Besides spectrally flowed representations, also indecomposable modules appear, which renders the CFT logarithmic.
 additional (indecomposable) representations will appear. 
 This can already be seen at the zero-mode level, see also Appendix~\ref{app:short representation}: $\mathscr{C}^j_\lambda$ is not irreducible for $\lambda=j$ (since it contains $\mathscr{D}^j_+$ as a subrepresentation), but indecomposable. Hence, we expect that $\lambda=\tfrac{1}{2}$ will play a special role. In fact, it turns out that $\mathscr{F}_{1/2}$ is \textit{not} separately part of the spectrum. Instead, $\sigma(\mathscr{F}_{1/2})$, two copies of the representation $\mathscr{F}_{1/2}$, as well as $\sigma^{-1}(\mathscr{F}_{1/2})$ join up to form one indecomposable representation, which we denote by $\mathscr{T}$. Thus, the representations appearing in the spectrum are in fact
\be \label{finalspectrum}
\sigma^w(\mathscr{F}_\lambda)\ , \quad \lambda \ne \tfrac{1}{2} \qquad\text{and}\qquad \sigma^{w}(\mathscr{T})\ , \qquad w \in \mathds{Z}\ .
\ee
While the emergence of the indecomposable representation $\mathscr{T}$ leads to many technical complications, it will turn out that the resulting physical spectrum is largely unaffected by this subtlety, see also \cite{Troost:2011fd,Gaberdiel:2011vf}. Furthermore, for many considerations (in particular, for the analysis of the partition function) we may work with the so-called 
%
%this should not distract too much this has not much influence on physics, thus we have described the details of this indecomposable module in Appendix~\ref{app:T details}. Indeed, $\lambda \in \mathds{Z}+\tfrac{1}{2}$ means from a dual CFT point of view that we are considering states with half-integer conformal weight (not necessarily chiral). These states should certainly not behave differently from the other states. We will consider the 
\textit{Grothendieck ring} of modules, where modules related by short exact sequences are identified, i.e.\ 
\be \label{Grothendieck}
\mathscr{C} \sim \mathscr{A} \oplus \mathscr{B} \quad \Longleftrightarrow \quad 0 \longrightarrow \mathscr{A} \longrightarrow \mathscr{C} \longrightarrow \mathscr{B} \longrightarrow 0\ .
\ee
This equivalence relation therefore forgets the indecomposablity of modules. 
On this, level $\mathscr{T}$ becomes then equivalent to $\sigma(\mathscr{F}_{1/2})\oplus 2\cdot \mathscr{F}_{1/2}\oplus \sigma^{-1}(\mathscr{F}_{1/2})$. (Similarly, $\mathscr{F}_{1/2}$ becomes equivalent to $\mathscr{G}_+\oplus2\cdot \mathscr{L}\oplus \mathscr{G}_-$, see the end of Appendix~\ref{app:short representation}.) There is no material difference between the $\lambda \neq  \tfrac{1}{2}$ contributions and that for $\lambda = \tfrac{1}{2}$ in (\ref{finalspectrum}), except potentially for a factor of $4$ that will also be resolved below, see eq.~(\ref{C.8}). A more careful treatment of the indecomposable representations is given in Appendix~\ref{app:T details}.

\subsection{The fusion rules of $\mathfrak{psu}(1,1|2)_1$} \label{subsec:fusion rules}
Let us next discuss the fusion rules of the theory. For this, we use the well-tested conjecture that spectral flow respects fusion \cite{Gaberdiel:2001ny}. 
%(We will come back to this issue below in Section~\ref{sec:fusion rules dual CFT}.) 
More precisely, for two modules $\mathscr{A}$ and $\mathscr{B}$, we have
\be 
\sigma^{w_1}(\mathscr{A}) \times \sigma^{w_2}(\mathscr{B}) \cong \sigma^{w_1+w_2}(\mathscr{A} \times \mathscr{B})\ . \label{eq:fusion spectral flow}
\ee
In particular, since $\mathscr{L}$ is the identity of the fusion ring, this determines the fusion of $\sigma^w(\mathscr{L})$ with any representation. Furthermore, it follows that it is sufficient to compute $\mathscr{A} \times \mathscr{B}$ without worrying about spectral flow.

In order to motivate our ansatz for the fusion of $\mathscr{F}_\lambda$, we note that, on the level of the Grothendieck ring, we have 
\be \label{F12}
\mathscr{F}_{1/2} \sim \mathscr{G}_+ \oplus 2\cdot \mathscr{L}\oplus \mathscr{G}_- \cong \sigma(\mathscr{L}) \oplus 2 \cdot \mathscr{L} \oplus \sigma^{-1}(\mathscr{L}) \ ,
\ee
and hence
\begin{align}\label{1212fus}
\mathscr{F}_{1/2} \times \mathscr{F}_{1/2} &\sim \Big(\sigma(\mathscr{L}) \oplus 2\cdot \mathscr{L}\oplus \sigma^{-1}(\mathscr{L})\Big) \times \mathscr{F}_{1/2}\\
&\cong \sigma(\mathscr{F}_{1/2}) \oplus 2\cdot \mathscr{F}_{1/2}\oplus\sigma^{-1}(\mathscr{F}_{1/2})\ .
\end{align}
Assuming that the general structure is similar for generic $\lambda$, this then suggests that 
%One can generalize this to compute the fusion of $\mathscr{F}_\lambda$ with $\mathscr{F}_\mu$ 
%%on the level of the Grothendieck ring 
%as follows. In the naive tensor product of the two representations, the $\mathfrak{sl}(2,\mathds{R})$ $J^3_0$-eigenvalues take values in $\tfrac{1}{2}\mathds{Z}+\lambda+\mu$. This suggests that the fusion is additive in $\lambda$ and $\mu$, and we expect  that 
(on the level of the Grothendieck ring, i.e.~ignoring indecomposability issues) 
\be 
\mathscr{F}_{\lambda} \times \mathscr{F}_{\mu}=\sigma(\mathscr{F}_{\lambda+\mu+\frac{1}{2}}) \oplus 2\cdot \mathscr{F}_{\lambda+\mu+\frac{1}{2}} \oplus \sigma^{-1}(\mathscr{F}_{\lambda+\mu+\frac{1}{2}})\ , \label{eq:psu112 fusion rules Grothendieck}
\ee
where the dependence on $\lambda$ and $\mu$ follows by requiring that the $J^3_0$ eigenvalues add up correctly --- this requires that the right-hand-side must only depend on $\lambda+\mu$ --- together with the requirement that (\ref{eq:psu112 fusion rules Grothendieck}) reduces to (\ref{1212fus}) for $\lambda=\mu=\frac{1}{2}$. Note that the $J^3_0$ charges of the middle term differ by $1/2$ with respect to those on the left-hand-side; since only the fermionic generators of $\mathfrak{psu}(1,1|2)$ have half-integer charges, it follows that the middle term has opposite fermion number relative to the left-hand-side (and indeed opposite fermion number relative to the other two terms on the right-hand-side, since one unit of spectral flow shifts the $J^3_0$ eigenvalue by $\frac{1}{2}$, see eq.~(\ref{eq:spectral flow a})). In terms of fusion rules, this means that the middle term arises in the `odd' fusion rules, while the other two terms are part of the `even' fusion rules, see e.g.\ \cite{Sotkov:1986bv}. This will play an important role in Section~\ref{sec:6.2}.

\subsection{A free field construction and the full fusion rules} \label{subsec:free field representation}

We can in fact deduce these fusion rules (including the correct indecomposable structure, see Appendix~\ref{app:T details}), using a free field realisation of $\mathfrak{psu}(1,1|2)_1$. To start with, we have the free field constructions
\begin{align} 
\mathfrak{su}(2)_1 \oplus \mathfrak{u}(1) &\cong \text{2 complex fermions}\ , \\ 
\mathfrak{sl}(2,\mathds{R})_1 \oplus \mathfrak{u}(1) &\cong \text{2 pairs of symplectic bosons} \ .
\end{align}
The first equivalence is well-known: if we denote the two complex fermions by $\psi^\alpha$ and $\bar{\psi}^\alpha$ with $\alpha=\pm$, then the $\mathfrak{su}(2)_1 \oplus \mathfrak{u}(1)_V$ generators come from the bilinears $\psi^\alpha \bar{\psi}^\beta$.
The second equivalence was first discussed in \cite{Goddard:1987td} and is probably less familiar. Recall that a pair of symplectic bosons consists of the two fields $\xi$ and $\bar{\xi}$, whose modes satisfy the commutation relations
\be 
[\bar{\xi}_m,\xi_n]=\delta_{m+n,0}\ ,\qquad [\xi_m,\xi_n]=[\bar{\xi}_m,\bar{\xi}_n]=0\ .
\ee
(Thus, the fields are bosons of spin $\tfrac{1}{2}$.) Considering two such pairs $\xi^\alpha$ and $\bar{\xi}^\alpha$ with $\alpha=\pm$, the bilinears $\xi^\alpha \, \bar{\xi}^\beta$ generate the Lie algebra $\mathfrak{sl}(2,\mathds{R})_1 \oplus \mathfrak{u}(1)_U$.\footnote{This is exactly the same construction as described in \cite{Gaberdiel:2018rqv}, except that it was interpreted there in terms of $\mathfrak{sl}(2,\mathds{R})_1\cong \mathfrak{su}(2)_{-1}$.} 
If we consider in addition the (neutral) bilinear generators involving one fermion and one symplectic boson, i.e.\ the generators $\psi^\alpha \bar{\xi}^\beta$ and $\bar{\psi}^\alpha \xi^\beta$, we obtain eight supercharges. Altogether, we thus generate the superalgebra $\mathfrak{u}(1,1|2)_1$
\be 
\mathfrak{u}(1,1|2)_1 \cong \text{2 pairs of symplectic bosons and 2 complex fermions}\ .
\ee
In order to reduce this to $\mathfrak{psu}(1,1|2)_1$, we thus only need to quotient out by the two $\mathfrak{u}(1)$ currents $\mathfrak{u}(1)_U$ and $\mathfrak{u}(1)_V$, i.e.\ we have\footnote{More abstractly,  
while $\mathfrak{su}(1,1|2)$ is a subalgebra of $\mathfrak{u}(1,1|2)$, $\mathfrak{psu}(1,1|2)$ is obtained from $\mathfrak{su}(1,1|2)$ by quotienting out the ideal generated by the identity, i.e.\ $\mathfrak{su}(1,1|2)$ is a central extension of $\mathfrak{psu}(1,1|2)$.} 
\begin{align} 
\mathfrak{psu}(1,1|2)_1 &\cong \frac{\mathfrak{u}(1,1|2)_1}{\mathfrak{u}(1)_U \oplus \mathfrak{u}(1)_V} \\
&\cong \frac{\text{2 pairs of symplectic bosons and 2 complex fermions}}{\mathfrak{u}(1)_U \oplus \mathfrak{u}(1)_V}\ . \label{eq:free field representation}
\end{align}
The details and our precise conventions for the free fields are summarised in Appendix~\ref{subapp:free field realization explicit}. As expected, this free field construction only has short representations of $\mathfrak{psu}(1,1|2)_1$, since it makes use of only four fermions.

While representations of complex fermions are standard, the fusion rules of the symplectic boson theory were worked out in detail in \cite{Ridout:2010jk}. One should note that even though this is a free field construction, the fusion rules are highly non-trivial. (In particular, the symplectic boson theory is also a logarithmic CFT.)
Translating the fusion rules of the free fields leads then to the fusion rules of the $\mathfrak{psu}(1,1|2)_1$-theory, see Appendix~\ref{app:free field representation} 
%and yields (including the module $\mathscr{T}$) the full fusion rules
\begin{subequations}
\begin{align}
\mathscr{F}_{\lambda} \times \mathscr{F}_{\mu}&=\begin{cases}
\ \sigma^{-1}(\mathscr{F}_{\lambda+\mu+\frac{1}{2}}) \oplus 2\cdot \mathscr{F}_{\lambda+\mu+\frac{1}{2}} \oplus \sigma(\mathscr{F}_{\lambda+\mu+\frac{1}{2}})\ , & \lambda+\mu\ne 0\ , \\
\ \mathscr{T}\ , & \lambda+\mu=0\ ,
\end{cases} \label{eq:psu112 fusion rules a}\\
\mathscr{F}_\lambda \times \mathscr{T} &=\sigma^{-2}(\mathscr{F}_\lambda) \oplus 4\cdot \sigma^{-1}(\mathscr{F}_\lambda)\oplus 6\cdot \mathscr{F}_\lambda \oplus 4\cdot \sigma(\mathscr{F}_\lambda)\oplus \sigma^{2}(\mathscr{F}_\lambda)\ , \label{eq:psu112 fusion rules b}\\
\mathscr{T} \times \mathscr{T} &= \sigma^{-2}(\mathscr{T}) \oplus 4\cdot \sigma^{-1}(\mathscr{T})\oplus 6\cdot \mathscr{T}\oplus 4 \cdot \sigma(\mathscr{T}) \oplus \sigma^{2}(\mathscr{T})\ .\label{eq:psu112 fusion rules c}
\end{align}
\end{subequations}
In particular, this argument shows that the set of representations given in (\ref{finalspectrum}) closes under fusion. 
As we shall see below, the chiral fields of the dual CFT come from the representation $\sigma(\mathscr{T})$, whose fusion with itself indeed contains $\sigma(\mathscr{T})$ again.

One can also obtain these fusion rules from a Verlinde formula; this is explained in Appendix~\ref{subapp:verlinde}. 

\subsection{The partition function and modular invariance} \label{subsec:psu112 partition function}

Next we will demonstrate that these representations give rise to a modular invariant spectrum, thus making the $\mathfrak{psu}(1,1|2)_1$ model also well-defined on the torus. The relevant modular invariant is the `diagonal modular invariant' with spectrum 
\be 
\mathcal{H} \cong  \bigoplus_{w \in \mathds{Z}}\ \int\limits_{[0,1) \setminus\{\frac{1}{2}\}}\hspace{-.86cm}\boldsymbol{\oplus}\, \mathrm{d}\lambda \ \sigma^w \big(\mathscr{F}_\lambda\big) \otimes \overline{\sigma^w\big(\mathscr{F}_\lambda\big)}\ . \label{eq:Hilbert space Grothendieck}
\ee
Including the indecomposable module $\mathscr{T}$ makes the structure of the Hilbert space slightly more complicated. In particular, an ideal has to be factored out to make the action of $L_0-\bar{L}_0$ diagonalisable and ensure locality \cite{Gaberdiel:1998ps}; this is again described in more detail in Appendix~\ref{app:T details}. 
Once this ideal is factored out, and working on the level of the Grothendieck ring (as appropriate for the discussion of the partition function), the above factor of $16=4\times 4$ is removed (see the discussion below eq.~(\ref{Grothendieck})), and the indecomposable representation just fills in the contribution for $\lambda=\frac{1}{2}$.

To show that (\ref{eq:Hilbert space Grothendieck}) is indeed modular invariant, we have to determine the characters of the representations $\mathscr{F}_\lambda$. This is done in Appendix~\ref{app:free field representation} with the help of the free field realization \eqref{eq:free field representation}, and leads to (see eq.~\eqref{eq:character second version})
\begin{align}
\mathrm{ch}[\sigma^w(\mathscr{F}_\lambda)](t,z;\tau)
&=q^{\frac{w^2}{2}} \sum_{r \in \mathds{Z}+\lambda} x^r q^{-rw} \frac{\vartheta_2\big(\frac{t+z}{2};\tau\big)\vartheta_2\big(\frac{t-z}{2};\tau\big)}{\eta(\tau)^4}\ , \label{eq:psu112 character}
\end{align}
where our conventions for theta-functions are spelled out in Appendix~\ref{app:theta}. 
Here $x=\mathrm{e}^{2\pi i t}$ is the chemical potential of $\mathfrak{sl}(2,\mathds{R})$, while $y=e^{2\pi i z}$ is the chemical potential of $\mathfrak{su}(2)$. In particular, $t$ will play the role of the modular parameter of the boundary torus of $\mathrm{AdS}_3$. The characters are treated as formal power series and not as meromorphic functions. Indeed, the sum over $r$ formally leads to the factor 
\be\label{modformal}
\sum_{r \in \mathds{Z}+\lambda} x^r q^{-rw} = e^{2\pi i \lambda(t - \tau w)} \, \sum_{m\in\mathds{Z}}  \delta(t-\tau w + m) \ ,
\ee
and thus modular invariance is a somewhat formal property. Note that this problem is not specific to $k=1$, but also arises for generic $k$ in the original discussion of \cite{Maldacena:2000hw}, see Appendix~B.4 of that paper. Incidentally, the delta-functions that appear in (\ref{modformal}) arise precisely at the points in the $\tau$-plane where the worldsheet torus can be mapped holomorphically to the boundary torus, see eq.~(74) of \cite{Maldacena:2000kv}. Unlike the situation described there (where for these values of $\tau$ there was a pole in partition function), the partition function localises in our case to these maps, thus suggesting that the ${\rm AdS}_3\times {\rm S}^3$ factor has become topological.

Under modular transformations, the characters transform into one another. Since invariance of \eqref{eq:Hilbert space Grothendieck} under the T-modular transformation is clear, we focus on the S-modular transformation. As usual in string partition functions, to get a good modular behaviour, we have to include a $(-1)^{\mathrm{F}}$ into the character. With this, the S-modular transformation of the characters is described by the formal S-matrix
\be \label{Smatrix}
S_{(w,\lambda),(w',\lambda')}=-i\, \mathrm{sgn}(\mathrm{Re}(\tau))\, \mathrm{e}^{2\pi i(w'\lambda+w\lambda')}\ ,
\ee
see Appendix~\ref{subapp:modular properties} for more details.
The fact that the S-matrix depends on $\tau$ is typical of logarithmic conformal field theories
\cite{Miyamoto:2002ar,Flohr:2005cm}, and it will cancel out once left- and right-movers are correctly combined. The S-matrix is formally unitary and symmetric. This allows us to deduce  that \eqref{eq:Hilbert space Grothendieck} is at least formally modular invariant.

Up to the zero modes, the character \eqref{eq:psu112 character} agrees precisely with the character of four R-sector fermions and two bosons (where the fermions transform in the $({\bf 2},{\bf 2})$ with respect to 
$\mathfrak{sl}(2,\mathds{R}) \oplus \mathfrak{su}(2)$). Morally, they originate from the four free bosons and fermions of the free field construction, of which two bosons have been factored out by the coset \eqref{eq:free field representation}. We should note that, for generic $k$, we should have expected to find six bosonic oscillators corresponding to the six-dimensional bosonic subalgebra (capturing the $6$-dimensional space ${\rm AdS}_3 \times {\rm S}^3$), and eight fermionic oscillators, one for each supercharge. Thus the character \eqref{eq:psu112 character} has four bosonic and fermionic oscillators fewer than in the generic case. (In particular, these representation have therefore many null-states!) This feature will carry through and is responsible for the fact that also in the final string theory answer, we will only have four bosonic and four fermionic oscillators. 

\section{The string theory spectrum} \label{sec:string theory partition function}

In the final step we now combine the $\mathfrak{psu}(1,1|2)_1$ WZW model with the other ingredients of the hybrid formalism and discuss the physical state conditions.

\subsection{Physical state conditions} \label{subsec:physical state conditions}
In addition to the $\mathfrak{psu}(1,1|2)_1$ WZW model, we have the sigma model corresponding to ${\cal M}_4$, as well as the ghosts. In this subsection we will first deal with the ghost contribution.

For any $k \ge 2$, we can obtain the physical string spectrum of the hybrid string by comparison to the RNS formalism.\footnote{One could also attempt to calculate the physical spectrum directly in the hybrid formalism, using the cohomological description of the physical state condition, but this calculation seems to be difficult. In fact, even for generic $k$, this has only been done for the first few energy levels and in the unflowed sector, see \cite{Gerigk:2012cq}.} The characters of the $\mathfrak{psu}(1,1|2)_k$ WZW model for $k \ge 2$ are known \cite{Gotz:2006qp}, and thus the spectrum before imposing the physical state conditions can be computed in the hybrid formalism. By comparison to the known physical spectrum as determined in the RNS formalism, we then conclude that the ghost contribution to the partition function in the hybrid formalism cancels four fermionic oscillators transforming in the $(\mathbf{2},\mathbf{2})$ of $\mathfrak{sl}(2,\mathds{R}) \oplus \mathfrak{su}(2)$, and two bosonic oscillators.\footnote{Note that, prior to imposing the physical state condition, the hybrid string has $8+4$ fermions ($8$ from the $\mathfrak{psu}(1,1|2)_k$ WZW model and $4$ from the $\mathbb{T}^4$) but only $6+4$ bosons. Thus this is the expected number.}
 The details of this computation are spelled out in Appendix~\ref{app:physical state conditions}. Thus, we have for $k \ge 2$
\be 
Z_\text{ghost}(t,z;\tau)=\Bigg| \frac{\eta(\tau)^4}{\vartheta_2\big(\frac{z+t}{2};\tau\big)\vartheta_2\big(\frac{z-t}{2};\tau\big)} \Bigg|^2\ .  \label{eq:ghost partition function}
\ee
Since this is independent of $k$ and the ghosts are free fields not interacting with the WZW model at the pure NS-NS point,  the ghost contribution should remain the same also for $k=1$.
Comparing with \eqref{eq:psu112 character}, we note that at the end of the day only the zero mode contribution survives after the physical state conditions have been imposed. This will, in turn, be fixed by the mass-shell condition. We should note  that this structure is strongly reminiscent of a topological theory.

\subsection{The sigma-model on $\mathbb{T}^4$} \label{subsec:sigma model on T4}
Let us concentrate in the following on the case where ${\cal M}_4$ is described by the sigma-model on $\mathbb{T}^4$ with small $\mathcal{N}=(4,4)$ supersymmetry. As discussed in \cite{Berkovits:1999im}, we actually need a topologically twisted version of the sigma-model. The topological twist will effectively amount to evaluating the partition function in the R-sector, for which we then find 
% The R-sector partition function of $\mathbb{T}^4$ reads
\be 
Z_{\mathbb{T}^4}^\text{R}(z,t;\tau)=\left|\frac{\vartheta_2\big(\frac{z+t}{2};\tau\big)\vartheta_2\big(\frac{z-t}{2};\tau\big)}{\eta(\tau)^6}\right|^2 \Theta_{\mathbb{T}^4}(\tau)\ . \label{eq:sigma model T4 partition function}
\ee
Here, the two theta-functions account for the four fermions in the R-sector, which transform in the $(\mathbf{2},\mathbf{2})$ with respect to $\mathfrak{sl}(2,\mathds{R})\oplus \mathfrak{su}(2)$, and the eta-functions in the denominator describe the four free bosons. We have also included the lattice theta-function 
\be
\Theta_{\mathbb{T}^4}(\tau)=\sum_{(p,\bar{p}) \in \Gamma_{4,4}} q^{\frac{1}{2}p^2} \bar{q}^{\frac{1}{2}\bar{p}^2}\ ,
\ee
which accounts for the non-zero winding and momentum states. Here $\Gamma_{4,4}$ is the Narain lattice of the torus.
%We also inserted the correct chemical potentials in the partition function. The four free fermions 

Combining the three ingredients \eqref{eq:psu112 character}, \eqref{eq:ghost partition function} and \eqref{eq:sigma model T4 partition function}, we see that the representation $\sigma^w(\mathscr{F}_\lambda)$ contributes altogether 
\be 
\left|\sum_{m \in \mathds{Z}+\lambda} x^m q^{-m w+\frac{w^2}{2}}\right|^2Z_{\mathbb{T}^4}^\text{R}(z,t;\tau) \label{eq:w spectrally flowed sector contribution}
\ee
to the worldsheet spectrum. Next we have to impose the mass-shell and level-matching conditions, i.e.\ we need to demand that 
\be 
h_{{\rm osc}}-m w+\frac{w^2}{2}=0\quad\Rightarrow\quad m=\frac{w}{2}+\frac{h_{{\rm osc}}}{w}\ , \label{eq:m solution}
\ee
where $h_{{\rm osc}}$ is the conformal dimension coming from the $\mathbb{T}^4$ sigma model, and similarly for the right-movers. Thus only one term in the sum survives, and the string partition function becomes 
\be 
Z_\text{string}(t,z)=\sum_{w=1}^\infty x^{\frac{w}{2}} \bar{x}^{\frac{w}{2}} Z_{\mathbb{T}^4}^{\text{R}'}\big(z,t;\tfrac{t}{w}\big) \ . \label{eq:string partition function}
\ee
Here, we have performed already the sum over the spectrally flowed sectors.\footnote{We have restricted the spectral flow to $w>0$, see Section~\ref{sec:fusion rules dual CFT} below for an interpretation of the other states.} 
We should note that there is one additional constraint coming from the physical state conditions: since the left- and right-movers are both in $\mathscr{F}_\lambda$ (for the same $\lambda$), we have to have 
\be \label{projection}
h_{{\rm osc}}-\bar{h}_{{\rm osc}} \equiv 0 \bmod w\ .
\ee
We have indicated this constraint by a prime in \eqref{eq:string partition function}.
%Thus, we obtain the complete string partition function by summing over the spectrally flowed sectors
%\begin{align} 
%Z_\text{string}(t,z)&=\sum_{w=1}^\infty x^{\frac{w}{2}} \bar{x}^{\frac{w}{2}}\left|\frac{\vartheta_2\big(\frac{z-t}{2};\frac{t}{w}\big)}{\vartheta_2\big(\frac{z+t}{2};\frac{t}{w}\big)}\right|^2 Z_{\mathbb{T}^4}^{\text{R}'}(z,t;\tfrac{t}{w}) \ ,
%\end{align}
%where the prime at the sigma-model partition function means that only terms satisfying (\ref{projection}) are kept. 
Next we can use the theta-function identity
\be 
\vartheta_2\big(\tfrac{z \pm t}{2};\tfrac{t}{w}\big)=y^{\mp \frac{w}{4}} x^{-\frac{w}{8}} \begin{cases}
\vartheta_2\big(\frac{z}{2};\frac{t}{w}\big)\ , &\quad w\text{ even}\ , \\
\vartheta_3\big(\frac{z}{2};\frac{t}{w}\big)\ , &\quad w\text{ odd}\ 
\end{cases}
\ee
to simplify the torus partition function \eqref{eq:sigma model T4 partition function}.
Thus, we finally arrive at the complete string partition function
\be 
Z_\text{string}(t,z)=\sum_{w=1, \text{ even}}^{\infty} x^{\frac{w}{4}} \bar{x}^{\frac{w}{4}}Z_{\mathbb{T}^4}^{\text{R}'}\big(z,0;\tfrac{t}{w}\big)+\sum_{w=1, \text{ odd}}^\infty x^{\frac{w}{4}} \bar{x}^{\frac{w}{4}}Z_{\mathbb{T}^4}^{\text{NS}'}\big(z,0;\tfrac{t}{w}\big) \ , \label{eq:final string spectrum}
\ee
where $Z_{\mathbb{T}^4}^{\text{NS}'}$ is the NS-sector version of (\ref{eq:sigma model T4 partition function}), for which the $\vartheta_2$ factors have been replaced by $\vartheta_3$.
This then reproduces precisely the single-particle partition function of the symmetric orbifold of $\mathbb{T}^4$, see \cite{Gaberdiel:2018rqv}. The spectral flow index $w$ is here identified with the length of the single cycle twisted sector of the orbifold CFT. We expect the analysis to work similarly for ${\cal M}_4= {\rm K3}$; details about this will be given elsewhere \cite{EGprog}. 

% We have thus found an explicit realisation of a higher spin symmetry from the worldsheet \cite{Gaberdiel:2014cha, Gaberdiel:2015mra}.

\subsection{The chiral fields} \label{subapp:chiral fields}

Given that $w$ corresponds to the length of the twisted sector cycle, the untwisted sector arises for $w=1$. In particular, the chiral fields of the dual CFT therefore come from the $w=1$ sector, as was already anticipated in \cite{Gaberdiel:2017oqg} and \cite{Ferreira:2017pgt}. We also see from \eqref{eq:m solution}, that for $w=1$ the quantum number $m$ must be a half-integer, i.e.\ that $\lambda=\tfrac{1}{2}$. 
Thus the chiral fields come in fact from  $\sigma(\mathscr{T})$. While $\mathscr{T}$ is the indecomposable representation discussed in Appendix~\ref{app:T details} in detail, this is largely invisible on the level of the physical spectrum. Indeed, as shown in the Appendix, we have to divide the space $\bigoplus_{w\in \mathds{Z}} \sigma^w(\mathscr{T}) \otimes \sigma^w(\mathscr{T})$ by an ideal to obtain the true atypical contribution to the Hilbert space. In the resulting quotient space, we can choose a gauge such that $\mathscr{T}$ becomes the moral analogue of $\mathscr{F}_{1/2}$, i.e. 
\be\label{Tstruct}
\left(\mathscr{T} \right)^{\rm gauge-fixed}  \sim \sigma^{-1}(\mathscr{L}) \oplus 2 \cdot \mathscr{L} \oplus \sigma(\mathscr{L}) \sim  \mathscr{F}_{1/2} \ ,
\ee
where $\mathscr{L}$ denotes the vacuum representation of $\mathfrak{psu}(1,1|2)_1$.
%Thus the physical state analysis is formally analogous to the calculation above, where $\mathscr{T}$ plays the role of $\mathscr{F}_{1/2}$. 
After the gauge-fixing, $\sigma(\mathscr{T})$ consists then of the modules $\mathscr{L}$, $2 \cdot \sigma(\mathscr{L})$ and $\sigma^2(\mathscr{L})$.

The vacuum module $\mathscr{L}$ yields exactly one physical state, namely the vacuum itself. (Any excited state has positive conformal weight on the worldsheet and hence cannot satisfy the worldsheet mass-shell condition.) This state has vanishing spacetime conformal dimension and hence corresponds to the spacetime vacuum. Thus, as one might have anticipated, the spacetime vacuum comes directly from a vacuum module on the worldsheet --- which sits however in a larger indecomposable module. 

Next, $\sigma(\mathscr{L})=\mathscr{G}_+$ is part of a continuous representation. In particular, its $L_0$ spectrum is bounded from below and hence only the ground states of the representation survive the mass-shell condition. In spacetime, $\sigma(\mathscr{L})$ hence yields exactly one $\mathfrak{psu}(1,1|2)$-representation, which corresponds to a $h=\tfrac{1}{2}$ BPS-representation. The only such representation in the vacuum sector of the symmetric orbifold are two of the fermions together with their superconformal descendants. Accounting for the multiplicity, we thus see that $2 \cdot \sigma(\mathscr{L})$ yields in spacetime the fundamental fields, i.e.\ the four fermions and the four bosons together with their derivatives. 

Finally, the remaining chiral fields come from the module $\sigma^2(\mathscr{L})$. Since its $L_0$ eigenvalue on the worldsheet is unbounded from below, there are many excited states which satisfy the mass-shell condition. In particular, this is the sector where the higher spin square (HSS) symmetry generators of \cite{Gaberdiel:2014cha,Gaberdiel:2015mra} sit. 

It is also instructive to understand where the exactly marginal operators come from. In the untwisted sector they sit in the sector 
\be
4\cdot \sigma(\mathscr{L}) \otimes \sigma(\mathscr{L}) \subset \sigma(\mathscr{T}) \otimes \sigma(\mathscr{T}) \ , 
\ee
corresponding to the $4 \times 4=16$ moduli which deform the 4-torus $\mathbb{T}^4$. The theory has one more 
exactly marginal operator that comes from the $2$-cycle twisted sector, and hence arises from 
\be
\sigma(\mathscr{L}) \otimes \sigma(\mathscr{L}) \subset \sigma^2(\mathscr{T}) \otimes \sigma^2(\mathscr{T}) \ . 
\ee
We should mention that each worldsheet representation $\sigma(\mathscr{L})$ gives only rise to one physical $\mathfrak{psu}(1,1|2)$ multiplet in the dual CFT, and hence each $\sigma(\mathscr{L}) \otimes \sigma(\mathscr{L})$ factor yields four moduli. This structure therefore reflects the $\mathrm{SO}(4,5)$ symmetry of the moduli space of the theory \cite{Seiberg:1988pf}.

Finally, it is worth mentioning that $\mathscr{T}$ plays also another special role in our construction: all spacetime (quarter)-BPS states come on the worldsheet from the spectrally flowed images of $\mathscr{T}$! This is simply a consequence of the fact that the spacetime conformal weight of the BPS states is in $\tfrac{1}{2}\mathds{Z}$, and hence they have to come from $\sigma^w(\mathscr{T})$ on the worldsheet. (A more careful argument shows that they cannot come from $\sigma^w(\mathscr{F}_0)$.) In particular, the chiral ring sits entirely in $\sigma^w(\mathscr{T})$, which ties together with the fact that $\bigoplus_{w \in \mathds{Z}}\sigma^w(\mathscr{T})$ closes under fusion on itself, see eq.~(\ref{eq:psu112 fusion rules c}). Furthermore, each summand in (\ref{Tstruct}) contains precisely one BPS state, and similarly for the $w$-flowed versions.
%The 16 terms of $\mathscr{T} \otimes \mathscr{T}$ after gauge-fixing contain then exactly one BPS state each.

\subsection{A subtlety at $w=0$}

Finally, we notice that there are in fact also `physical' states for $w=0$. Looking back at \eqref{eq:m solution} we see that for $w=0$ the ground states of $\mathfrak{psu}(1,1|2)$ satisfy the physical state condition, without any excitation along either $\mathfrak{psu}(1,1|2)$ or $\mathbb{T}^4$. (This is a direct consequence of (\ref{eq:psu112 Casimir vanishing}), and therefore independent of $\lambda$.)  From the perspective of the spacetime CFT, these states therefore transform as in (\ref{shortrep}). In particular, this representation is non-unitary since it contains the summand 
\be\label{sick} 
\bigl( \mathscr{C}^{0}_{\lambda+\frac{1}{2}} ,{\bf 1} \bigr) \ , 
\ee
which is a non-unitary representation of $\mathfrak{sl}(2,\mathds{R})$ (unless $\lambda=\frac{1}{2}$). Since (\ref{sick}) is the `zero-momentum' ground state of ${\rm AdS}_3 \times {\rm S}^3$, these states are the natural analogue of the state in, say, bosonic string theory given by 
\be
\alpha^0_{-1} |p=0\rangle \ , 
\ee
which is also physical despite having negative norm. (Here $\alpha^0_{-1}$ denotes the time-like oscillator, and $|p=0\rangle$ is the ground state with zero momentum.) These states should therefore be discarded. We note that this is consistent since the $w=0$ states can never be produced in OPEs of physical states with $w>0$, which follows, as we shall see, from the fusion rules (\ref{wfusion1}).

\section{The fusion rules} \label{sec:fusion rules dual CFT}

In the previous sections we have shown that the spectrum of the hybrid string on ${\rm AdS}_3 \times {\rm S}^3 \times \mathbb{T}^4$ with a single unit of NS-NS flux ($k=1$) agrees precisely with that of the symmetric orbifold of $\mathbb{T}^4$. However, on the face of it, it seems that the fusion rules of the worldsheet theory, see in particular eq.~(\ref{eq:fusion spectral flow}), are not compatible with those of the symmetric orbifold theory. (This issue was already alluded to in \cite{Giribet:2018ada}.) Indeed, the above dictionary implies that we should identify $w$ with the length of the twisted cycle $\ell$ in the symmetric orbifold. However, as was shown in \cite{Jevicki:1998bm,Pakman:2009zz,Pakman:2009ab}, the fusion rules of single cycle twisted sectors at leading order in $1/N$ take the form 
\be
[\ell_1] \times [\ell_2]  = \bigoplus_{\ell=|\ell_1-\ell_2|+1}^{\ell_1+\ell_2-1} \, [\ell] \ , \label{eq:symmetric product fusion rules}
\ee
where $\ell$ denotes the length of the cycle, and the sum on the right-hand-side runs  over every other value, i.e.\ $\ell_1+\ell_2+\ell$ is odd. 

\subsection{The $x$-basis}

In order to see how to reconcile this with the worldsheet description, we recall from Section~\ref{subsec:representations sl2R} and Appendix~\ref{app:groupreps} that the representation theory of the M\"obius group of the dual CFT requires us to work in the so-called $x$-basis of \cite{Maldacena:2000hw}.\footnote{\label{foot:19}As is also explained there, this is not just a basis change of a given representation, but rather considers certain direct sums of representations. In fact, the $x$-basis only depends on $j$, i.e.\ the value of the Casimir $\mathcal{C}=-j(j-1)$, but does not fix $\lambda$; it therefore includes all representations corresponding to the different values of $\lambda$. We should also note that in the $m$-basis the fusion rules preserve the $J^3_0$ eigenvalue modulo integers. If the $m$-basis was the correct basis for the description of the dual CFT, this would imply that conformal dimensions in OPEs of the dual CFT would add modulo integers, which is not true in general.} In that `basis' the fusion rules were worked out at the end of Appendix~D of \cite{Maldacena:2001km}, where it was argued that they only lead to the constraint (see eq.~(D.8) of \cite{Maldacena:2001km})
\be\label{weakerfusion}
w \leq w_1+w_2 + 1 \ , 
\ee
but that the lower bound (that is visible in the $m$-basis, see for example eq.~(D.7) of that paper) is not present any longer. As a consequence, the fusion rules of the symmetric orbifold are then compatible with those of the worldsheet theory. 
\smallskip

In the following we want to explain the fusion rules in the $x$-basis more conceptually. To start with we recall that the spectral flow in eq.~(\ref{eq:spectral flow a}) and (\ref{eq:spectral flow b}) can be understood as arising from conjugation by a loop in the $J^3_0$ direction \cite{PS,Gaberdiel:1995yk}. Under the action of the spacetime M\"obius group ${\rm SL}(2,\mathds{R})$, a state in a spectrally flowed representation is hence mapped to one in which the spectral flow direction has been conjugated, see also the paragraph below eq.~(D.8) in \cite{Maldacena:2001km}. Thus there is really a moduli space of (isomorphic) spectrally flowed representations that are characterised by the direction of the spectral flow. This direction is described by an element in the Lie algebra of $\mathfrak{sl}(2,\mathds{R})$, and hence transforms in the ${\bf 3}$ of $\mathfrak{sl}_2$. 

The above analysis applies directly to the case of unit spectral flow ($w=1$). If we combine two spectral flow automorphisms corresponding to $w=1$, we may in principle choose them to point in different directions. The resulting spectral flow automorphism thus transforms in the tensor product 
\be
{\bf 3} \otimes {\bf 3} = {\bf 1} \oplus {\bf 3} \oplus {\bf 5} \ . 
\ee
Note that the special case where the two spectral flow directions point in the same direction describes the highest weight state of this tensor product, and hence lies in the ${\bf 5}$ of this $\mathfrak{sl}_2$; this is therefore the spectral flow that should be identified with the $w=2$ sector. Recursively proceeding in this manner we thus conclude that the $w$-spectrally flowed sector is characterised by transforming in the ${\bf 2w+1}$ dimensional representation of $\mathfrak{sl}_2$.

We note in passing that this prescription naturally incorporates the constraint $w > 0$:  the states with negative $w$ lie in the same representation of the spacetime M\"obius group ${\rm SL}(2,\mathds{R})$ since the conformal transformation $\gamma(x)= - \tfrac{1}{x}$ inverts the sign of $w$. Indeed, $\gamma$ induces the inner automorphism of $\mathfrak{sl}(2,\mathds{R})$ corresponding to 
\be
J^3 \mapsto - J^3 \ , \qquad J^\pm \mapsto J^\mp \ , 
\ee
which inverts the sign of $w$ in (\ref{eq:spectral flow a}) and (\ref{eq:spectral flow b}). This also ties in with the fact that, as already argued in \cite{Maldacena:2001km}, $w>0$ describes the  `in'-states at $x=0$, while $w<0$ corresponds to the `out'-states that are inserted at $x=\infty=\gamma(0)$. 

The fusion rules of the spectral flow (in the $x$-basis) are therefore constrained by the representation theory of this $\mathfrak{sl}_2$, and hence take the form
\be\label{wfusion}
[w_1] \times [w_2] = \bigoplus_{w=|w_1-w_2|}^{w_1 + w_2} \, [w] \ . 
\ee
This selection rule replaces (\ref{eq:fusion spectral flow}), which was derived under the assumption that all spectral flow automorphisms point in the same direction --- this is the situation that arises for $w=w_1 + w_2$. 
Together with the shift by $\pm 1$ in spectral flow that comes directly from the fusion rules, see eq.~(\ref{eq:psu112 fusion rules a}), the upper limit of (\ref{wfusion}) thus reproduces (\ref{weakerfusion}).

Since the different spectral flow directions lead to isomorphic representations --- they are related to one another by an inner automorphism --- one may wonder why one cannot always choose them to lie in the same direction. The reason why this is not possible is that this is a singular `gauge' choice; indeed, from the viewpoint of the dual CFT, the spectral flow direction is related to the position $x$ in the dual CFT since the M\"obius group that maps the `in' states at $x=0$ to some generic point also rotates the spectral flow direction. Thus requiring the spectral flow directions to align corresponds to the (singular) configuration where the points in the dual CFT coincide, see also the discussion leading to (\ref{wfusion1}) below.

\subsection{Symmetric orbifold fusion rules}\label{sec:6.2}

While these fusion rules are now compatible with the symmetric orbifold answer of (\ref{eq:symmetric product fusion rules}), they do not quite match precisely yet. In particular, the upper and lower bounds on $w$ are $w=w_1+w_2+1$ and $w=|w_1-w_2|-1$, respectively, while from the symmetric orbifold we would expect $w=w_1+w_2-1$ and $w=|w_1-w_2|+1$, respectively. In addition, in the symmetric orbifold we have the parity constraint that $\ell_1+\ell_2+\ell \equiv w_1 + w_2 + w$ is odd, which is not visible in the above fusion rules.

As regards the second point, we note that spectral flow by one unit changes the fermion number by one, see e.g.\ \cite{Ferreira:2017pgt}. If we use the convention that the ground state of the vacuum representation $\mathscr{L}$ is bosonic (as is natural), then the ground state of $\sigma(\mathscr{L})=\mathscr{G}_+ \subset \mathscr{F}_{1/2}$ is fermionic, and hence the ground state of $\sigma^w( \mathscr{F}_{\lambda}  )$ has fermion number 
\be\label{fermnum}
\hbox{fermion number} \Bigl[ \sigma^w\bigl( \mathscr{F}_{\lambda}  \bigr)  \Bigr] = (-1)^{w+1} \ . 
\ee
Here by the `ground state' of $\sigma^w(\mathscr{F}_{\lambda})$ we mean the (spectral flow) of the affine primaries transforming in the $(\mathscr{C}^{\frac{1}{2}}_\lambda,\mathbf{2})$ of $\mathscr{F}_{\lambda}$, see eq.~(\ref{shortrep}). Incidentally, the fermion number may also be read off from the $\mathfrak{su}(2)$ spin: since the fermionic generators transform in the ${\bf 2}$ with respect to $\mathfrak{su}(2)$, we note that the states in ${\bf 1}$ are bosonic, while those in ${\bf 2}$ are fermionic. Together with the fact that each single spectral flow exchanges the two $\mathfrak{su}(2)$ representations, this also leads to (\ref{fermnum}).

Thus the term $\sigma^w( \mathscr{F}_{\lambda})$ has the same fermion number as 
$\sigma^{w_1}( \mathscr{F}_{\lambda}) \times \sigma^{w_2}( \mathscr{F}_{\mu})$ (and hence appears in the `even' fusion rules) provided that 
\be
(-1)^{w_1+1} \, (-1)^{w_2+1} = (-1)^{w+1} \ , \qquad \hbox{i.e.\ if $w_1 + w_2 + w = $ odd} \ . 
\ee
On the other hand, as we have seen in Section~\ref{sec:string theory partition function} above,  the ghost contribution removes the entire $\mathfrak{psu}(1,1|2)_1$ descendants, including the fermionic zero modes, since only the bosonic zero modes survive, see eq.~(\ref{eq:w spectrally flowed sector contribution}). Thus the physical states all come from `bosonic' fields in $\mathfrak{psu}(1,1|2)_1$, and hence in correlation functions of physical states only the even fusion rules contribute. This then implies that for the OPE of physical states we need to have $w_1 + w_2 + w = $ odd. 

This leaves us with understanding the extremal values $w=w_1+w_2+1$ and $w=|w_1-w_2|-1$. Looking back at (\ref{wfusion}), it is clear that $w=w_1+w_2+1$ can only arise if the two spectral flow directions point in exactly the same direction, i.e.\ it comes from the $w=w_1+w_2$ term in (\ref{wfusion}).\footnote{The analysis for $w=|w_1-w_2|-1$ is similar since again the three spectral flow directions all have to align.} In terms of the dual CFT this means that two of the fields are inserted at the same point, which we should exclude. Incidentally, the same issue is also visible from the viewpoint of \cite{Pakman:2009zz, Pakman:2009ab}, where it is assumed that the branched coverings are regular in the sense that the branch points do not coincide; if we relaxed this condition, $w=w_1+w_2$ would also appear in the analogue of (\ref{wfusion}). Thus we conclude that the even fusion rules (that are relevant for the physical states) are
\be\label{wfusion1}
\sigma^{w_1}(\mathscr{F}) \times \sigma^{w_2}(\mathscr{F})
=  \bigoplus_{w=|w_1-w_2|+1}^{w_1 + w_2-1} \, 
\ \sigma^{w}(\mathscr{F}) \ ,
\ee
where $w_1+w_2+w$ is odd, and we have suppressed the dependence on $\lambda$ and $\mu$ --- as we have mentioned before, see footnote~\ref{foot:19}, the actual representations of the spacetime M\"obius group involve all values of $\lambda$. 
This is then in precise agreement with the symmetric product fusion rules  (\ref{eq:symmetric product fusion rules}).

%\footnote{The only difference is that the worldsheet theory also seems to allow fusion to $w=w_1+w_2+1$ and $w = |w_1 - w_2|-1$, and that the %selection rule on the parity of $w$ is not directly visible.} In any case, t

\section{Conclusion and Outlook}

We have brought together several of the observations in  \cite{Gaberdiel:2017oqg, Ferreira:2017pgt},  and more specifically \cite{Gaberdiel:2018rqv}, into a coherent picture of the tensionless limit of 
${\rm AdS}_3$ superstring theory (with NS-NS flux) and its precise equivalence to the free symmetric product CFT. The evidence for this equivalence consists of matching, not only the spectra, but also the fusion rules governing vertex operators on the worldsheet with the selection rules for correlators in the  orbifold CFT. It will be nice to actually compare a set of three point correlators on both sides such as for the extremal ones of \cite{Pakman:2009ab} but especially those of non-BPS operators. We focussed here on the particular case of  ${\rm AdS}_3\times {\rm S}^3\times \mathbb{T}^4$, but there seem to be no barriers for the considerations to go through to the other maximally supersymmetric  backgrounds as well \cite{EGprog}.

The resulting picture that emerges fits in with many of the expectations one has on the tensionless limit of ${\rm AdS}$ string theory. The match of the spectra directly implies that there are enhanced unbroken symmetries: they arise from the massless higher spin gauge fields which are dual to the additional conserved currents in the dual free CFT \cite{Gross:1988ue, Sundborg:2000wp, Witten, Mikhailov:2002bp}, \cite{Gaberdiel:2014cha}. Indeed, as explained in  \cite{Gaberdiel:2015mra, Gaberdiel:2015wpo}, the tensionless string theory dual to the symmetric product CFT must have an enlarged stringy symmetry --- the Higher Spin Square (HSS) --- which organises the entire perturbative spectrum. 
We identified, in Section~\ref{subapp:chiral fields},  the sector of the worldsheet spectrum which corresponds to the chiral currents generating the HSS. Given the explicit description that we have now proposed for the string theory, it should be possible to investigate the properties of the HSS and its representations from the worldsheet viewpoint. We find it quite striking that the dual of the spacetime free CFT is also given in terms of free fields on the worldsheet. Note that the free field description on the worldsheet arises precisely at $k=1$ whereas the supergroup sigma model is generically an interacting theory for $k\geq 2$. This is perhaps a reflection of the general phenomenon whereby additional symmetries in spacetime are mirrored on the worldsheet. 

There have been indications from several directions that the tensionless limit in AdS is a topological string theory. For instance, this is the natural way in which one can reproduce correlators of a dual free CFT.  Thus, in the proposed general scheme of \cite{Gopakumar:2003ns, Gopakumar:2004qb, Gopakumar:2005fx} to obtain the string worldsheet theory from the free CFT, one sees signatures of localisation of correlators on the worldsheet \cite{Aharony:2007fs, Razamat:2008zr, Gopakumar:2011ev, Gopakumar:2012ny}, a property common to topological string theories. In particular, a similar feature was also noticed in \cite{Pakman:2009zz} in their attempt to construct worldsheets dual to the symmetric orbifold CFT. It will be very interesting to connect our worldsheet description with these approaches that start from the field theory.     

As remarked at several places, we see independent signatures of an underlying topological string description of our worldsheet theory. We have only short representations of the worldsheet CFT contributing and as a result there are no net string oscillator degrees of freedom in the  ${\rm AdS}_3\times {\rm S}^3$ directions after including the ghost contributions (see eq.~(\ref{eq:w spectrally flowed sector contribution}) which has only the zero mode contributions in these directions). Yet another indication comes from the 
worldsheet partition function of the $\mathfrak{psu}(1,1|2)_1$ theory which is formally a sum of contact terms ---  see eq.~(\ref{modformal}). This kind of localisation to maps which are holomorphic from the worldsheet to the boundary spacetime torus is seen in A-model topological string theories \cite{Bershadsky:1993ta}. It will be interesting to relate our worldsheet description to proposals made by Berkovits et.al.\ for an A-model topological string dual for the free ${\cal N}=4$ Super Yang-Mills theory \cite{Berkovits:2007zk, Berkovits:2007rj, Berkovits:2008qc, Berkovits:2008ga} (see also \cite{Bonelli:2008us} for the ${\rm AdS}_4$ case). These works are similar in spirit to the present case in that they start from the corresponding supercoset sigma models. We also note the topological sector of ${\rm AdS}_3$ superstring theory studied in \cite{Sugawara:1999fq, Rastelli:2005ph} though their specific proposal, based on the RNS formalism, appears to be for $k\geq 2$. 
  
Other directions also open up through having a worldsheet dual to the spacetime CFT.  These include the possibility of much more refined tests of the AdS/CFT correspondence in this background, going beyond tree level in string coupling. One should also be able to study specific marginal perturbations away from the free theory and study their possible integrability, thus connecting with the growing literature of integrable spin chain descriptions for this system --- see the recent works \cite{Baggio:2018gct, OhlssonSax:2018hgc, Dei:2018mfl} and references therein. Finally, studying the effect of the specific marginal perturbation which corresponds to the blowup mode of the symmetric orbifold can shed light on the stringy higgsing of the higher spin symmetries --- see \cite{Gaberdiel:2015uca} for a study from the orbifold CFT point of view.   

%Finally, one very tantalising possibility is that of recovering the Ramond sector of the symmetric orbifold CFT from D-brane states in the $\mathfrak{psu}(1,1|2)_1$ sigma model. This would give a worldsheet handle on the Strominger-Vafa microstates \cite{Strominger:1996sh} which give rise to the the Bekenstein-Hawking entropy for the three charge black hole. We also hope to come back to this idea in the near future.
%  
%We need to decide whether to mention the idea of D-branes in the WZW model and matching with the R-sector of the orbifold CFT.

\section*{Acknowledgements}

We thank Andrea Dei, Gian Michele Graf, Kevin Ferreira, Shiraz Minwalla, Sylvain Ribault, and Ashoke Sen for useful discussions. LE would like to thank Imperial College, London for hospitality, where this work was started. The work of LE is supported by the Swiss National Science Foundation, and LE and MRG acknowledge support by the NCCR SwissMAP, which is also funded by the Swiss National Science Foundation. RG's research was supported in part by the J.C.~Bose Fellowship of the SERB, Govt.~of India, and the Infosys Excellence Grant to the ICTS.  He would like to specially acknowledge the overarching framework of support for the basic sciences from the people of India.

\appendix

\section{Representations of $\boldsymbol{{\rm SL}(2,\mathds{R})}$}\label{app:groupreps}

In this appendix we describe the representations of (the universal cover of) $\mathrm{SL}(2,\mathds{R})$ that are relevant for the description of the Maldacena \& Ooguri theory \cite{Maldacena:2000hw}. We begin by reviewing the `usual' construction of unitary representations of $\mathrm{SL}(2,\mathds{R})$, see e.g.\ \cite{Sugiura}. 

\subsection{Representations of ${\rm SU}(1,1)$}

The unitary representations of $\mathrm{SL}(2,\mathds{R})$ are most easily constructed in terms of the group ${\rm SU}(1,1)$, which is isomorphic to $\mathrm{SL}(2,\mathds{R})$. The group ${\rm SU}(1,1)$ consists of the complex $2\times 2$ matrices of the form 
\be\label{su11}
D=\begin{pmatrix} a \ & b \\ \bar{b} \ & \bar{a} \end{pmatrix}
\ee
with $|a|^2-|b|^2=1$. These matrices have a natural action on the unit disc $|z|\leq 1$ via 
\be\label{gammaD}
z \mapsto \gamma_D(z) = \frac{az + b}{\bar{b} z + \bar{a}} \ , 
\ee
which in particular maps the unit circle, $|z|=1$, to itself.
The irreducible unitary (continuous) representations of ${\rm SU}(1,1)$ can be constructed on the Hilbert space 
\be 
\mathrm{L}_{j,\lambda}^2(\mathrm{S}^1)=\bigl\{f:\mathrm{S}^1 \to \mathds{C}\,|\, f(\phi+2\pi)=\mathrm{e}^{- 2\pi i(j+\lambda)} f(\phi) \bigr\}\ ,
\ee
where the action of a group element $D \in \mathrm{SU}(1,1)$ is defined via
\be 
(D \cdot f)(z)=(\gamma_D^{-1})'(z)^{j} \, f(\gamma_D^{-1}(z))\ ,
\ee
and $\gamma_D(z)$ is given in (\ref{gammaD}). (Here we identify $\mathrm{S}^1$ with the set $|z|=1$, and write $z=\mathrm{e}^{i\phi}$.) 
We have denoted the spin by $j$ (or equivalently the quadratic Casimir by ${\cal C}=-j(j-1)$). 
 
It is not difficult to show that the whole group $\mathrm{SU}(1,1)$ acts on 
$\mathrm{L}_{j,\lambda}^2(\mathrm{S}^1)$. There is a natural set of basis functions $f_m(\phi)$, $m\in \mathds{Z}+\lambda$, of $\mathrm{L}_{j,\lambda}^2(\mathrm{S}^1)$, given by 
\be
f_m(\phi) = \mathrm{e}^{-i(j +m)\phi} \ , 
\ee
and on them the Lie algebra generator 
\be
\Lambda_{0}=\frac{i}{2}\begin{pmatrix} 1 & 0 \\ 0 & -1 \end{pmatrix} \in \mathfrak{su}(1,1)
\ee
acts diagonally with eigenvalue 
\be
(\Lambda_0 \cdot f_m)(\phi) =  i m \, f_m(\phi) \ . 
\ee
Note that  $\Lambda_0$ generates the compact Cartan torus $\mathrm{U}(1) \subset \mathrm{SU}(1,1)$.
Depending on the value of $j$ and $\lambda$, these representations therefore define the usual continuous and discrete representations of ${\rm SU}(1,1)$. 

\subsection{Representations of ${\rm SL}(2,\mathds{R})$}

On the other hand, the Lie group $\mathrm{SL}(2,\mathds{R})$  consists of the real $2\times 2$ matrices 
\be\label{sl2}
M = \begin{pmatrix} a \ & b \cr c\  & d \end{pmatrix}  \ , \qquad 
ad - bc = 1 \ . 
\ee
This group acts naturally on the upper half-plane via 
\be\label{gammaM}
\tau \mapsto \gamma_M(\tau) = \frac{a\tau + b}{ c\tau +d} \ , 
\ee
and it fixes the real line $\tau\in\mathds{R}$. Its Lie algebra is generated by the elements $L_0$, $L_{\pm 1}$ with 
\be
L_{0}=\frac{1}{2}\begin{pmatrix} 1  & 0 \\ 0 & -1 \end{pmatrix}\ , \qquad 
L_{-1}=\begin{pmatrix} 0\ & 1 \\ 0 \ & 0 \end{pmatrix}\ , \qquad 
L_{1}=\begin{pmatrix} 0 \ & 0 \\ -1\ & 0 \end{pmatrix}\ , 
\ee
that satisfy the Lie algebra 
\be 
[L_m,L_n]=(m-n)L_{m+n}\ , \qquad m,n=-1,0,1\ . \label{eq:sl2R commutation relations}
\ee
Note that the Cartan torus corresponding to $L_0$ is, in this case, the noncompact subgroup of diagonal matrices of $\mathrm{SL}(2,\mathds{R})$. 

The Lie group $\mathrm{SL}(2,\mathds{R})$ is isomorphic to ${\rm SU}(1,1)$: the isomorphism is 
achieved by the Cayley transform that maps the upper half-plane to the unit disc, 
\be
C = \frac{\mathrm{e}^{-\frac{i\pi}{4}}}{\sqrt{2}}\, \begin{pmatrix} 1 & -i \cr 1 & i \end{pmatrix} \ , \qquad c(\tau) = z = \frac{\tau-i}{\tau+i} \ .
\ee
%with inverse
%\be
%C^{-1} = \frac{\mathrm{e}^{\frac{i\pi}{4}}}{\sqrt{2}}\, \begin{pmatrix} 1  & 1 \cr i & -i \end{pmatrix} \ , \qquad c^{-1} (z) = \tau = - i \, \frac{z+1}{z-1} \ .
%\ee
Indeed, for any $M \in {\rm SL}(2,\mathds{R})$, 
\be
D = C \, M \, C^{-1} \in {\rm SU}(1,1) \ . 
\ee
As a consequence, any irreducible unitary representation of ${\rm SU}(1,1)$ gives rise to such a representation of ${\rm SL}(2,\mathds{R})$, and vice versa. Under this isomorphism, the generator $\Lambda_0$ that acts diagonally on the unitary representations of ${\rm SU}(1,1)$ becomes 
\be
L_0' = C^{-1} \, \Lambda_0 \,  C = \frac{1}{2} \bigl( L_{-1} + L_{1} \bigr) \ . 
\ee
In the analysis of \cite{Maldacena:2000hw} $J^3_0$ is taken to be the (compact) $L_0'$ generator of the WZW algebra $\mathfrak{sl}(2,\mathds{R})$ (which acts diagonally on the standard representations of $\mathfrak{sl}(2,\mathds{R})$). However, from the viewpoint of the spacetime theory, $J^3_0$ is identified with the (non-compact) M\"obius generator $L_0$ (rather than $L_0'$) of the dual CFT. This has important consequences since the corresponding representations are not isomorphic as representations of $\mathrm{SL}(2,\mathds{R})$. This is a consequence of the fact that $L_0$ and $L_0'$ are {\it not} conjugate to one another in $\mathrm{SL}(2,\mathds{R})$, but only in $\mathrm{SL}(2,\mathds{C})$. 

In order to understand the structure of the representations for which $L_0$ acts diagonally, we consider the 
vector space of functions 
\be\label{Hdef}
{\cal H} = \bigl\{f:\mathds{R} \to \mathds{C} \bigr\} \ , 
\ee
on which $M\in {\rm SL}(2,\mathds{R})$ acts via 
\be \label{Mact}
(M \cdot f)(x)=(\gamma_M^{-1})'(x)^{j} \, f(\gamma_M^{-1}(x))\ ,
\ee
where 
\be
\gamma_M(x) = \frac{ax+b}{cx+d} \ , \qquad \hbox{for} \qquad
M = \begin{pmatrix} a \ &  b \cr c\ & d \end{pmatrix} \in {\rm SL}(2,\mathds{R}) \ . 
\ee
In this representation the Lie algebra generators $L_m$ of $\mathfrak{sl}(2,\mathds{R})$ are the differential operators
\be
L_1=-\frac{\partial}{\partial x}\ , \qquad L_0= -x \frac{\partial}{\partial x}-j\ , \qquad
L_{-1}=-x^2 \frac{\partial}{\partial x}-2jx\ ,
\ee
and the Casimir equals $\mathcal{C}=-j(j-1)$. The Lie algebra generators act on the subspace of functions that is generated by 
\be
g_m(x) = x^{-j -m} \ , \qquad m\in\mathds{Z} +\lambda \ ,
\ee
where $L_0$ acts diagonally with eigenvalue
\be
(L_0 \cdot g_m)(x) = m\, g_m(x) \ .
\ee
For generic $\lambda$, the Lie algebra generators map the different $g_m(x)$ into one another, and hence realise a representation of the Lie algebra that is isomorphic to the usual continuous representation $\mathscr{C}^j_\lambda$. On the other hand, for $\lambda=-j$, we can restrict to the subspace of functions with $m+j \in \mathds{Z}_{\le 0}$ which leads to $\mathscr{D}^-_j$, while for $\lambda=j$ we get $\mathscr{D}^+_j$ from the functions with $m-j\in \mathds{Z}_{\ge 0}$. 

However, unlike the situation described above in the context of ${\rm SU}(1,1)$, this realisation does {\it not} actually lead to the corresponding representation of the Lie group ${\rm SL}(2,\mathds{R})$. Indeed, for the inversion element of ${\rm SL}(2,\mathds{R})$, 
\be
M_{\rm inv} = \begin{pmatrix} 0 \ & 1 \cr -1 \ & 0 \end{pmatrix} \ , \qquad \gamma_{M_{\rm inv}}(x) = - \frac{1}{x} \ , 
\ee
we find 
\be
(M_{\rm inv} \cdot g_m )(x)  = \mathrm{e}^{-\pi i(j+m)} x^{-j +m} \ , 
\ee
which cannot (in general, i.e.\ if $\lambda\not\in \frac{1}{2}\mathds{Z}$) be written in terms of the $g_m(x)$ functions. (Similarly, the discrete representations $\mathscr{D}^{\pm}_j$ do not define a representation of the Lie group by themselves, but the inversion element of ${\rm SL}(2,\mathds{R})$ maps $\mathscr{D}^+_j$ to $\mathscr{D}^-_j$, and vice versa.)
%
%Thus only the direct sum of $\mathscr{C}^j_\lambda$ and $\mathscr{C}^j_{-\lambda}$ can form a representation of the covering group in the continuous case (i.e.\ for generic $\lambda$), and similarly the direct sum of  $\mathscr{D}^-_j$ and $\mathscr{D}^+_j$ in the discrete case. 
%
Other group elements in ${\rm SL}(2,\mathds{R})$ (in particular, translations) map $g_m(x)$ also to functions that have branch-cuts originating from other points $x\neq 0$ on the real line.  While one can formally write these functions in terms of Laurent polynomials around $x=0$, the more natural way to describe the space on which (the universal covering group of) ${\rm SL}(2,\mathds{R})$ acts, is as the full space (\ref{Hdef}), with the action being given by (\ref{Mact}). This is then the so-called $x$-basis of \cite{Maldacena:2000hw}.\footnote{Incidentally, the fact that both $\mathscr{D}^+_j$ and $\mathscr{D}^-_j$ appear in this basis was also already noticed there.} In particular, the representation of this ${\rm SL}(2,\mathds{R})$ incorporates all representations of $\mathfrak{sl}(2,\mathds{R})$ with a given value of $j$ (and hence of the Casimir $\mathcal{C}=-j(j-1)$), but with all values of $\lambda$, including the two discrete representations $\mathscr{D}^+_j$ and $\mathscr{D}^-_j$, see also \cite{Ribault:2009ui} for a related observation.
%Thus if we want to consider representations of the spacetime ${\rm SL}(2,\mathds{R})$, we have to consider this $x$-basis rather than the $m$-basis. 
This observation plays an important role in Section~\ref{sec:fusion rules dual CFT}.

\section{The short representation of $\boldsymbol{\mathfrak{psu}(1,1|2)}$} \label{app:short representation}

In this appendix we describe the short representation described by (\ref{eq:psu112 short multiplet}) explicitly. We label the states as
\begin{eqnarray}
& & \ket{m,\uparrow,0}\ , \ \ket{m,\downarrow,0}  \in (\mathscr{C}^{\frac{1}{2}}_\lambda ,{\bf 2}) \\
& & \ket{m,0,\uparrow} \in (\mathscr{C}^{0}_{\lambda +\frac{1}{2}},{\bf 1}) \ , \quad
\ket{m,0,\downarrow} \in (\mathscr{C}^{1}_{\lambda +\frac{1}{2}},{\bf 1}) \ . 
\end{eqnarray}
Thus $m \in \mathds{Z}+\lambda$ in the first line, while $m \in \mathds{Z}+\lambda+\frac{1}{2}$ for the states in the second line. In each case the third entry keeps track of the quantum numbers with respect to the outer automorphism $\mathfrak{su}(2)$. In particular, the two states in the second line transform actually in a doublet under the outer automophism.\footnote{Note that $\mathscr{C}^{0}_{\lambda +1/2} \cong \mathscr{C}^{1}_{\lambda +1/2}$ since both have the same Casimir $\mathcal{C}=0$. On the other hand, the Casimir of the representation $\mathscr{C}^{1/2}_{\lambda }$ equals $\mathcal{C}=\tfrac{1}{4}$.}
 The bosonic subalgebra $\mathfrak{sl}(2,\mathds{R}) \oplus \mathfrak{su}(2)$ acts on these states according to the representation theory we described in Subsection~\ref{subsec:representations sl2R} and the usual representation theory of $\mathfrak{su}(2)$. In our conventions, this takes the form
\begin{subequations}
\begin{align}
J^3_0 \ket{m,\updownarrow,0}&=m \ket{m,\updownarrow,0}\ , & J^3_0 \ket{m,0,\updownarrow}&=m \ket{m,0,\updownarrow}\ , \\
J^+_0 \ket{m,\updownarrow,0}&=\ket{m+1,\updownarrow,0}\ , & J^+_0 \ket{m,0,\updownarrow}&=\ket{m+1,0,\updownarrow}\ , \\
J^-_0 \ket{m,\updownarrow,0}&=\big(m-\tfrac{1}{2}\big)^2\ket{m-1,\updownarrow,0}\ , \!\!\!\! & J^-_0 \ket{m,0,\updownarrow}&=m(m-1)\ket{m-1,0,\updownarrow}\ , \\
K^3_0 \ket{m,\uparrow,0}&=\tfrac{1}{2}\ket{m,\uparrow,0}\ , & K^3_0 \ket{m,\downarrow,0}&=-\tfrac{1}{2}\ket{m,\downarrow,0}\ ,\\
K^+_0 \ket{m,\downarrow,0}&=\ket{m,\uparrow,0}\ , & K^-_0 \ket{m,\uparrow,0}&=\ket{m,\downarrow,0}\ .
\end{align}
Here, by $\updownarrow$ we mean either of the two states corresponding to $\uparrow$ or $\downarrow$. On the other hand, the 8 supercharges act as 
\begin{align}
S_0^{---}\ket{m,\uparrow,0}&=-\big(m-\tfrac{1}{2}\big)\ket{m-\tfrac{1}{2},0,\downarrow}\ , & S_0^{+--}\ket{m,\uparrow,0}&=\ket{m+\tfrac{1}{2},0,\downarrow}\ , \\
S_0^{---}\ket{m,0,\uparrow}&=m\ket{m-\tfrac{1}{2},\downarrow,0}\ , & S_0^{+--}\ket{m,0,\uparrow}&=-\ket{m+\tfrac{1}{2},\downarrow,0}\ , \\
S_0^{--+}\ket{m,\uparrow,0}&=-\big(m-\tfrac{1}{2}\big)\ket{m-\tfrac{1}{2},0,\uparrow}\ , & S_0^{+-+}\ket{m,\uparrow,0}&=\ket{m+\tfrac{1}{2},0,\uparrow}\ , \\
S_0^{--+}\ket{m,0,\downarrow}&=-m\ket{m-\tfrac{1}{2},\downarrow,0}\ , & S_0^{+-+}\ket{m,0,\downarrow}&=\ket{m+\tfrac{1}{2},\downarrow,0}\ , \\
S_0^{-+-}\ket{m,\downarrow,0}&=\big(m-\tfrac{1}{2}\big)\ket{m-\tfrac{1}{2},0,\downarrow}\ , & S_0^{++-}\ket{m,\downarrow,0}&=-\ket{m+\tfrac{1}{2},0,\downarrow}\ , \\
S_0^{-+-}\ket{m,0,\uparrow}&=m\ket{m-\tfrac{1}{2},\uparrow,0}\ , & S_0^{++-}\ket{m,0,\uparrow}&=-\ket{m+\tfrac{1}{2},\uparrow,0}\ , \\
S_0^{-++}\ket{m,\downarrow,0}&=\big(m-\tfrac{1}{2}\big)\ket{m-\tfrac{1}{2},0,\uparrow}\ , & S_0^{+++}\ket{m,\downarrow,0}&=-\ket{m+\tfrac{1}{2},0,\uparrow}\ , \\
S_0^{-++}\ket{m,0,\downarrow}&=-m\ket{m-\tfrac{1}{2},\uparrow,0}\ , & S_0^{+++}\ket{m,0,\downarrow}&=\ket{m+\tfrac{1}{2},\uparrow,0}\ ,
\end{align}
\end{subequations}
while all other actions are zero. This characterises the representation completely. As we have indicated in the main text, for $\lambda=\tfrac{1}{2}$, the states
\be\label{A.4}
\quad \ket{m,\updownarrow,0}\ , \quad m\ge \tfrac{1}{2}\qquad\text{and}\qquad \ket{m,0,\updownarrow}\ , \quad m \ge 1
\ee
form an irreducible subrepresentation which is obtained from (\ref{eq:psu112 short multiplet}) upon replacing 
$\mathscr{C}^{1/2}_{1/2}$ by its subrepresentation $\mathscr{D}^{1/2}_+$, and $\mathscr{C}^{1}_{1} \cong\mathscr{C}^{0}_{1}$ by its subrepresentation $\mathscr{D}^{1}_+$.
%, see the discussion below eq.~(\ref{Grothendieck}).  
Similarly, the states 
\be\label{A.5}
\ket{m,\updownarrow,0}\ , \quad m\ge \tfrac{1}{2}\qquad\text{and}\qquad \ket{m,0,\updownarrow}\ , \quad m \ge 0
\ee
form a slightly bigger indecomposable subrepresentation. The quotient of (\ref{A.5}) by (\ref{A.4}) 
consists of the two states $\ket{0,0,\updownarrow}$, which is isomorphic to twice the trivial representation.

\section{The free field representation of $\boldsymbol{\mathfrak{psu}(1,1|2)_1}$} \label{app:free field representation}

In this appendix we provide details about the free field realisation of $\mathfrak{psu}(1,1|2)_1$. We begin by reviewing the free field realisation of $\mathfrak{sl}(2,\mathds{R})_{1/2}$ in terms of a pair of symplectic bosons  \cite{Lesage:2002ch, Ridout:2010jk}. 

\subsection{The symplectic boson theory} \label{subapp:symplectic boson}
%We recall the (almost) free field realization \eqref{eq:free field representation}. We want to use this representation to determine the character of $\mathfrak{psu}(1,1|2)_1$-representations and their fusion rules. 

We begin with a single pair of symplectic bosons $\xi_m$ and $\bar{\xi}_m$, satisfying the commutation relations 
\be
{}[ {\xi}_m, \bar{\xi}_n] = \delta_{m,-n}\ . 
\ee
It gives rise to an $\mathfrak{sl}(2,\mathds{R})_{1/2}$ affine algebra since we have 
\be
J^{3}_m=-\tfrac{1}{2} (\xi\bar{\xi})_m \ , \qquad J^{+}_m=\tfrac{1}{2}(\xi\xi)_m \ , \qquad 
J^{-}_m=\tfrac{1}{2}(\bar{\xi}\bar{\xi})_m \ .
\ee
Both $\xi_m$ and $\bar{\xi}_m$ are spin-$\tfrac{1}{2}$ fields and possess therefore NS- and R-sector representations. The NS-sector highest weight representation is described by
\be 
\xi_r\, \ket{0}=0\ , \qquad \bar{\xi}_r\, \ket{0}=0\ , \qquad r \ge \tfrac{1}{2} \ ,
\ee
and gives the vacuum representation of the theory, which we denote by $\mathscr{K}$. 
On the other hand, the R-sector representations of the symplectic boson pair have a zero-mode representation on the states $|m\rangle$ with action 
\be\label{xiaction}
\xi_0 \, |m\rangle = \sqrt{2}\, |m+\tfrac{1}{2} \rangle \ , \qquad 
\bar{\xi}_0 \, |m\rangle = \sqrt{2} (m-\tfrac{1}{4})\,  |m-\tfrac{1}{2} \rangle \ , 
\ee
so that, in terms of the $\mathfrak{sl}(2,\mathds{R})$ generators we have 
\be
J^3_0 \, |m\rangle = m \, |m\rangle \ , \qquad \mathcal{C}^{\mathfrak{sl}(2,\mathds{R})} \, |m\rangle = \frac{3}{16}\, |m\rangle \ . 
\ee
Thus the R-sector representations of the symplectic boson are labelled by $\lambda \in \mathds{R}/\tfrac{1}{2}\mathds{Z}$, describing the eigenvalues of $J^3_0$ mod $\tfrac{1}{2}\mathds{Z}$; these representations are denoted by $\mathscr{E}_\lambda$.  At $\lambda=\tfrac{1}{4}$, $\mathscr{E}_{1/4}$ becomes reducible, but indecomposable (as follows from the second term in (\ref{xiaction})).
%The Hilbert space features instead an indecomposable. 
It is not separately part of the Hilbert space, but rather combines together with other representations into an indecomposable representation $\mathscr{S}$ \cite{Ridout:2010jk}. Its structure is best described in terms of the so-called composition series, which takes for $\mathscr{S}$ the form 
\be 
\mathscr{S} : \quad 
\begin{tikzpicture}[baseline={([yshift=-.5ex]current bounding box.center)}]
\node (top) at (0,1.5) {$\mathscr{K}$};
\node (right) at (1.5,0) {$\sigma^2(\mathscr{K})$};
\node (left) at (-1.5,0) {$\sigma^{-2}(\mathscr{K})$};
\node (bottom) at (0,-1.5) {$\mathscr{K}$};
\draw[thick,->] (right) to (bottom);
\draw[thick,->] (left) to  (bottom);
\draw[thick,->] (top) to  (left);
\draw[thick,->] (top) to (right);
%\node (center) at (0,0) {$\mathscr{S}$};
\end{tikzpicture}
\ee
Here the bottom line is the irreducible vacuum representation $\mathscr{K}$, and it forms a proper subrepresentation of $\mathscr{S}$. Since $\mathscr{S}$ is indecomposable, the complement of $\mathscr{K}$ does {\it not} form another subrepresentation of $\mathscr{S}$. However, the quotient space $\mathscr{S}/\mathscr{K}$ contains  subrepresentations, namely the two irreducible representations described by the middle line of $\mathscr{S}$. Again, their complement is not another subrepresentation, so one needs to quotient again by the representations in the middle line. The resulting space is then the irreducible vacuum representation $\mathscr{K}$ appearing at the top of the diagram. 
In this language, the direction of the arrows indicates the symplectic boson action: symplectic bosons can map from top to bottom, but not back.
The top element of the composition series is called the ``head", whereas the bottom element is called the ``socle".

The representation $\mathscr{S}$ is closely related to $\mathscr{E}_{1/4}$ since, on the level of the Grothen\-dieck ring, we have
\be 
\mathscr{E}_{1/4} \sim \sigma(\mathscr{K}) \oplus \sigma^{-1}(\mathscr{K})\quad \Longrightarrow \quad 
\mathscr{S} \sim \sigma (\mathscr{E}_{1/4}) \oplus \sigma^{-1}(\mathscr{E}_{1/4}) \ . 
\ee
Here $\sigma$ denotes the spectral flow of the symplectic boson theory which acts via
\be
\sigma (\xi_r)=\xi_{r-\frac{1}{2}}\ , \qquad 
\sigma (\bar{\xi}_r) = \bar{\xi}_{r+\frac{1}{2}} \ . 
\ee
The fusion rules of this theory were worked out in \cite[eq.\ (5.8)]{Ridout:2010jk},
\begin{subequations}
\begin{align}
\mathscr{E}_\lambda \times \mathscr{E}_\mu &\cong \begin{cases}
\sigma(\mathscr{E}_{\lambda+\mu+\frac{1}{4}}) \oplus \sigma^{-1}(\mathscr{E}_{\lambda+\mu+\frac{1}{4}})\ , \quad & \lambda+\mu \ne 0\ , \\
\mathscr{S}\ , \quad &\lambda+\mu=0\ ,
\end{cases} \label{eq:symplectic boson fusion rules a}\\
\mathscr{E}_\lambda \times \mathscr{S} &\cong \sigma^{2}(\mathscr{E}_\lambda) \oplus 2\cdot \mathscr{E}_\lambda \oplus \sigma^{-2}(\mathscr{E}_\lambda)\ , \label{eq:symplectic boson fusion rules b}\\
\mathscr{S} \times \mathscr{S}&\cong \sigma^{2}(\mathscr{S}) \oplus 2\cdot \mathscr{S} \oplus \sigma^{-2}(\mathscr{S})\ .\label{eq:symplectic boson fusion rules c}
\end{align}
\end{subequations}

\subsection{The explicit form of the free field representation} \label{subapp:free field realization explicit}

In order to describe the free field realisation of $\mathfrak{psu}(1,1|2)_1$ we combine together two such pairs of symplectic bosons, together with $2$ complex fermions. More explicitly, let us denote the fermions by $\psi^\alpha$, $\bar{\psi}^\alpha$, and the symplectic bosons by $\xi^\alpha$ and $\bar{\xi}^\alpha$ with $\alpha=\pm$ and (anti)-commutation relations
\begin{align}
\{\bar{\psi}^\alpha_r,\psi^\beta_s\}=\epsilon^{\alpha\beta} \delta_{r, -s}\ , \qquad [\bar{\xi}^\alpha_r,\xi^\beta_s]=\epsilon^{\alpha\beta} \delta_{r,-s}\ .
\end{align}
(Anti)-commutators of barred with barred oscillators vanish, and similarly for the unbarred combinations. As is clear from the previous section, we can  construct two $\mathfrak{sl}(2,\mathds{R})_{1/2}$ algebras out of the symplectic bosons, which are explicitly given as 
\begin{equation}\label{Jpm}
\begin{array}{rclrcl}
J^{(+)3}_m&=-\tfrac{1}{2} (\xi^+\bar{\xi}^-)_m \ , & \qquad J^{(-)3}_m&=-\tfrac{1}{2} (\xi^-\bar{\xi}^+)_m \ , \\[4pt]
J^{(+)+}_m&=\tfrac{1}{2}(\xi^+\xi^+)_m \ , & \qquad J^{(-)+}_m&=\tfrac{1}{2}(\bar{\xi}^+\bar{\xi}^+)_m \ , \\[4pt] 
J^{(+)-}_m&=\tfrac{1}{2}(\bar{\xi}^-\bar{\xi}^-)_m \ , & \qquad J^{(-)+}_m&=\tfrac{1}{2}(\xi^-\xi^-)_m \ .
\end{array}
\end{equation}
We define the two spectral flow symmetries via %and the $\mathfrak{sl}(2,\mathds{R})_{1/2}\oplus \mathfrak{sl}(2,\mathds{R})_{1/2}$ fields:
\begin{subequations}
\begin{align}
\sigma^{(+)} (\xi^+_r)&=\xi^+_{r-\frac{1}{2}}\ , & \sigma^{(-)} (\bar{\xi}^+_r)&=\bar{\xi}^+_{r-\frac{1}{2}}\ , \label{eq:spectral flow 1}\\
\sigma^{(+)} (\bar{\xi}^-_r)&=\bar{\xi}^-_{r+\frac{1}{2}}\ , & \sigma^{(-)} (\xi^-_r)&=\xi^-_{r+\frac{1}{2}}\ , \label{eq:spectral flow 2}
\end{align}
\end{subequations}
so that $\sigma^{(+)}$ only acts on $\xi^+$ and $\bar{\xi}^-$ (that appear in the $J^{(+)a}$ generators), while $\sigma^{(-)}$ only acts on $\xi^-$ and $\bar{\xi}^+$. We also define their action on the fermions via 
\begin{subequations}
\begin{align}
\sigma^{(+)} (\psi^+_r)&=\psi^+_{r+\frac{1}{2}}\ , & \sigma^{(-)} (\bar{\psi}^+_r)&=\bar{\psi}^+_{r+\frac{1}{2}}\ , \label{eq:spectral flow 1f}\\
\sigma^{(+)} (\bar{\psi}^-_r)&=\bar{\psi}^-_{r-\frac{1}{2}}\ , & \sigma^{(-)} (\psi^-_r)&=\psi^-_{r-\frac{1}{2}}\ .\label{eq:spectral flow 2f}
\end{align}
\end{subequations}
We now realise the $\mathfrak{u}(1,1|2)_1$ algebra via the combinations 
\begin{subequations}
\begin{align}
J^3_m&=-\tfrac{1}{2} (\xi^+\bar{\xi}^-)_m-\tfrac{1}{2} (\xi^-\bar{\xi}^+)_m\ , & K^3_m&=-\tfrac{1}{2} (\psi^+\bar{\psi}^-)_m-\tfrac{1}{2} (\psi^-\bar{\psi}^+)_m\ , \\
J^\pm_m&=(\xi^\pm\bar{\xi}^\pm)_m\ , & K^\pm_m&=\pm(\psi^\pm\bar{\psi}^\pm)_m\ , \\
S_m^{\alpha\beta+}&=(\psi^\beta \bar{\xi}^\alpha)_m\ , & S_m^{\alpha\beta-}&=-(\xi^\alpha\bar{\psi}^\beta)_m \ ,  \\
U_m&=-\tfrac{1}{2} (\xi^+\bar{\xi}^-)_m+\tfrac{1}{2} (\xi^-\bar{\xi}^+)_m\ , & V_m&=-\tfrac{1}{2} (\psi^+\bar{\psi}^-)_m+\tfrac{1}{2} (\psi^-\bar{\psi}^+)_m\ .
\end{align}
\end{subequations}
Here the generators $J^a_m$ define $\mathfrak{sl}(2,\mathds{R})_1$, but they are {\it not} just the direct sums of the $J^{(\pm)a}_m$ generators from (\ref{Jpm}). 
It is convenient to define the two linear combinations
\be 
Z_m=U_m+V_m\qquad\text{and}\qquad Y_m=U_m-V_m\ .
\ee
Then the modes $J^a_m$, $K^a_m$ and $S^{\alpha\beta\gamma}_m$ satisfy \eqref{eq:psu112 commutation relations a} -- \eqref{eq:psu112 commutation relations i} with $k=1$, except that $Z_m$ appears as a central extension in \eqref{eq:psu112 commutation relations i},
\begin{multline} 
\{S^{\alpha\beta\gamma}_m,S^{\mu\nu\rho}_n\}=km \epsilon^{\alpha\mu}\epsilon^{\beta\nu}\epsilon^{\gamma\rho}\delta_{m+n,0}-\epsilon^{\beta\nu}\epsilon^{\gamma\rho} c_a\tensor{\sigma}{_a^{\alpha\mu}} J^a_{m+n}\\
+\epsilon^{\alpha\mu}\epsilon^{\gamma\rho} \tensor{\sigma}{_a^{\beta\nu}} K^a_{m+n}+\epsilon^{\alpha\mu}\epsilon^{\beta\nu}\delta^{\gamma,-\rho} Z_{m+n}\ . 
\end{multline}
Here $\delta^{\gamma,-\rho}$ denotes as usual the Kronecker delta.
$J^a_m$, $K^a_m$, the supercharges $S^{\alpha\beta\gamma}_m$ and the central extension $Z_m$ generate then the superalgebra $\mathfrak{su}(1,1|2)_1$. In particular, $Z_m$ commutes with the modes of $J^a_m$, $K^a_m$, $S^{\alpha\beta\gamma}_m$ and its own modes.
The remaining commutators involving $Y_m$ are given by
\begin{align}
[Y_m,J^a_n]&=0\ , & [Y_m,K^a_n]&=0\ , & [Y_m,S^{\alpha\beta\gamma}_n]&=-\gamma S^{\alpha\beta\gamma}\ , \\
[Y_m,Y_n]&=0\ , & [Y_m,Z_n]&=-m \delta_{m+n,0}\ .
\end{align}
From these commutation relations, we see that only the zero-charge sector of the central extension $Z$ lifts to representations of $\mathfrak{psu}(1,1|2)_1$. Furthermore, in order to obtain complete $\mathfrak{psu}(1,1|2)_1$-representations, we have to sum over all charges of $\mathfrak{u}(1)_Y$, because the supercharges carry charge with respect to $\mathfrak{u}(1)_Y$.

We should mention that the combination of spectral flow $\sigma^{(+)}\, \circ\, (\sigma^{(-)})^{-1}$ keeps the bosonic generators $J^a_m$ and $K^a_m$ invariant. On the other hand, the spectral flow $\sigma$ of  $\mathfrak{psu}(1,1|2)_1$ can be identified with 
\be
\sigma \equiv \sigma^{(+)}\, \circ\,\sigma^{(-)}\ . 
\ee
Its action on the generators coincides with \eqref{eq:spectral flow a} -- \eqref{eq:spectral flow e}.

\subsection{The fusion rules} \label{subapp:fusion rules}

The coset representations of \eqref{eq:free field representation} are labelled by
\be 
(\sigma^{(+)m}(\mathscr{E}_\lambda),\, \sigma^{(-)n}(\mathscr{E}_\mu);\, Y,\, Z)\ ,\qquad m+n \in 2\mathds{Z}\ ,
\ee
where the first entry $\sigma^{(+)m}(\mathscr{E}_\lambda)$ denotes the representation of the pair of symplectic bosons $\xi^+$ and $\bar{\xi}^-$, while the second entry $\sigma^{(-)n}(\mathscr{E}_\mu)$ is a representation of the pair of symplectic bosons $\xi^-$ and $\bar{\xi}^+$. The condition $m+n \in 2\mathds{Z}$ imposes the constraint that all four symplectic bosons are equally moded, see \eqref{eq:spectral flow 1} and \eqref{eq:spectral flow 2}. Since the supercharges are bilinear expressions of one symplectic boson and one fermion, and since we require them to be integer-moded, the moding of the symplectic bosons also fixes that of the fermions. Thus the fermions will be in the R sector if $m$ is even, and in the NS sector if $m$ is odd. In particular, this thereby fixes the representations of the $\mathfrak{su}(2)_1$ algebra. Finally,  $Y$ and $Z$ denote the eigenvalues of $Y_0$ and $Z_0$. 

We recall that $Z_0$ is the central extension of $\mathfrak{psu}(1,1|2)$, while $Y_0$ is the other $\mathfrak{u}(1)$-charge extending $\mathfrak{su}(1,1|2)$ to $\mathfrak{u}(1,1|2)$.
With these conventions, the symplectic bosons and free fermions transform 
%Thus, the fundamental fields transform in the following representations 
with respect to $\mathfrak{sl}(2,\mathds{R}) \oplus \mathfrak{su}(2) \oplus \mathfrak{u}(1)_Y \oplus \mathfrak{u}(1)_Z$ as
\begin{align}
&\text{symplectic bosons}: & &(\mathbf{2},\mathbf{1})_{1,1} \oplus (\mathbf{2},\mathbf{1})_{-1,-1}\ , \label{eq:symplectic boson charges}\\
&\text{fermions}: & &(\mathbf{1},\mathbf{2})_{1,-1} \oplus (\mathbf{1},\mathbf{2})_{-1,1}\ . \label{eq:fermion charges}
\end{align}
We have explained above that for $\mathfrak{psu}(1,1|2)$-representations we have to require $Z$ to vanish, but should sum over all possible charges of $\mathfrak{u}(1)_Y$.
%The constraint $m+n \in 2\mathds{Z}$ comes from the fact that all symplectic bosons have to be equally moded. 
%In order for these representations to define $\mathfrak{psu}(1,1|2)$-representations, the central extension $Z$ has to vanish. 
Furthermore, we have the selection rules that both $Z$ and $Y$ must satisfy
\be 
Z,\, Y \in \tfrac{1}{2}\mathds{Z}+\lambda-\mu\ .
\ee
Since $\lambda,\mu \in \mathds{R}/\tfrac{1}{2}\mathds{Z}$, the condition $Z=0$ allows us to choose, without loss of generality, $\lambda=\mu$. We have the identifications 
\begin{subequations}
\begin{align} 
\mathscr{F}_\lambda & \cong \bigoplus_{Y \in \mathds{Z}}\, (\sigma^{m}(\mathscr{E}_{\frac{\lambda}{2}}),\, \sigma^{-m}(\mathscr{E}_{\frac{\lambda}{2}});\, Y,\, 0)\ ,  \label{eq:coset continuous representation} \\
\mathscr{L} & \cong \bigoplus_{Y \in \mathds{Z}}\, (\sigma^m(\mathscr{K}),\, \sigma^{-m}(\mathscr{K});\, Y,\, 0)  \ , \label{eq:coset trivial representation}
\end{align}
\end{subequations}
where the spectral flow parameter $m$ is arbitrary (and we have suppressed the indices $(\pm)$ on the spectral flows). Since the spectral flow $(\sigma^{w}\, ,\, \sigma^{-w})$ leaves the bosonic subalgebra of $\mathfrak{psu}(1,1|2)_1$-algebra invariant, \eqref{eq:coset continuous representation} and \eqref{eq:coset trivial representation} define indeed highest weight representations.
We can then readily compute the fusion rules: for $\lambda+\mu \ne 0$ we find from (\ref{eq:symplectic boson fusion rules a})
\begin{align}
(\mathscr{E}_{\frac{\lambda}{2}},\, \mathscr{E}_{\frac{\lambda}{2}};\, Y_1,\, 0) \otimes (\mathscr{E}_{\frac{\mu}{2}},\, \mathscr{E}_{\frac{\mu}{2}};\, Y_2,\, 0) &\cong (\sigma(\mathscr{E}_{\frac{\lambda+\mu}{2}+\frac{1}{4}}),\, \sigma(\mathscr{E}_{\frac{\lambda+\mu}{2}+\frac{1}{4}});\, Y_1+Y_2,\, 0)\nonumber\\
&\qquad\!\! \oplus (\sigma^{-1}(\mathscr{E}_{\frac{\lambda+\mu}{2}+\frac{1}{4}}),\,\sigma^{-1}(\mathscr{E}_{\frac{\lambda+\mu}{2}+\frac{1}{4}});\, Y_1+Y_2,\, 0)\nonumber\\
&\qquad\!\!\oplus (\sigma(\mathscr{E}_{\frac{\lambda+\mu}{2}+\frac{1}{4}}),\,\sigma^{-1}(\mathscr{E}_{\frac{\lambda+\mu}{2}+\frac{1}{4}});\, Y_1+Y_2,\, 0)\nonumber\\
&\qquad\!\!\oplus (\sigma^{-1}(\mathscr{E}_{\frac{\lambda+\mu}{2}+\frac{1}{4}}),\,\sigma(\mathscr{E}_{\frac{\lambda+\mu}{2}+\frac{1}{4}});\, Y_1+Y_2,\, 0)\ . \label{C.25}
\end{align}
Summing over the $Y$-charge yields then the fusion rules for the $\mathfrak{psu}(1,1|2)_1$ representations
\be 
\mathscr{F}_\lambda \times \mathscr{F}_\mu \cong \sigma(\mathscr{F}_{\lambda+\mu+\frac{1}{2}})\oplus 2\cdot \mathscr{F}_{\lambda+\mu+\frac{1}{2}} \oplus\sigma^{-1}(\mathscr{F}_{\lambda+\mu+\frac{1}{2}})\ ,
\ee
where the middle term arises from the last two lines in (\ref{C.25}).
\smallskip

Let us also consider the exceptional case, where $\lambda+\mu=0$. Then the symplectic boson language predicts the appearance of the module $(\mathscr{S},\mathscr{S};Y_1+Y_2,0)$. This module has a composition series of 16 terms. When summing over the $Y$-charge, we get by definition the module $\mathscr{T}$. Its composition series is (we have not specified how the arrows act on the different summands that have non-trivial multiplicities)
\be 
\mathscr{T}:\quad
\begin{tikzpicture}[baseline={([yshift=-.5ex]current bounding box.center)}]
\node (top) at (0,3) {$\mathscr{L}$};
\node (right) at (3,0) {$\sigma^2(\mathscr{L})$};
\node (left) at (-3,0) {$\sigma^{-2}(\mathscr{L})$};
\node (bottom) at (0,-3) {$\mathscr{L}$};
\node (topright) at (1.5,1.5) {$2 \sigma(\mathscr{L})$};
\node (topleft) at (-1.5,1.5) {$2 \sigma^{-1}(\mathscr{L})$};
\node (bottomright) at (1.5,-1.5) {$2 \sigma(\mathscr{L})$};
\node (bottomleft) at (-1.5,-1.5) {$2 \sigma^{-1}(\mathscr{L})$};
\node (center) at (0,0) {$4\mathscr{L}$};
\draw[thick,->] (right) to (bottomright);
\draw[thick,->] (left) to  (bottomleft);
\draw[thick,->] (bottomright) to (bottom);
\draw[thick,->] (bottomleft) to  (bottom);
\draw[thick,->] (top) to  (topleft);
\draw[thick,->] (top) to (topright);
\draw[thick,->] (topright) to  (right);
\draw[thick,->] (topleft) to (left);
\draw[thick,->] (topleft) to (center);
\draw[thick,->] (topright) to (center);
\draw[thick,->] (center) to (bottomleft);
\draw[thick,->] (center) to (bottomright);
\end{tikzpicture} \label{eq:composition diagram T}
\ee
for the complete description, see Appendix~\ref{app:T details}.
In particular, because of (\ref{F12}), we have in the Grothendieck ring
\be 
\mathscr{T} \sim \sigma(\mathscr{F}_{\frac{1}{2}}) \oplus 2\cdot \mathscr{F}_{\frac{1}{2}} \oplus \sigma^{-1}(\mathscr{F}_{\frac{1}{2}})\ . \label{eq:T Grothendieck ring}
\ee
%It now follows
%\be 
%\mathscr{F}_\lambda \times \mathscr{F}_{-\lambda} \cong \mathscr{T}\ .
%\ee
Next, we compute the fusion rules of $\mathscr{F}_\lambda$ with $\mathscr{T}$. Because of \eqref{eq:T Grothendieck ring}, we expect that only $\mathscr{F}_\lambda$ and spectrally flowed images can appear on the right hand side, and this indeed follows from the symplectic boson fusion rules \eqref{eq:symplectic boson fusion rules b}, leading to 
\be 
\mathscr{F}_\lambda \times \mathscr{T} \cong \sigma^2(\mathscr{F}_\lambda) \oplus 4\cdot \sigma(\mathscr{F}_\lambda)\oplus 6\cdot \mathscr{F}_\lambda \oplus 4\cdot \sigma^{-1}(\mathscr{F}_\lambda) \oplus \sigma^{-2}(\mathscr{F}_\lambda)\ .
\ee
Finally, the fusion of $\mathscr{T}$ with itself follows from (\ref{eq:symplectic boson fusion rules c}), and we find 
\be 
\mathscr{T} \times \mathscr{T} \cong \sigma^2(\mathscr{T}) \oplus 4 \cdot \sigma(\mathscr{T}) \oplus 6 \cdot \mathscr{T} \oplus 4 \cdot \sigma^{-1}(\mathscr{T})\oplus \sigma^{-2}(\mathscr{T})\ .
\ee
This reproduces \eqref{eq:psu112 fusion rules a} -- \eqref{eq:psu112 fusion rules c}.

\subsection{The characters} \label{subapp:characters}

We can also use the free field realisation to compute the character of $\mathfrak{psu}(1,1|2)_1$-representations. For this, we recall that four free fermions in the R-sector have the character
\be 
\frac{\vartheta_2\big(\frac{z+\mu-\nu}{2};\tau\big)\vartheta_2\big(\frac{z-\mu+\nu}{2};\tau\big)}{\eta(\tau)^2}=\frac{\vartheta_2(z;2\tau)\vartheta_3(\mu-\nu;2\tau)+\vartheta_3(z;2\tau)\vartheta_2(\mu-\nu;2\tau)}{\eta(\tau)^2}\ . \label{eq:character fermion contribution}
\ee
Our conventions regarding modular functions are collected in Appendix~\ref{app:theta}. Here, we have introduced the chemical potentials $\mu$ and $\nu$, which are associated to $Y$ and $Z$, respectively, while $z$ is the chemical potential with respect to the $\mathfrak{su}(2)$ algebra, see eq.~\eqref{eq:fermion charges}. The right hand side of the equation expresses the character in terms of $\mathfrak{su}(2)_1 \oplus \mathfrak{su}(2)_1$-characters, which separates the chemical potentials, see eq.~(\ref{eq:equivalence four fermions and su2su2}). Similarly, the character of the continuous representation of two pairs of symplectic bosons $(\mathscr{E}_{\lambda/2},\mathscr{E}_{\lambda/2})$ is given by
\begin{multline} 
\sum_{m,\, n \, \in \frac{1}{2}\mathds{Z}+\frac{\lambda}{2}} x^{m+n} \mathrm{e}^{2\pi i(\mu+\nu)(m-n)} \frac{1}{\eta(\tau)^4}\\
=\Bigg(\sum_{r \in \mathds{Z}+\lambda,\, s\in \mathds{Z}}+\sum_{r \in \mathds{Z}+\lambda+\frac{1}{2},\, s\in \mathds{Z}+\frac{1}{2}}\Bigg) x^r \mathrm{e}^{2\pi i(\mu+\nu)s}\frac{1}{\eta(\tau)^4}\ , \label{eq:character boson contribution}
\end{multline}
where $t$ is the chemical potential of $\mathfrak{sl}(2,\mathds{R})_1$ and we have set $x=\mathrm{e}^{2\pi i t}$, see eq.~\eqref{eq:symplectic boson charges}. 
The oscillator part of the character is uncharged, since any charge can be absorbed into the zero-modes of the representation. \eqref{eq:character fermion contribution} and \eqref{eq:character boson contribution} can be multiplied to obtain the character of the numerator algebra in \eqref{eq:free field representation}.  It is then straightforward to obtain the coset character of $(\mathscr{E}_{\lambda/2},\mathscr{E}_{\lambda/2},Y,Z)$ by restricting to the respective exponents of $\mathrm{e}^{2\pi i\mu}$ and $\mathrm{e}^{2\pi i\nu}$. In particular, we find for $Z=0$
\begin{align} 
\mathrm{ch}[(\mathscr{E}_{\lambda/2},\mathscr{E}_{\lambda/2},Y,0)](t,z;\tau)=\begin{cases} q^{Y^2/4}\sum_{r \in \mathds{Z}+\lambda} x^r \frac{\vartheta_2(z,2\tau)}{\eta(\tau)^4} \quad & Y\text{ even}\ ,\\
q^{Y^2/4}\sum_{r \in \mathds{Z}+\lambda+\frac{1}{2}} x^r \frac{\vartheta_3(z,2\tau)}{\eta(\tau)^4} \quad & Y\text{ odd}\ .
\end{cases}
\end{align}
With the identification \eqref{eq:coset continuous representation}, we finally deduce the character of the continuous $\mathfrak{psu}(1,1|2)_1$-representations
\begin{align} 
\mathrm{ch}[\mathscr{F}_\lambda](t,z;\tau)&=\sum_{r \in \mathds{Z}+\lambda} x^r \frac{\vartheta_3(2\tau)\vartheta_2(z;2\tau)}{\eta(\tau)^4}+\sum_{r \in \mathds{Z}+\lambda+\frac{1}{2}} x^r \frac{\vartheta_2(2\tau)\vartheta_3(z;2\tau)}{\eta(\tau)^4}\\
&=\sum_{r \in \mathds{Z}+\lambda} x^r \frac{\vartheta_3(t;2\tau)\vartheta_2(z;2\tau)}{\eta(\tau)^4}+\sum_{r \in \mathds{Z}+\lambda} x^r \frac{\vartheta_2(t;2\tau)\vartheta_3(z;2\tau)}{\eta(\tau)^4} \\
&=\sum_{r \in \mathds{Z}+\lambda} x^r  \frac{\vartheta_2\big(\frac{t+z}{2};\tau\big)\vartheta_2\big(\frac{t-z}{2};\tau\big)}{\eta(\tau)^4}\ ,
\end{align}
where $x=\mathrm{e}^{2\pi i t}$, and we have used that
\be
\sum_{n\in\mathds{Z}} q^{n^2} = \vartheta_3(2\tau) \ , \qquad
\sum_{n\in\mathds{Z}+\frac{1}{2}} q^{n^2} = \vartheta_2(2\tau) \ .
\ee
We have also used the identity \eqref{eq:equivalence four fermions and su2su2}.
Thus, the character formally looks like 4 fermions in the $(\mathbf{2},\mathbf{2})$ of $\mathfrak{sl}(2,\mathds{R}) \oplus \mathfrak{su}(2)$ together with two free bosons and the zero-modes from the continuous representation. 

The spectrally flowed character can be obtained from this by following the rules \eqref{eq:spectral flow a} -- \eqref{eq:spectral flow e} and \eqref{eq:spectral flow f}. This gives
\begin{align}
\mathrm{ch}[\sigma^w(\mathscr{F}_\lambda)](t,z;\tau)
%q^{-\frac{w^2}{4}} x^{\frac{w}{2}} \Bigg(
%\sum_{r \in \mathds{Z}+\lambda} x^r q^{-rw} \frac{\vartheta_3(2\tau)}{\eta(\tau)^4}\left\{\begin{matrix}
%\vartheta_3(z;2\tau) \\ \vartheta_2(z;2\tau)
%\end{matrix}\right\} \nonumber\\
%&\qquad+\sum_{r \in \mathds{Z}+\lambda+\frac{1}{2}} x^r q^{-rw} \frac{\vartheta_2(2\tau)}{\eta(\tau)^4}
%\left\{\begin{matrix}
%\vartheta_2(z;2\tau) \\ \vartheta_3(z;2\tau)
%\end{matrix}\right\}\Bigg)  \\
%&=q^{\frac{w^2}{4}} \Bigg(
%\sum_{r \in \mathds{Z}+\lambda+\frac{w}{2}} x^r q^{-rw} \frac{\vartheta_3(2\tau)}{\eta(\tau)^4}\left\{\begin{matrix}
%\vartheta_3(z;2\tau) \\ \vartheta_2(z;2\tau)
%\end{matrix}\right\} \nonumber\\
%&\qquad+\sum_{r \in \mathds{Z}+\lambda+\frac{w+1}{2}} x^r q^{-rw} \frac{\vartheta_2(2\tau)}{\eta(\tau)^4}
%\left\{\begin{matrix}
%\vartheta_2(z;2\tau) \\ \vartheta_3(z;2\tau)
%\end{matrix}\right\}\Bigg) \label{eq:character first version}\\
&=q^{\frac{w^2}{2}} \sum_{r \in \mathds{Z}+\lambda} x^r q^{-rw} \frac{\vartheta_2\big(\frac{t+z}{2};\tau\big)\vartheta_2\big(\frac{t-z}{2};\tau\big)}{\eta(\tau)^4}\ . \label{eq:character second version}
\end{align}
\subsection{Modular properties} \label{subapp:modular properties}
We now calculate the modular behaviour of the characters \eqref{eq:character second version}. We stress that we are treating the character as a formal object and not as a meromorphic function. Hence the following manipulations will be somewhat formal. Our calculations are inspired by \cite{Maldacena:2000hw,Creutzig:2012sd}. To obtain good modular properties, we include a $(-1)^{\mathrm{F}}$ into the character --- the new characters will be denoted by $\tilde{\mathrm{ch}}$ ---  which corresponds to the replacement $\vartheta_2 \longrightarrow \vartheta_1$ in \eqref{eq:character second version}. We first note that we can write (recall that $x=\mathrm{e}^{2\pi i t}$ and $q=\mathrm{e}^{2\pi i \tau}$)
\begin{align} 
\tilde{\mathrm{ch}}[\sigma^w(\mathscr{F}_\lambda)](t,z;\tau)&=q^{\frac{w^2}{2}} \mathrm{e}^{2\pi i(t-w \tau)\lambda}\sum_{r \in \mathds{Z}} \mathrm{e}^{2\pi i(t-w\tau)r} \, \frac{\vartheta_1\big(\frac{t+z}{2};\tau\big)\vartheta_1\big(\frac{t-z}{2};\tau\big)}{\eta(\tau)^4}\\
&=q^{\frac{w^2}{2}} \mathrm{e}^{2\pi i(t-w \tau)\lambda} \sum_{m \in \mathds{Z}}\delta(t-w \tau-m) \, \frac{\vartheta_1\big(\frac{t+z}{2};\tau\big)\vartheta_1\big(\frac{t-z}{2};\tau\big)}{\eta(\tau)^4} \\
&=q^{\frac{w^2}{2}} \sum_{m \in \mathds{Z}} \mathrm{e}^{2\pi i m \lambda} \delta(t-w \tau-m)\, \frac{\vartheta_1\big(\frac{t+z}{2};\tau\big)\vartheta_1\big(\frac{t-z}{2};\tau\big)}{\eta(\tau)^4}\ .
\end{align}
With this at hand, it is straightforward to compute the S-modular transformation of the characters from
\begin{align}
\tilde{\mathrm{ch}}[\sigma^w(\mathscr{F}_\lambda)](t,z;\tau)\rightarrow\ &\mathrm{e}^{\frac{\pi i}{2\tau}(t^2-z^2)}\tilde{\mathrm{ch}}[\sigma^w(\mathscr{F}_\lambda)]\big(\tfrac{t}{\tau},\tfrac{z}{\tau};-\tfrac{1}{\tau}\big) \\
=&\frac{\mathrm{e}^{\frac{\pi i}{\tau}(t^2-w^2)}}{i \tau}\sum_{m \in \mathds{Z}} \mathrm{e}^{2\pi i m \lambda} \delta \big(\tfrac{t+w-m \tau}{\tau}\big) \frac{\vartheta_1\big(\frac{t+z}{2};\tau\big)\vartheta_1\big(\frac{t-z}{2};\tau\big)}{\eta(\tau)^4} \\
=&-i\, \mathrm{sgn}(\mathrm{Re}(\tau)) \sum_{m \in \mathds{Z}} q^{\frac{m^2}{2}} \mathrm{e}^{2\pi i m \lambda} \delta(t+w-m \tau)\nonumber\\
&\qquad\times\frac{\vartheta_1\big(\frac{t+z}{2};\tau\big)\vartheta_1\big(\frac{t-z}{2};\tau\big)}{\eta(\tau)^4}\ . \label{eq:S modular transform character}
\end{align}
Here the prefactor $\mathrm{e}^{\frac{\pi i}{2\tau}(t^2-z^2)}$ comes from the general transformation property of weak Jacobi forms of index 1 and $-1$, respectively, see e.g.\ \cite{Gaberdiel:2012yb}, and we have used \eqref{eq:theta 1 transformation}. In the final step we have also set $t=m\tau -w$ (because of the $\delta$ function), and used that both $m$ and $w$ are integers. Finally, we have inserted the formal identity
\be 
\delta\big(\tfrac{x}{\tau}\big)=\tau\, \mathrm{sgn}(\mathrm{Re}(\tau))\, \delta(x)\ , \label{eq:delta reparametrization}
\ee
which follows by writing
\be 
\delta(x)=\lim_{\varepsilon \to 0} \frac{1}{\sqrt{2\pi \varepsilon}} \mathrm{e}^{-\frac{x^2}{2\varepsilon}}\ .
\ee
By definition, we put the branch cut of the square root on the imaginary axis, which is the reason for the jump in  \eqref{eq:delta reparametrization} at this point. Of course, other choices for the branch cut give in the end the same physical result.\footnote{In particular, our formula differs from \cite{Creutzig:2012sd}. We feel that the holomorphic prescription we are using is more adequate, since in particular no $\bar{\tau}$ should appear in the S-modular transformation of (formally) holomorphic characters.} In particular, the sign cancels out once we combine the left-movers with the right-movers.

The expression \eqref{eq:S modular transform character} can now be written as
\be 
\sum_{w'\in \mathds{Z}} \int_0^1 \mathrm{d}\lambda' \, S_{(w,\lambda),(w',\lambda')}\, \, \tilde{\mathrm{ch}}[\sigma^{w'}(\mathscr{F}_{\lambda'})](t,z;\tau)\ ,
\ee
with
\be 
S_{(w,\lambda),(w',\lambda')}=-i\, \mathrm{sgn}(\mathrm{Re}(\tau)) \, \mathrm{e}^{2\pi i(w'\lambda+w\lambda')}\ .
\ee
Thus, we have derived the formal S-matrix of the S-modular transformation, see eq.~(\ref{Smatrix}).
As in \cite{Creutzig:2012sd}, it is not entirely independent of $\tau$. This dependence cancels out in every physical calculation. The S-matrix is (formally) unitary since
\begin{align} 
\sum_{w''\in \mathds{Z}} \int_0^1 \mathrm{d}\lambda'' \, S^\dag_{(w,\lambda),(w'',\lambda'')} S_{(w'',\lambda''),(w',\lambda')}&=\sum_{w''\in \mathds{Z}} \int_0^1 \mathrm{d}\lambda'' \, \mathrm{e}^{2\pi i (w''\lambda'+w'\lambda''-w \lambda''-w''\lambda)} \\
&=\delta_{w,w'}\sum_{w''\in \mathds{Z}} \mathrm{e}^{2\pi i (w''\lambda'-w''\lambda)}\\
&=\delta_{w,w'}\, \delta(\lambda-\lambda'\bmod 1)\ .
\end{align}
Moreover, it is clearly symmetric. These properties suffice to deduce that the combination \eqref{eq:Hilbert space Grothendieck} is indeed modular invariant.

\subsection{The Verlinde formula} \label{subapp:verlinde}
We now use the formal S-matrix to derive the typical fusion rules a third time by using a continuum version of the Verlinde formula. For this, we also need the S-matrix element of the vacuum with a continuous representation. To this end, we notice that on the level of the Grothendieck ring
\be 
\sigma(\mathscr{F}_{1/2}) \sim\mathscr{L} \oplus 2\cdot \sigma(\mathscr{L} )\oplus \sigma^2(\mathscr{L})\ .
\ee
Using this identification repeatedly, one proves by induction
\begin{multline} 
\mathrm{ch}[\mathscr{L}]=\sum_{m=1}^n (-1)^{m+1} m\, \mathrm{ch}[\sigma^m(\mathscr{F}_{1/2})]\\+(-1)^n(n+1)\,\mathrm{ch}[\sigma^n(\mathscr{L})] +(-1)^n n\, \mathrm{ch}[\sigma^{n+1}(\mathscr{L})]
\end{multline}
for any $n$. Thus by taking $n \to \infty$ we conclude
\be 
\mathrm{ch}[\mathscr{L}]=\sum_{m=0}^\infty  (-1)^{m+1} m\, \mathrm{ch}[\sigma^{m}(\mathscr{F}_{1/2})]\ .
\ee
We can then calculate the S-matrix element
\be 
S_{\mathrm{vac},(w,\lambda)}=\sum_{m=0}^\infty (-1)^{m+1} m\, S_{(m,\frac{1}{2}),(w,\lambda)}=(-1)^w\frac{-i\, \mathrm{sgn}(\mathrm{Re}(\tau))}{\big(\mathrm{e}^{i\pi \lambda}+\mathrm{e}^{-i\pi \lambda}\big)^2}\ .
\ee
Finally, we use the Verlinde formula to calculate the fusion rules:
\begin{align} 
\mathcal{N}_{(w_1,\lambda_1)(w_2,\lambda_2)}^{(w_3,\lambda_3)}&=\sum_{w \in \mathds{Z}} \int_0^1 \mathrm{d}\lambda\ \frac{S_{(w_1,\lambda_1)(w,\lambda)} S_{(w_2,\lambda_2)(w,\lambda)}S_{(w_3,\lambda_3)(w,\lambda)}^*}{S_{\mathrm{vac},(w,\lambda)}} \\
&=\sum_{w\in \mathds{Z}} \int_0^1 \mathrm{d}\lambda\ \mathrm{e}^{2\pi i((w_1+w_2-w_3)\lambda+w(\lambda_1+\lambda_2-\lambda_3))} (-1)^w\big(\mathrm{e}^{\pi i \lambda}+\mathrm{e}^{-\pi i \lambda}\big)^2\\
&=\Big(\delta_{w_3,w_1+w_2-1}+2\delta_{w_3,w_1+w_2}+\delta_{w_3,w_1+w_2+1}\Big) \ \delta\Big(\lambda_3=\lambda_1+\lambda_2+\tfrac{1}{2}\Big)\ .
\end{align}
This reproduces \eqref{eq:psu112 fusion rules Grothendieck}.

\section{The indecomposable module $\boldsymbol{\mathscr{T}}$} \label{app:T details}
In this appendix, we discuss the indecomposable module $\mathscr{T}$ that appears in the fusion rules \eqref{eq:psu112 fusion rules a} -- \eqref{eq:psu112 fusion rules c} in some more detail. In particular, we discuss how it modifies the structure of the Hilbert space. The composition diagram of $\mathscr{T}$ was given in \eqref{eq:composition diagram T}. 

In a refined version of \eqref{eq:Hilbert space Grothendieck}, we should not include $\mathscr{F}_{1/2}$ in the Hilbert space, but rather $\mathscr{T}$. Thus, the naive ansatz for the Hilbert space would be
\be 
\mathcal{H}^\text{naive} \cong  \bigoplus_{w \in \mathds{Z}}\, \Bigg(\sigma^w(\mathscr{T}) \otimes \overline{\sigma^w(\mathscr{T})} \oplus \int\limits_{[0,1) \setminus\{\frac{1}{2}\}}\hspace{-.86cm}\boldsymbol{\oplus}\, \mathrm{d}\lambda \ \sigma^w \big(\mathscr{F}_\lambda\big) \otimes \overline{\sigma^w\big(\mathscr{F}_\lambda\big)}\Bigg)\ . \label{eq:Hilbert space naive}
\ee
We refer to the first summand as the atypical Hilbert space $\mathcal{H}_\text{atyp}^\text{naive}$, and the second term as the typical Hilbert space $\mathcal{H}_\text{typ}$.
While this contains now only modules which close under fusion, there are two problems with this proposal. First, locality requires that $L_0-\bar{L}_0$ acts diagonalisably, since otherwise the complete correlation functions would be multi-valued. In addition, \eqref{eq:Hilbert space naive} does not agree with \eqref{eq:Hilbert space Grothendieck} on the level of the Grothendieck ring, and hence would not be modular invariant. As explained in \cite{Quella:2007hr,Gaberdiel:2007jv}, the true Hilbert space is obtained by quotienting out an ideal $\mathcal{I} \subset \mathcal{H}_\text{atyp}^\text{naive}$ from $\mathcal{H}_\text{atyp}^\text{naive}$.

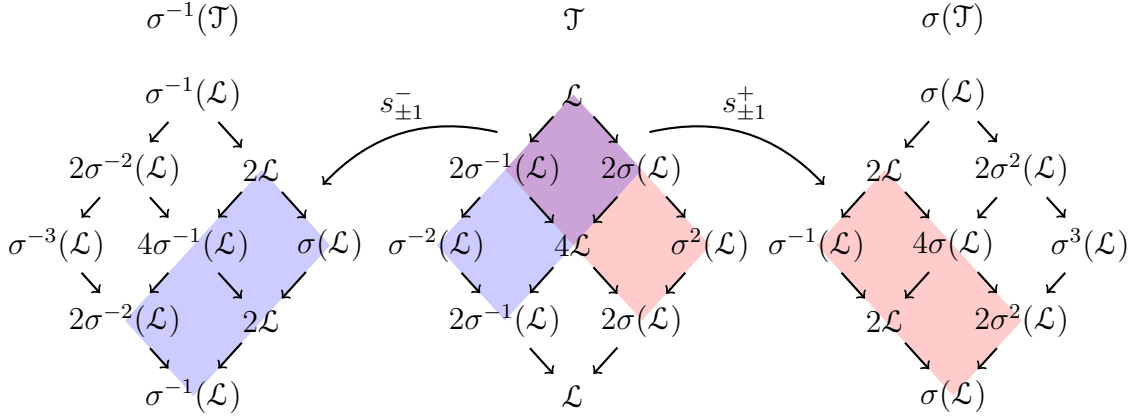
\begin{figure}
\begin{center}
\begin{tikzpicture}[baseline={([yshift=-.5ex]current bounding box.center)}]
\fill[red,opacity=.2] (0,2) -- (-.9,1) -- (.9,-1) -- (1.8,0); 
\fill[red,opacity=.2] (4.1,1) -- (3.2,0) -- (5,-2) -- (5.9,-1); 
\fill[blue,opacity=.2] (0,2) -- (.9,1) -- (-.9,-1) -- (-1.8,0); 
\fill[blue,opacity=.2] (-4.1,1) -- (-3.2,0) -- (-5,-2) -- (-5.9,-1); 
\node at (-5,3) {$\sigma^{-1}(\mathscr{T})$};
\node (top2) at (-5,2) {$\sigma^{-1}(\mathscr{L})$};
\node (right2) at (-3.2,0) {$\sigma(\mathscr{L})$};
\node (left2) at (-6.8,0) {$\sigma^{-3}(\mathscr{L})$};
\node (bottom2) at (-5,-2) {$\sigma^{-1}(\mathscr{L})$};
\node (topright2) at (-4.1,1) {$2 \mathscr{L}$};
\node (topleft2) at (-5.9,1) {$2 \sigma^{-2}(\mathscr{L})$};
\node (bottomright2) at (-4.1,-1) {$2 \mathscr{L}$};
\node (bottomleft2) at (-5.9,-1) {$2\sigma^{-2}(\mathscr{L})$};
\node (center2) at (-5,0) {$4\sigma^{-1}(\mathscr{L})$};
\draw[thick,->] (right2) to (bottomright2);
\draw[thick,->] (left2) to  (bottomleft2);
\draw[thick,->] (bottomright2) to (bottom2);
\draw[thick,->] (bottomleft2) to  (bottom2);
\draw[thick,->] (top2) to  (topleft2);
\draw[thick,->] (top2) to (topright2);
\draw[thick,->] (topright2) to  (right2);
\draw[thick,->] (topleft2) to (left2);
\draw[thick,->] (topleft2) to (center2);
\draw[thick,->] (topright2) to (center2);
\draw[thick,->] (center2) to (bottomleft2);
\draw[thick,->] (center2) to (bottomright2);
\node at (0,3) {$\mathscr{T}$};
\node (top) at (0,2) {$\mathscr{L}$};
\node (right) at (1.8,0) {$\sigma^2(\mathscr{L})$};
\node (left) at (-1.8,0) {$\sigma^{-2}(\mathscr{L})$};
\node (bottom) at (0,-2) {$\mathscr{L}$};
\node (topright) at (.9,1) {$2 \sigma(\mathscr{L})$};
\node (topleft) at (-.9,1) {$2 \sigma^{-1}(\mathscr{L})$};
\node (bottomright) at (.9,-1) {$2 \sigma(\mathscr{L})$};
\node (bottomleft) at (-.9,-1) {$2 \sigma^{-1}(\mathscr{L})$};
\node (center) at (0,0) {$4\mathscr{L}$};
\draw[thick,->] (right) to (bottomright);
\draw[thick,->] (left) to  (bottomleft);
\draw[thick,->] (bottomright) to (bottom);
\draw[thick,->] (bottomleft) to  (bottom);
\draw[thick,->] (top) to  (topleft);
\draw[thick,->] (top) to (topright);
\draw[thick,->] (topright) to  (right);
\draw[thick,->] (topleft) to (left);
\draw[thick,->] (topleft) to (center);
\draw[thick,->] (topright) to (center);
\draw[thick,->] (center) to (bottomleft);
\draw[thick,->] (center) to (bottomright);
\node at (5,3) {$\sigma(\mathscr{T})$};
\node (top1) at (5,2) {$\sigma(\mathscr{L})$};
\node (right1) at (6.8,0) {$\sigma^3(\mathscr{L})$};
\node (left1) at (3.2,0) {$\sigma^{-1}(\mathscr{L})$};
\node (bottom1) at (5,-2) {$\sigma(\mathscr{L})$};
\node (topright1) at (5.9,1) {$2 \sigma^2(\mathscr{L})$};
\node (topleft1) at (4.1,1) {$2 \mathscr{L}$};
\node (bottomright1) at (5.9,-1) {$2 \sigma^2(\mathscr{L})$};
\node (bottomleft1) at (4.1,-1) {$2\mathscr{L}$};
\node (center1) at (5,0) {$4\sigma(\mathscr{L})$};
\draw[thick,->] (right1) to (bottomright1);
\draw[thick,->] (left1) to  (bottomleft1);
\draw[thick,->] (bottomright1) to (bottom1);
\draw[thick,->] (bottomleft1) to  (bottom1);
\draw[thick,->] (top1) to  (topleft1);
\draw[thick,->] (top1) to (topright1);
\draw[thick,->] (topright1) to  (right1);
\draw[thick,->] (topleft1) to (left1);
\draw[thick,->] (topleft1) to (center1);
\draw[thick,->] (topright1) to (center1);
\draw[thick,->] (center1) to (bottomleft1);
\draw[thick,->] (center1) to (bottomright1);

\draw[thick,->,bend left=30] (1,1.5) to node[above] {$s_{\pm 1}^+$} (3.3,.8); 
\draw[thick,->,bend right=30] (-1,1.5) to node[above] {$s_{\pm 1}^-$} (-3.3,.8); 
\end{tikzpicture} 
\end{center}
\caption{The definition of the maps $s^\pm_{(i)}$.} \label{fig:definition of spm}
\end{figure} 

There is a very natural way of defining this ideal \cite{Gaberdiel:2011vf}. For this, let us describe the module $\mathscr{T}$ a bit more conceptually. $\mathscr{T}$ has 16 terms in its composition series, which we shall denote by $(\varepsilon_1,\varepsilon_2,\varepsilon_3,\varepsilon_4)$ for $\varepsilon_i \in \{0,1\}$. The first line of the composition series \eqref{eq:composition diagram T} corresponds to $(0,0,0,0)$, the second line to $(\varepsilon_1,\varepsilon_2,\varepsilon_3,\varepsilon_4)$ with $\sum_i \varepsilon_i=1$, the third line to $(\varepsilon_1,\varepsilon_2,\varepsilon_3,\varepsilon_4)$ with $\sum_i \varepsilon_i=2$, etc. $(\varepsilon_1,\varepsilon_2,\varepsilon_3,\varepsilon_4)$ will correspond to a term $\sigma^{-\varepsilon_1-\varepsilon_2+\varepsilon_3+\varepsilon_4}(\mathscr{L})$ in the composition series \eqref{eq:composition diagram T}. It is useful to introduce a $\mathds{Z}$-grading of the terms in the composition series to lift the degeneracy of the various modules, see \cite{Gaberdiel:2011vf}; it takes the form 
\be 
\mathrm{grad}(\varepsilon_1,\varepsilon_2,\varepsilon_3,\varepsilon_4)=-\varepsilon_1+\varepsilon_2-\varepsilon_3+\varepsilon_4\ .
\ee
Furthermore, we note that there is an arrow in the composition series \eqref{eq:composition diagram T} from $(\varepsilon_1,\varepsilon_2,\varepsilon_3,\varepsilon_4)$ to $(\varepsilon_1',\varepsilon_2',\varepsilon_3',\varepsilon_4')$ if $\varepsilon_i'=\varepsilon_i+1$ for exactly one $i$ and $\varepsilon_j'=\varepsilon_j$ for all remaining $j \ne i$.\footnote{Geometrically, this corresponds to drawing a 4-dimensional hypercube, where the modules sit on the vertices and the arrows represent the edges. They are given by all possible shortest paths from one vertex to the opposite vertex.} We can define the intertwiner maps 
\be 
s^\pm_{\rho}: \sigma^w(\mathscr{T}) \to \sigma^{w\pm 1}(\mathscr{T})\ ,
\ee
where $\rho\in \{-1,1\}$ denotes the grade of the map, by
\begin{subequations}
\begin{align}
s^+_{-1}\sigma^w(\varepsilon_1,\varepsilon_2,\varepsilon_3,\varepsilon_4)&=\sigma^{w+1}(\varepsilon_1+1,\varepsilon_2,\varepsilon_3,\varepsilon_4)\ , \\
s^+_{+1}\sigma^w(\varepsilon_1,\varepsilon_2,\varepsilon_3,\varepsilon_4)&=\sigma^{w+1}(\varepsilon_1,\varepsilon_2+1,\varepsilon_3,\varepsilon_4)\ , \\
s^-_{-1}\sigma^w(\varepsilon_1,\varepsilon_2,\varepsilon_3,\varepsilon_4)&=\sigma^{w-1}(\varepsilon_1,\varepsilon_2,\varepsilon_3+1,\varepsilon_4)\ , \\
s^-_{+1}\sigma^w(\varepsilon_1,\varepsilon_2,\varepsilon_3,\varepsilon_4)&=\sigma^{w-1}(\varepsilon_1,\varepsilon_2,\varepsilon_3,\varepsilon_4+1)\ .
\end{align}
\end{subequations}
By definition, the right hand side is zero if one of its argument is bigger than one.
These maps preserve the grading, 
%the amount of spectral flow of the module $\sigma^w(\varepsilon_1,\varepsilon_2,\varepsilon_3,\varepsilon_4)$ and its grading.  
and we have given a graphical representation in Figure~\ref{fig:definition of spm}. 
In fact, they generate the long exact sequences
\be 
\begin{tikzpicture}[baseline={([yshift=-.5ex]current bounding box.center)}]
\node (p1) at (-6,0) {$\sigma^{w-1}(\mathscr{T})$};
\node (p2) at (-3,0) {$\sigma^{w}(\mathscr{T})$};
\node (p3) at (0,0) {$\sigma^{w+1}(\mathscr{T})$};
\node (p4) at (3,0) {$\sigma^{w+2}(\mathscr{T})$};
\draw[thick,->,bend left=30] (p1) to node[above] {$s^+_{\rho}$} (p2);
\draw[thick,->,bend left=30] (p2) to node[above] {$s^+_{\rho}$} (p3);
\draw[thick,->,bend left=30] (p3) to node[above] {$s^+_{\rho}$} (p4);
%\draw[thick,->,bend left=30] (p4) to node[above] {$s^+$} (p5);
%\draw[thick,->,bend left=30] (p5) to node[below] {$s^-$} (p4);
\draw[thick,->,bend left=30] (p4) to node[below] {$s^-_{\rho}$} (p3);
\draw[thick,->,bend left=30] (p3) to node[below] {$s^-_{\rho}$} (p2);
\draw[thick,->,bend left=30] (p2) to node[below] {$s^-_{\rho}$} (p1);
\node at (-7.5,0) {$\cdots$};
\node at (4.5,0) {$\cdots$};
\end{tikzpicture}
\ee
We stress that the maps are intertwiners, i.e.\ morphisms of $\mathfrak{psu}(1,1|2)_1$ modules. With this at hand, we define the ideal $\mathcal{I}$ as the ideal generated by $\mathcal{I}^\pm_\rho$ with
\be 
\mathcal{I}^\pm_\rho \equiv \bigoplus_{w \in \mathds{Z}}\, \Big(s^\pm_{\rho} \otimes \overline{\mathds{1}} -\mathds{1} \otimes \overline{s^\mp_{-\rho}} \Big) \big(\sigma^w(\mathscr{T}) \otimes \sigma^{w\pm 1}(\mathscr{T})\big)\ .
\ee
Then the Hilbert spaces becomes 
\be 
\mathcal{H}\equiv \mathcal{H}_\text{atyp}\oplus \mathcal{H}_\text{typ} \ , \qquad \hbox{with} \qquad 
\mathcal{H}_\text{atyp} \equiv \mathcal{H}^\text{naive}_\text{atyp}/\mathcal{I}\ .
\ee
Hence we have the equivalence relations
\begin{subequations}
\begin{align}
\sigma^w(\varepsilon_1+1,\varepsilon_2,\varepsilon_3,\varepsilon_4;\bar{\varepsilon}_1,\bar{\varepsilon}_2,\bar{\varepsilon}_3,\bar{\varepsilon}_4) \sim \sigma^{w-1}(\varepsilon_1,\varepsilon_2,\varepsilon_3,\varepsilon_4;\bar{\varepsilon}_1,\bar{\varepsilon}_2,\bar{\varepsilon}_3,\bar{\varepsilon}_4+1)\ , \label{eq:T equivalence relation a}\\
\sigma^w(\varepsilon_1,\varepsilon_2+1,\varepsilon_3,\varepsilon_4;\bar{\varepsilon}_1,\bar{\varepsilon}_2,\bar{\varepsilon}_3,\bar{\varepsilon}_4) \sim \sigma^{w-1}(\varepsilon_1,\varepsilon_2,\varepsilon_3,\varepsilon_4;\bar{\varepsilon}_1,\bar{\varepsilon}_2,\bar{\varepsilon}_3+1,\bar{\varepsilon}_4)\ , \label{eq:T equivalence relation b}\\
\sigma^w(\varepsilon_1,\varepsilon_2,\varepsilon_3+1,\varepsilon_4;\bar{\varepsilon}_1,\bar{\varepsilon}_2,\bar{\varepsilon}_3,\bar{\varepsilon}_4) \sim \sigma^{w+1}(\varepsilon_1,\varepsilon_2,\varepsilon_3,\varepsilon_4;\bar{\varepsilon}_1,\bar{\varepsilon}_2+1,\bar{\varepsilon}_3,\bar{\varepsilon}_4)\ , \label{eq:T equivalence relation c}\\
\sigma^w(\varepsilon_1,\varepsilon_2,\varepsilon_3,\varepsilon_4+1;\bar{\varepsilon}_1,\bar{\varepsilon}_2,\bar{\varepsilon}_3,\bar{\varepsilon}_4) \sim \sigma^{w+1}(\varepsilon_1,\varepsilon_2,\varepsilon_3,\varepsilon_4;\bar{\varepsilon}_1+1,\bar{\varepsilon}_2,\bar{\varepsilon}_3,\bar{\varepsilon}_4)\ , \label{eq:T equivalence relation d}
\end{align}
\end{subequations}
where we have included the right-moving modules in a hopefully obvious way.
In particular, these relations imply that the following modules are null:
\begin{align}
&\sigma^w(1,\varepsilon_2,\varepsilon_3,\varepsilon_4;\bar{\varepsilon}_1,\bar{\varepsilon}_2,\bar{\varepsilon}_3,1)\sim \sigma^w(\varepsilon_1,1,\varepsilon_3,\varepsilon_4;\bar{\varepsilon}_1,\bar{\varepsilon}_2,1,\bar{\varepsilon}_4)\\
 & \sim \ \sigma^w(\varepsilon_1,\varepsilon_2,1,\varepsilon_4;\bar{\varepsilon}_1,1,\bar{\varepsilon}_3,\varepsilon_4)\sim \sigma^w(\varepsilon_1,\varepsilon_2,\varepsilon_3,1;1,\bar{\varepsilon}_2,\bar{\varepsilon}_3,\bar{\varepsilon}_4)\sim 0\ .
\end{align}
Thus, out of the 256 terms in the composition series of $\mathscr{T} \times \overline{\mathscr{T}}$, 175 are set to zero. 
If $\varepsilon_1=0$ and $\bar{\varepsilon}_4=1$ or vice versa, \eqref{eq:T equivalence relation a} tells us that the values of $\varepsilon_1$ and $\bar{\varepsilon}_4$ can be interchanged, if we compensate in the spectral flow,
\be 
\sigma^w(1,\varepsilon_2,\varepsilon_3,\varepsilon_4;\bar{\varepsilon}_1,\bar{\varepsilon}_2,\bar{\varepsilon}_3,0) \sim \sigma^{w-1}(0,\varepsilon_2,\varepsilon_3,\varepsilon_4;\bar{\varepsilon}_1,\bar{\varepsilon}_2,\bar{\varepsilon}_3,1)\ .
\ee
Hence, we can choose a representative by fixing either $\varepsilon_1=0$ or $\bar{\varepsilon}_4=0$.
An analogous statement holds for the other pairs of indices from \eqref{eq:T equivalence relation b}--\eqref{eq:T equivalence relation d}. In total, the `gauge freedom' is fixed by setting either $\varepsilon_i=0$ or $\bar{\varepsilon}_{5-i}=0$ for $i=1,\dots,4$.

To see the structure of the resulting representation of $\mathfrak{psu}(1,1|2)_1 \times \mathfrak{psu}(1,1|2)_1$ most clearly, we first fix the representatives such that $\bar{\varepsilon}_i=0$ for $i=1,\dots,4$. Thus, equivalence classes will be labelled by $[\sigma^w(\varepsilon_1,\varepsilon_2,\varepsilon_3,\varepsilon_4)]$. The left-moving action acts on these equivalence classes in the obvious way. 
%The right-moving action has the following structure. 
There is an arrow for the right-moving action
\be 
[\sigma^w(\varepsilon_1,\varepsilon_2,\varepsilon_3,\varepsilon_4)] \longrightarrow [\sigma^{w'}(\varepsilon_1',\varepsilon_2',\varepsilon_3',\varepsilon_4')]
\ee
if $\varepsilon_i'=\varepsilon_i+1$ for some $i$ and $\varepsilon_j'=\varepsilon_j$ for the other $j \ne i$. Moreover, $w'=w+1$ for $i\in \{1,2\}$ and $w'=w-1$ for $i\in \{3,4\}$. Obviously, we can also choose ${\varepsilon}_i=0$ for $i=1,\dots,4$, and obtain the same structure with the roles of left- and right-movers interchanged. Thus, the structure of the atypical Hilbert space is 
\be 
\mathcal{H}_\text{atyp}\cong\bigoplus_{w \in \mathds{Z}}\, \sigma^w(\mathscr{T})\otimes \sigma^w(\mathscr{L})\qquad\text{and}\qquad \mathcal{H}_\text{atyp}\cong\bigoplus_{w \in \mathds{Z}}\, \sigma^w(\mathscr{L})\otimes \sigma^w(\mathscr{T})
\ee
with respect to the left (right) action of $\mathfrak{psu}(1,1|2)_1$. The module $\sigma^w(\mathscr{L})$ appearing here is always the head of the module $\sigma^w(\mathscr{T})$.

To investigate the structure of physical states in string theory, it is more convenient to choose the gauge as $\varepsilon_2=\varepsilon_4=\bar{\varepsilon}_2=\bar{\varepsilon}_4=0$. Then equivalence classes of terms are labelled by $[\sigma^w(\varepsilon_1,\varepsilon_3;\bar{\varepsilon}_1,\bar{\varepsilon}_3)]$. Then the structure is of the form 
\be 
\begin{tikzpicture}[baseline={([yshift=-.5ex]current bounding box.center)}]
\node (top) at (0,1.5) {$\sigma^w(\mathscr{L})$};
\node (left) at (-1.5,0) {$\sigma^{w-1}(\mathscr{L})$};
\node (right) at (1.5,0) {$\sigma^{w+1}(\mathscr{L})$};
\node (bottom) at (0,-1.5) {$\sigma^w(\mathscr{L})$};
\draw[thick,->] (top) to (right);
\draw[thick,->] (top) to (left);
\draw[thick,->] (left) to (bottom);
\draw[thick,->] (right) to (bottom);
\end{tikzpicture}
\quad
\otimes 
\quad
\begin{tikzpicture}[baseline={([yshift=-.5ex]current bounding box.center)}]
\node (top) at (0,1.5) {$\sigma^w(\mathscr{L})$};
\node (left) at (-1.5,0) {$\sigma^{w-1}(\mathscr{L})$};
\node (right) at (1.5,0) {$\sigma^{w+1}(\mathscr{L})$};
\node (bottom) at (0,-1.5) {$\sigma^w(\mathscr{L})$};
\draw[thick,->] (top) to (right);
\draw[thick,->] (top) to (left);
\draw[thick,->] (left) to (bottom);
\draw[thick,->] (right) to (bottom);
\end{tikzpicture}\ , \label{eq:symmetric gauge}
\ee
where we drew only the `obvious' arrows of the $\mathfrak{psu}(1,1|2)_1 \times \mathfrak{psu}(1,1|2)_1$ action. Since on the level of the Grothendieck ring, $\mathscr{F}_{1/2} \sim \sigma^{-1}(\mathscr{L}) \oplus 2 \mathscr{L} \oplus \sigma(\mathscr{L})$, this equals $\sigma^w(\mathscr{F}_{1/2}) \otimes \sigma^w(\mathscr{F}_{1/2})$ in the Grothendieck ring.
Thus, $\mathscr{T}$ becomes the moral analogue of $\mathscr{F}_{1/2}$ and in particular the character analysis does not differ in the atypical case from the typical case.
Summarizing,
\be \label{C.8}
\mathcal{H}_\text{atyp} \sim \bigoplus_{w \in \mathds{Z}}\, \sigma^w(\mathscr{F}_{1/2}) \otimes \overline{\sigma^w(\mathscr{F}_{1/2})}\ ,
\ee
i.e.\ the quotient has removed the factor of $16=4 \times 4$ that was mentioned below eq.~(\ref{Grothendieck}).
Thus the atypical contribution precisely fills the gaps of the typical contribution in \eqref{eq:Hilbert space naive}, so that in total we retrieve \eqref{eq:Hilbert space Grothendieck}, which is modular invariant.
Finally, one may check that $L_0-\bar{L}_0$ now acts indeed diagonalisably, and thus correlation functions are single-valued. The resulting Hilbert space therefore defines a local consistent CFT.

%Similarly, the module $\sigma^{-1}(\mathscr{T})$ yields the left-moving physical modules $\mathscr{L}$, $2 \cdot \sigma^{-1}(\mathscr{L})$ and $\sigma^{-2}(\mathscr{L})$, which yield the out-states of the chiral fields we discussed above. 
%Furthermore, the gauge-fixed module $\mathscr{T}$ consists of the modules $\sigma^{-1}(\mathscr{L})$, $2 \cdot \mathscr{L}$ and $\sigma(\mathscr{L})$, which is equivalent to $\mathscr{F}_{1/2}$ on the Grothendieck ring. The ground states of this module satisfy the mass-shell condition and hence survive in string theory. This is the $\lambda=\tfrac{1}{2}$ analogue of the additional states from the unflowed sector we mentioned in Section~\ref{sec:fusion rules dual CFT}.

\section{Physical states of the hybrid formalism for $\boldsymbol{k \ge 2}$} \label{app:physical state conditions}
Let us spell out the arguments leading to \eqref{eq:ghost partition function} for $k \ge 2$. The RNS character for the continuous representation (without insertion of $(-1)^{\mathrm{F}}$) is obtained as follows. We have the contributions from the bosonic pieces $\mathfrak{sl}(2,\mathds{R})_{k+2}$, $\mathfrak{su}(2)_{k-2}$ and the $\mathbb{T}^4$ partition function. Combining them gives
\be 
Z_\text{bos}(t,z;\tau)=\left|\sum_{m \in \mathds{Z}+\lambda} x^m \frac{\mathrm{ch}[\mathscr{M}_\ell^{k-2}](z;\tau)}{\eta(\tau)^7}\right|^2 \Theta_{\mathbb{T}^4}(\tau)\ ,
\ee
where we have denoted the spin-$\ell$ module of $\mathfrak{su}(2)_{k-2}$ by $\mathscr{M}^{k-2}_\ell$.
Imposing physical state conditions amounts to multiplying this by $\left|\eta(\tau)^2\right|^2$, which removes two neutral oscillators. Similarly, we obtain the fermionic contribution. Summing over spin structures and using the Jacobi abstruse identity \eqref{eq:Jacobi abstruse identity} we find for the physical contribution of the fermions
\be 
Z_\text{ferm}^\text{(phys)}(t,z;\tau)=\left| \frac{\vartheta_2(\frac{z+t}{2};\tau)^2 \vartheta_2(\frac{z-t}{2};\tau)^2}{\eta(\tau)^4}\right|^2\ .
\ee
Combining the two results, we thus obtain for the complete physical contribution
\begin{align} 
Z^{\text{(phys)}}(t,z;\tau)&=\left|\sum_{m \in \mathds{Z}+\lambda} x^m \frac{\mathrm{ch}[\mathscr{M}_\ell^{k-2}](z;\tau)\vartheta_2(\frac{z+t}{2};\tau)^2 \vartheta_2(\frac{z-t}{2};\tau)^2}{\eta(\tau)^9}\right|^2 \Theta_{\mathbb{T}^4}(\tau)\\
&=\left|\sum_{m \in \mathds{Z}+\lambda} x^m \frac{\mathrm{ch}[\mathscr{M}_\ell^{k-2}](z;\tau)\vartheta_2(\frac{z-t}{2};\tau)\vartheta_2(\frac{z+t}{2};\tau)}{\eta(\tau)^3}\right|^2 Z_{\mathbb{T}^4}^\text{R}(z,t;\tau) \ . \label{eq:RNS formalism physical contribution}
\end{align}
Here we have inserted the partition function of $\mathbb{T}^4$ (in the R-sector with $(-1)^\mathrm{F}$ insertions, or alternatively the topologically twisted partition function).

Let us compare this to the hybrid formalism. The $\mathfrak{psu}(1,1|2)_k$-characters for $k \ge 2$ can be found in \cite{Gotz:2006qp}. They are essentially based on the equivalence \cite{Bars:1990hx, Berkovits:1999im, Gotz:2006qp} 
\be 
\mathfrak{psu}(1,1|2)_k \cong \mathfrak{sl}(2,\mathds{R})_{k+2} \oplus \mathfrak{su}(2)_{k-2} \oplus \text{8 topologically twisted fermions}\ .
\ee
Correspondingly, the continuous character contribution to the partition function reads in the hybrid formalism
\be 
Z(t,z;\tau)=\left|\sum_{m \in \mathds{Z}+\lambda} x^m \frac{\mathrm{ch}[\mathscr{M}_\ell^{k-2}](z;\tau)\vartheta_2(\frac{z-t}{2};\tau)^2\vartheta_2(\frac{z+t}{2};\tau)^2}{\eta(\tau)^7}\right|^2 Z_{\mathbb{T}^4}^\text{R}(z,t;\tau) \ . \label{eq:hybrid formalism unphysical contribution}
\ee
Comparing \eqref{eq:hybrid formalism unphysical contribution} and \eqref{eq:RNS formalism physical contribution}, we thus conclude that the ghost contribution in the hybrid formalism amounts to \eqref{eq:ghost partition function}. As it should be, this is independent of the character we are considering and continues to hold also in the spectrally flowed sectors. 
\smallskip

Alternatively, one can deduce the result for the ghost contribution also directly from the hybrid formalism. The $\rho$-ghost of the hybrid formalism together with a pair of topologically twisted fermions can be `rebosonised', meaning that these are precisely the fields needed to bosonise a $\beta\gamma$-system with $\lambda=2$. The contribution of this $\beta\gamma$-system to the partition function can be computed and cancels another pair of topologically twisted fermions. Thus, the $\rho$-ghost cancels in total two pairs of topologically twisted fermions. The $\sigma$-ghost is the bosonised conformal ghost and cancels as always two bosonic oscillators. This again reproduces \eqref{eq:ghost partition function}.

\section{Theta functions}\label{app:theta}
We follow the notation of \cite{Blumenhagen:2013fgp} and define the theta functions as
\begin{align}
\vartheta\!\begin{bmatrix}
  \alpha \\
  \beta
  \end{bmatrix}(z;\tau)&\equiv\sum_{n\in\mathds Z}\mathrm{e}^{\pi i(n+\alpha)^2\tau+2\pi i(n+\alpha)(z+\beta)}\\
  &=\mathrm{e}^{2\pi i\alpha (z+\beta)} q^{\frac{\alpha^2}{2}} \prod_{n=1}^\infty \big(1-q^n \big) \big(1+q^{n+\alpha-\frac{1}{2}} \mathrm{e}^{2\pi i(z+\beta)} \big)\big(1+q^{n-\alpha-\frac{1}{2}} \mathrm{e}^{-2\pi i(z+\beta)} \big)\ .
\end{align}
The four Jacobi theta functions are the special cases
\be
\vartheta_1\equiv\vartheta\!\begin{bmatrix}
  \frac{1}{2} \\
  \frac{1}{2}
  \end{bmatrix}\ ,\qquad
\vartheta_2\equiv\vartheta\!\begin{bmatrix}
  \frac{1}{2} \\
  0
  \end{bmatrix}\ ,\qquad
\vartheta_3\equiv\vartheta\!\begin{bmatrix}
  0 \\
  0
  \end{bmatrix}\ ,\qquad
\vartheta_4\equiv\vartheta\!\begin{bmatrix}
  0 \\
  \frac{1}{2}
  \end{bmatrix}\ .
\ee
In particular, we have the identity 
\be 
\vartheta_2\big(\tfrac{z+t}{2};\tau)\vartheta_2\big(\tfrac{z-t}{2};\tau\big)=\vartheta_2(z;2\tau)\vartheta_3(t;2\tau)+\vartheta_3(z;2\tau)\vartheta_2(z;2\tau)\ , \label{eq:equivalence four fermions and su2su2}
\ee
which expresses the equivalence of four free fermions to $\mathfrak{su}(2)_1 \oplus \mathfrak{su}(2)_1$.

We also make use of the behaviour under modular transformations, in particular of $\vartheta_1(z,\tau)$ and $\eta(\tau)$, 
\be 
\vartheta_1\left(\frac{z}{\tau};-\frac{1}{\tau}\right)=-i \sqrt{-i \tau} \, \mathrm{e}^{\frac{\pi i z^2}{\tau}} \vartheta_1(z;\tau)\  , \qquad 
\eta\left(-\frac{1}{\tau}\right) =  \sqrt{-i \tau} \, \eta(\tau) \ .
 \label{eq:theta 1 transformation}
\ee
Finally, we use the Jacobi abstruse identity in Appendix~\ref{app:physical state conditions}
\begin{multline}
\frac{1}{2}\Big(\vartheta_3(t;\tau)\vartheta_3(z;\tau)\vartheta_3(\tau)^2-\vartheta_4(t;\tau)\vartheta_4(z;\tau)\vartheta_4(\tau)^2+\vartheta_2(t;\tau)\vartheta_2(z;\tau)\vartheta_2(\tau)^2\Big)\\
=\vartheta_2\big(\tfrac{t+z}{2};\tau\big)^2\vartheta_2\big(\tfrac{t-z}{2};\tau\big)^2 \ . \label{eq:Jacobi abstruse identity}
\end{multline}

\end{document}